\documentclass[10pt]{article}

\usepackage[left=.75in,right=.75in,top=.75in,bottom=.75in]{geometry}

\usepackage[utf8]{inputenc}

\usepackage{amssymb,amsmath,latexsym,amsthm,amsfonts,mathtools}

\usepackage{mathrsfs}

\usepackage{multirow}
\usepackage{graphicx}
\usepackage{textcomp,siunitx}
\usepackage{float}
\usepackage{subcaption}

\usepackage{booktabs}

\usepackage[hyperref]{xcolor}
\definecolor{ao(english)}{rgb}{0.0, 0.5, 0.0}

\usepackage{hyperref}
\hypersetup{colorlinks, breaklinks, citecolor=blue, linkcolor=ao(english), urlcolor=blue}

\usepackage[nameinlink,noabbrev,sort&compress]{cleveref}

\crefname{assumption}{Assumption}{Assumptions}

\usepackage{setspace}

\theoremstyle{plain}% Theorem-like structures provided by amsthm.sty
\newtheorem{theorem}{Theorem}%[section]
\newtheorem{lemma}{Lemma}

\newtheorem{objective}{Objective}

\theoremstyle{remark}
\newtheorem{remark}{Remark}

\theoremstyle{definition}
\newcommand{\sign}{\mathrm{sign}}

\usepackage{scalerel}
\usepackage{tikz}
\usetikzlibrary{svg.path}

\definecolor{orcidlogocol}{HTML}{A6CE39}
\tikzset{
	orcidlogo/.pic={
		\fill[orcidlogocol] svg{M256,128c0,70.7-57.3,128-128,128C57.3,256,0,198.7,0,128C0,57.3,57.3,0,128,0C198.7,0,256,57.3,256,128z};
		\fill[white] svg{M86.3,186.2H70.9V79.1h15.4v48.4V186.2z}
		svg{M108.9,79.1h41.6c39.6,0,57,28.3,57,53.6c0,27.5-21.5,53.6-56.8,53.6h-41.8V79.1z M124.3,172.4h24.5c34.9,0,42.9-26.5,42.9-39.7c0-21.5-13.7-39.7-43.7-39.7h-23.7V172.4z}
		svg{M88.7,56.8c0,5.5-4.5,10.1-10.1,10.1c-5.6,0-10.1-4.6-10.1-10.1c0-5.6,4.5-10.1,10.1-10.1C84.2,46.7,88.7,51.3,88.7,56.8z};
	}
}

\newcommand\orcidicon[1]{\href{https://orcid.org/#1}{\mbox{\scalerel*{
				\begin{tikzpicture}[yscale=-1,transform shape]
					\pic{orcidlogo};
				\end{tikzpicture}
			}{|}}}}

\title{Three-agent Time-constrained Cooperative Pursuit-Evasion}

\author{Abhinav~Sinha\textsuperscript{\orcidicon{0000-0001-6419-2353}}
	\thanks{Corresponding author.\newline A. Sinha and S.R. Kumar are with the Intelligent Systems \& Control Lab, Department of Aerospace Engineering, Indian Institute of Technology Bombay, Powai- 400076, Mumbai, India. e-mails: \{abhinavsinha,srk\}@aero.iitb.ac.in.\newline D. Mukherjee is with Department of Electrical Engineering, Indian Institute of Technology Bombay, Powai- 400076, Mumbai, India. e-mail: dm@ee.iitb.ac.in.\newline \textbf{The Version of Record of this article is published in the Journal of Intelligent \& Robotic Systems, and is available online at https://doi.org/10.1007/s10846-022-01570-y}}	
	\and Shashi~Ranjan~Kumar\textsuperscript{\orcidicon{0000-0001-6446-7281}}
	\and Dwaipayan~Mukherjee\textsuperscript{\orcidicon{0000-0001-6993-9305}}
}

\date{}

\begin{document}

\maketitle
\doublespacing

\begin{abstract}
	This paper considers a pursuit-evasion scenario among three agents-- an evader, a pursuer, and a defender. We design cooperative guidance laws for the evader and the defender team to safeguard the evader from an attacking pursuer. Unlike differential games, optimal control formulations, and other heuristic methods, we propose a novel perspective on designing effective nonlinear feedback control laws for the evader-defender team using a time-constrained guidance approach. The evader lures the pursuer on the collision course by offering itself as bait. At the same time, the defender protects the evader from the pursuer by exercising control over the engagement duration. Depending on the nature of the mission, the defender may choose to take an aggressive or defensive stance. Such consideration widens the applicability of the proposed methods in various three-agent motion planning scenarios such as aircraft defense, asset guarding, search and rescue, surveillance, and secure transportation. We use a fixed-time sliding mode control strategy to design the control laws for the evader-defender team and a nonlinear finite-time disturbance observer to estimate the pursuer's maneuver. Finally, we present simulations to demonstrate favorable performance under various engagement geometries, thus vindicating the efficacy of the proposed designs.\medskip
	
\end{abstract}

%\keywords{
	%Pursuit-evasion, Unmanned vehicles, Impact time control guidance, Time-constrained cooperative control, Multiagent systems.}

\section{Introduction}\label{sec:introduction}
There has been a rapid increase in the use of unmanned vehicles with various levels of autonomy. With a rise in their operational utility, unmanned vehicles are now being deployed as teams in missions such as guarding a territory \cite{Raslan2016}, border patrol \cite{Lau2014}, target capture \cite{8588382,Manzoor2017}, and pursuit-evasion \cite{Zhu2020,4660318,1512346}. Due to a multitude of possible applications and variations, the problem of multi-vehicular pursuit-evasion remains an important aspect in the areas of robotics, aerospace, and control engineering.

Asset defending scenarios describe a setting wherein a protective unmanned vehicle (the \emph{defender}) is tasked to safeguard a high-value asset (the \emph{evader}) against an incoming threat (the \emph{pursuer}). In essence, the evader and the defender are a team and the pursuer is the opposition. In the event the protective unmanned vehicle is captured by the incoming threat, the relative cost incurred is much lower than the loss of the high-value asset or the failure of the mission. The high-value asset may be a stationary point or a mobile vehicle. Significant efforts have been devoted to extend the protection capabilities of the high-value asset, which may respond to the incoming threat by performing evasive maneuvers, deploying electronic countermeasures, flares, and decoys. While the aforementioned survival tactics employed by the high-value asset are passive, an active protection strategy would involve launching a defender against the adversary. Having such a protective strategy is useful, particularly in scenarios where a mobile high-value asset has some constraints on its route or carries a load, such as goods or people, search and rescue, etc. 

Depending on the mission, the protective vehicle may assume a defensive or an aggressive role. For instance, scenarios involving asset guarding may require the protective vehicle to shield the high-value asset against an incoming threat by reaching the asset before the adversary can do that. In this situation, the protective vehicle may take a defensive stance. On the other hand, if the incoming threat is required to be neutralized before it can come in the vicinity of the high-value asset, then the protective vehicle may take an aggressive role. The resulting multi-body engagement scenarios \cite{LIANG2020105529,SHALUMOV2020105996,YAN2020105787,9274339} are different from the the traditional one-to-one engagements \cite{SINHA2021106776,YE2020105715,SINHA2021106824}, and present a formidable challenge in the design of the vehicles' strategies.

Some of the early results on the kinematics of three-agent problem were presented in \cite{4101686,4102335}, which provided closed-form solutions of the problem under the asymptotic approximation of constant bearing trajectories. The relation was derived under consideration of a simple ratio of the pursuer-evader range at the time instant when the defender is launched to that at the time instant when the defender captures the pursuer. Requirements on defender's launch angle and distance covered by it to intercept the pursuer were analyzed in \cite{4102335}, as a function of the engagement geometry and the time instant at which the defender was launched. The study in \cite{4101810} provided some simplifications over \cite{4101686}, and presented compact expressions for closing velocities. Although the results in \cite{4101686,4102335} provided insights into the kinematics of a three-agent engagement, these strategies could not ensure satisfactory performance for the initial conditions far from the collision course. A \emph{triangle guidance} concept was proposed in \cite{doi:10.2514/6.2010-7876,9301417}, wherein the defender was required to remain on the line joining the evader and the pursuer, to capture the pursuer before it reaches the evader. This concept is closely related to the line-of-sight (LOS) guidance, which is a \emph{three-point} guidance strategy. Three points in such strategy were usually a pursuer, an evader, and a reference from which the evader was observed. In the context of the three-agent problem \cite{doi:10.2514/1.58566}, the evader was deemed as one of three points in place of the reference, where the other two points were the defender and the pursuer. Authors in \cite{doi:10.2514/1.58566} proposed an {airborne-command-to-line-of-sight} (ACLOS) guidance strategy based on optimal control formulation and a velocity feedback term. By manipulating two LOS rates associated with three agents, the study in \cite{doi:10.2514/1.58566} exhibited benefits over the classical three-point guidance strategies. However, none of the aforementioned strategies accounted for evader-defender cooperation, except \cite{9301417}, in which instantaneous lateral accelerations of the evader-defender team, rather than the total lateral acceleration, were minimized.

Introducing cooperation between the evader and the defender may improve the performance of the agents' strategies. Authors in \cite{doi:10.2514/1.50572} proposed LOS guidance as the defender's strategy. The kinematic relations were derived considering the evader as a moving launch platform, and the cooperative guidance strategy for the evader was designed by maximizing the pursuer to defender lateral acceleration ratio. However, LOS guidance may induce an oscillatory behavior in the guidance command, as demonstrated in \cite{doi:10.2514/1.G000659}. For a cooperating evader-defender team, the work in \cite{doi:10.2514/1.56924} provided further insights into the problem from the pursuer and defender's perspectives by analyzing launch envelops and acceleration ratio of the adversaries, using various classical guidance schemes.

Since optimal control based formulations yield guidance commands with lesser energy cost, such formulations were used to design cooperative guidance strategies in \cite{doi:10.2514/1.G001083,doi:10.2514/1.51765,doi:10.2514/1.49515,doi:10.2514/1.58531}. Authors in \cite{doi:10.2514/1.G001083} proposed a cooperative optimal guidance strategy in conjunction with a differential game formulation, wherein the separation between the pursuer and the evader was maximized, as a solution to a two-point boundary value problem. Optimal cooperative evasion and pursuit strategies for the evader-defender team were derived in \cite{doi:10.2514/1.51765} assuming arbitrary order linearized dynamics of the adversaries. While the pursuer's guidance strategy was assumed to be known in \cite{doi:10.2514/1.51765}, a multiple-model adaptive estimator approach was presented in \cite{doi:10.2514/1.49515} as a cooperative information sharing mechanism between the evader and the defender to estimate the probable linear class of guidance strategy used by the pursuer. Three different levels of cooperation were presented in \cite{doi:10.2514/1.58531}, and the effects of information sharing between the evader and the defender were analyzed. The three-agent problem studied in \cite{doi:10.2514/6.2012-4910} constructed a guaranteed miss distance to ensure that the pursuer did not intercept the evader. In \cite{doi:10.2514/1.61832}, some algebraic conditions were derived for the pursuer to capture the evader, while evading the defender. Most of the aforementioned studies used linearized formulations. While such an approach simplifies the guidance design, the applicability of the strategy may be restricted to limited operating regions.

Since the adversaries have conflicting goals to maximize their chances of success, the three-agent problem can also be cast as a linear quadratic differential game. In \cite{doi:10.2514/6.2010-8057}, the three-agent problem was posed as a dynamic game involving three agents-- the lady, the bandit, and the bodyguard. The objective of the bandit was to capture the lady, while that of the bodyguard was to prevent this from happening. The bodyguard tried to capture the bandit before it could come in the vicinity of the lady. The solutions to the game presented in \cite{doi:10.2514/6.2010-8057} were obtained using multi-objective optimization and differential game. The complexity of obtaining the optimal solutions might be prohibitive in guidance applications. As a result, suboptimal solutions were presented in \cite{doi:10.2514/6.2010-8057}, involving two-point boundary value problems with backward integration. Such numerical solutions might be computationally intensive. Another linear quadratic differential game strategy was proposed in \cite{doi:10.2514/1.51611}, wherein arbitrary order linearized dynamics of the adversaries were considered, and the pursuer's guidance strategy was not required to be known. Analytical and discrete formulations of the three-agent problem were discussed in \cite{doi:10.2514/1.51611}, and a terminal-projection transformation was used to reduce the order of the problem. The results in \cite{RUSNAK20119349} dealt with the limiting values of the agents' optimal strategies, as the quadratic weight on the defender's acceleration approached zero.

The situation of guarding a stationary asset wherein an agent attempts to reach and destroy the target while its opponent tries to safeguard the same was also introduced in the seminal work of Isaacs \cite{isaacs}. This problem was further investigated in \cite{doi:10.2514/1.G002652} as a reachability game, and the defender tried to intercept the pursuer as far as possible from the evader. The strategy proposed in \cite{8718039} allowed a single evader to reach the target's location against a group of pursuers. The asset guarding game was revisited in \cite{8304800} from the perspective of a real-time implementation for the optimal solution. In \cite{5751240}, a linear quadratic approach was presented to analyze the case of asset guarding. A numerical approach to finding optimal trajectories for the agents in an asset-guarding game was proposed in \cite{sin2020iterative} using an iterative best response scheme. Authors in \cite{polak2016} proposed a receding horizon control technique for the defenders to guard a harbor, while fuzzy-logic combined with reinforcement learning was used in \cite{Raslan2016} in the context of guarding a territory.

A vast majority of the previously discussed works used linearized kinematics, hence they are valid only for engagements with small heading angle errors. A guidance strategy developed by considering nonlinear kinematics may guarantee superior performance over linearized approaches as the assumption of the vehicle's small heading angle errors can be eliminated. Using a three-dimensional sliding mode approach, a terminal intercept guidance and autopilot were designed in \cite{6315051} for a defender trying to protect an evader from a pursuer. In \cite{6315051}, higher-order sliding mode control was used to design the guidance strategy, requiring four parameters to be tuned. Another sliding mode based nonlinear guidance strategy, for the evader-defender team, was proposed in \cite{doi:10.2514/1.G000659}, which required only a single parameter to be tuned. The triangle guidance strategy was presented within a nonlinear framework in \cite{9301417}. A hybrid cooperative guidance law for defender was developed in \cite{doi:10.2514/1.G003059} by integrating inertial delay control, and prescribed performance control with sliding mode control. While this guidance scheme provided flexibility and guaranteed performance that may not be ensured with only a single technique, the guidance strategy required switching between several guidance schemes.

In this paper, we present a novel approach to design strategies for the evader-defender team to protect the evader from the pursuer. In a three-agent scenario, the time of interception is a crucial parameter and cannot be ignored while designing effective guidance strategies against the opposition. As the main contribution, we propose an impact time control based cooperative guidance approach to defend the evader from the pursuer. If the defender safeguards the evader using aggressive or defensive strategies before the pursuer can come in the vicinity of the evader, the mission can be deemed a success. The evader-defender team can use various levels of cooperation to achieve this objective, depending on the availability of communication resources and desired simplicity during the implementation. The evader aligns itself on the collision course with the pursuer to lure the latter and launches the defender against it. Once on the collision course, the pursuer is expected to become non-maneuvering, and the time of interception for the pursuer-evader engagement can be estimated with reasonable accuracy. Depending on the mission and using the information about the time of interception, the defender either rendezvous with the evader or the pursuer in a time which is sufficiently lesser than the estimated time-to-go of the pursuer-evader engagement, even if the evader does not continuously communicate its states to the defender. On the other hand, if the evader communicates with the defender, then explicit time-to-go information may not be necessary, and the defender can rendezvous with the evader or the pursuer with a pre-specified time margin.

In light of the aforementioned works, the contributions of this paper can now be summarized as follows.
\begin{itemize}
	\item This paper offers a novel perspective to design guidance strategies for the evader-defender team against a pursuer using a time-constrained approach. Compared to heuristic and numerical methods, the proposed approach is not computationally demanding. Moreover, the proposed work also paves the way for existing impact time guidance strategies to be applied in three-agent pursuit-evasion scenarios.
	\item The strategies of the evader-defender team are derived using nonlinear engagement kinematics and are also valid for engagements that commence far from the collision course. This eliminates the possibilities of errors that may arise due to linearization of the kinematic equations.
	\item Deviated pursuit guidance, which was originally developed against mobile adversaries, is used to design the strategies of the defender. This renders the use of predicted interception point unnecessary while intercepting a mobile adversary.
	\item Based on the availability of communication resources and desired simplicity during implementation, two different modes of cooperation are proposed. Further, the defender may take an aggressive or a defensive stance in either mode of cooperation to safeguard the evader by choosing to rendezvous with the pursuer or with the evader.
	\item The strategies of the evader-defender team are based on fixed-time convergent sliding mode control using which the rate of convergence of the relevant error variables are made independent of the agents' initial engagement geometries. Further, a nonlinear finite-time disturbance observer is used to estimate the information about the pursuer's actual strategy.
\end{itemize}

The remainder of this paper is organized as follows. In \Cref{sec:problem}, some necessary background has been presented, followed by the problem statement. Several guidance strategies for the evader-defender team have been derived in \Crefrange{sec:evaderLaw}{sec:defenderLawDef}, followed by \Cref{sec:simulations}, in which various numerical simulations have been presented establishing the applicability of the proposed strategies. \Cref{sec:conclusion} provides some concluding remarks, and proposes outlines for future investigations.

\section{Three-agent Steered Vehicle Model}\label{sec:problem}
\begin{figure}[h!]
	\centering
	\includegraphics[width=.7\linewidth]{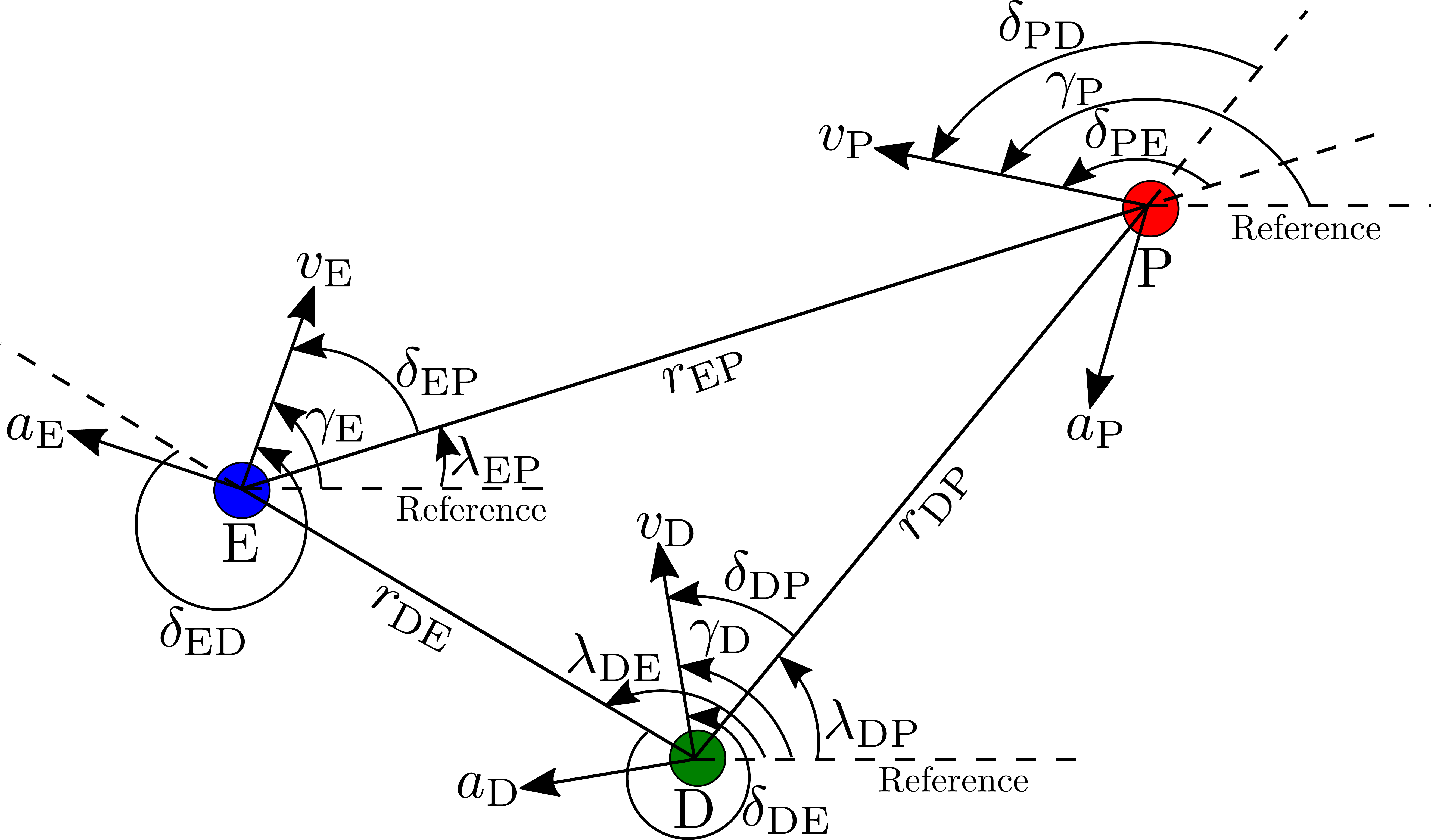}
	\caption{Three-agent planar engagement geometry.}
	\label{fig:enggeo}
\end{figure}

We consider a three-agent planar kinematic model of point-mass vehicles that move at constant speeds, and are subject to steering controls to alter their orientations. As depicted in \Cref{fig:enggeo}, the speeds of the evader (E) and the pursuer (P) are denoted by $v_\mathrm{E}$ and $v_\mathrm{P}$, respectively. 
%The speeds of the evader and the pursuer are comparable while
The defender (D), whose speed is represented by $v_\mathrm{D}$, has a speed advantage over them. The heading angles and the lateral accelerations (steering controls) of the pursuer, the evader, and the defender are denoted by $\left(\gamma_\mathrm{P}, \gamma_\mathrm{E}, \gamma_\mathrm{D}\right)$, and $\left(a_\mathrm{P}, a_\mathrm{E}, a_\mathrm{D}\right)$, respectively. The relative distance and the line-of-sight (LOS) angle for the pursuer-evader engagement are $\left(r_\mathrm{EP},\lambda_\mathrm{EP}\right)$, while those for the defender-pursuer engagement are $\left(r_\mathrm{DP},\lambda_\mathrm{DP}\right)$. Similarly, variables for the defender-evader engagement are denoted by $\left(r_\mathrm{DE},\lambda_\mathrm{DE}\right)$.

The equations of motion for the three pairs of engagement can be succinctly expressed in the polar coordinates as
\begin{subequations}\label{eq:engdyn}
	\begin{align}
		\dot{r}_\mathrm{EP} =&~ v_\mathrm{P} \cos\left(\gamma_\mathrm{P}-\lambda_\mathrm{EP}\right) - v_\mathrm{E}\cos\left(\gamma_\mathrm{E}-\lambda_\mathrm{EP}\right) = v_\mathrm{P} \cos\delta_\mathrm{PE} - v_\mathrm{E}\cos\delta_\mathrm{EP} = v_{r_\mathrm{EP}}, \label{eq:rEPdot} \\
		{r}_\mathrm{EP}\dot{\lambda}_\mathrm{EP} =&~ v_\mathrm{P} \sin\left(\gamma_\mathrm{P}-\lambda_\mathrm{EP}\right) - v_\mathrm{E}\sin\left(\gamma_\mathrm{E}-\lambda_\mathrm{EP}\right) =v_\mathrm{P} \sin\delta_\mathrm{PE} - v_\mathrm{E}\sin\delta_\mathrm{EP}=v_{\lambda_\mathrm{EP}}, \label{eq:lambdaEPdot} \\
		\dot{r}_\mathrm{DP} =&~v_\mathrm{P} \cos\left(\gamma_\mathrm{P}-\lambda_\mathrm{DP}\right) - v_\mathrm{D}\cos\left(\gamma_\mathrm{D}-\lambda_\mathrm{DP}\right) =v_\mathrm{P}\cos\delta_\mathrm{PD}-v_\mathrm{D}\cos\delta_\mathrm{DP}=v_{r_\mathrm{DP}},\label{eq:rDPdot}\\		
		{r}_\mathrm{DP}\dot{\lambda}_\mathrm{DP} =&~v_\mathrm{P} \sin\left(\gamma_\mathrm{P}-\lambda_\mathrm{DP}\right) - v_\mathrm{D}\sin\left(\gamma_\mathrm{D}-\lambda_\mathrm{DP}\right) =v_\mathrm{P}\sin\delta_\mathrm{PD}-v_\mathrm{D}\sin\delta_\mathrm{DP}=v_{\lambda_\mathrm{DP}},\label{eq:lambdaDPdot}\\
		\dot{r}_\mathrm{DE} =&~v_\mathrm{E} \cos\left(\gamma_\mathrm{E}-\lambda_\mathrm{DE}\right) - v_\mathrm{D}\cos\left(\gamma_\mathrm{D}-\lambda_\mathrm{DE}\right) =v_\mathrm{E}\cos\delta_\mathrm{ED}-v_\mathrm{D}\cos\delta_\mathrm{DE}=v_{r_\mathrm{DE}},\label{eq:rDEdot}\\		
		{r}_\mathrm{DE}\dot{\lambda}_\mathrm{DE} =&~v_\mathrm{E} \sin\left(\gamma_\mathrm{E}-\lambda_\mathrm{DE}\right) - v_\mathrm{D}\sin\left(\gamma_\mathrm{D}-\lambda_\mathrm{DE}\right) =v_\mathrm{E}\sin\delta_\mathrm{ED}-v_\mathrm{D}\sin\delta_\mathrm{DE}=v_{\lambda_\mathrm{DE}},\label{eq:lambdaDEdot}
	\end{align}
\end{subequations}
where $v_{r_\mathrm{EP}}$, $v_{r_\mathrm{DP}}$, $v_{r_\mathrm{DE}}$, $v_{\lambda_\mathrm{EP}}$, $v_{\lambda_\mathrm{DP}}$, and $v_{\lambda_\mathrm{DE}}$, represent the components of relative velocities of the relevant agents along and perpendicular to the corresponding LOS. Note that the engagement between the evader and the pursuer is governed by \eqref{eq:rEPdot}--\eqref{eq:lambdaEPdot} while the defender-pursuer engagement is governed by \eqref{eq:rDPdot}--\eqref{eq:lambdaDPdot}. The defender-evader cooperative engagement is described using \eqref{eq:rDEdot}--\eqref{eq:lambdaDEdot}. The steering control of the $i$\textsuperscript{th} agent is assumed to be bounded and is governed by 
\begin{equation}\label{eq:gammadot}
	\dot{\gamma}_i = \dfrac{a_i}{v_i};\;\left\vert a_i\right\vert \leq a_i^\mathrm{max},\,\forall\,i\in\left\{\mathrm{P},\mathrm{E},\mathrm{D}\right\}.
\end{equation} 

We define the time of capture (also known as the time of interception, or impact time), $T_f$, as the time at which the position of any two agents coincide. At any instant of time, the remaining time till capture (also known as the time-to-go), $t_\mathrm{go}$, can be defined as the difference between $T_f$ and the current time, $t$, that is $t_\mathrm{go} = T_f - t$. 

Note that $T_f$, and hence $t_\mathrm{go}$, is crucial to the design of time-constrained feedback control laws for the evader and the defender. The evader and the defender can use various modes of cooperation by exercising control over the agents' respective engagement duration to safeguard the former from the pursuer. The defender is said to have an \emph{aggressive} stance if it rendezvous with the pursuer before the pursuer can come in some vicinity of the evader, thereby neutralizing the incoming threat. On the other hand, the defender is said to adopt a \emph{defensive} stance if it rendezvous with the evader before the pursuer can capture the latter, thereby shielding the evader from the incoming threat.

Based on whether the defender is required to take a defensive or an aggressive stance to safeguard the evader, the objectives of the current paper can now be stated in the following manner.
\begin{objective}
	Design the evader's control law (also known as the guidance strategy of the evader) to assist the defender in safeguarding the evader.
\end{objective}
\begin{objective}
	Design the \emph{aggressive} strategy of the defender to neutralize (intercept or capture) the pursuer before the latter can capture the evader.
\end{objective}
\noindent\textbf{Objective 2 (Alternate).} \emph{Design the \emph{defensive} strategy of the defender to shield the evader from the pursuer before the pursuer can capture the evader.}
%\begin{objective}
%	Design the \emph{defensive} strategy of the defender to shield the evader from the pursuer before the pursuer can capture the evader.
%\end{objective}

It is also desirable that the developed strategies should remain valid for all $\delta_\mathrm{EP}\in\left(-\dfrac{\pi}{2},\dfrac{\pi}{2}\right)$, $\delta_\mathrm{DP}\in\left(-\dfrac{\pi}{2},\dfrac{\pi}{2}\right)$, and $\delta_\mathrm{DE}\in\left(-\dfrac{\pi}{2},\dfrac{\pi}{2}\right)$ even if engagements commence far from the collision course \cite{shneydor1998missile}. Dealing with the problem in a nonlinear framework circumvents the possible errors that arise due to linearization, and ensures that the designed strategies remain applicable even over larger operating regions.

\section{Strategy of the Evader}\label{sec:evaderLaw}
In the existing literature on three-agent scenario, parameters such as miss distance and relative separation between the adversaries have been considered while designing the strategies of the evader-defender team. Observe that interception time is also an important parameter to be considered from the viewpoint of the evader. If the evader obtains the information about the time it will take for the pursuer to capture it, the evader can cooperate with the defender by passing this information to the defender, which must then safeguard the evader within a lesser time. In essence, the cooperative defense problem can be posed as an impact time control problem, in which the impact time for the defender-pursuer engagement or that for the defender-evader engagement is shaped in accordance with the pursuer-evader engagement.

From the equations of motion, \eqref{eq:engdyn}, and the geometry in \Cref{fig:enggeo}, it follows that at any instant in the pursuer-evader engagement, the time-to-go for the pursuer-evader engagement (the remaining time required by the pursuer to capture the evader) can be estimated as 
\begin{equation}\label{eq:tgoEP}
	t_\mathrm{go}^\mathrm{EP} =  -\dfrac{r_\mathrm{EP}}{\dot{r}_\mathrm{EP}},
\end{equation}
whose dynamics can be expressed as
\begin{equation}\label{eq:tgoEPdot}
	\dot{t}_\mathrm{go}^\mathrm{EP} = -1 + \dfrac{r_\mathrm{EP}^2\dot{\lambda}_\mathrm{EP}^2}{\dot{r}_\mathrm{EP}^2} + \dfrac{{r}_\mathrm{EP}\sin\delta_\mathrm{EP}}{\dot{r}_\mathrm{EP}^2}a_\mathrm{E} - \dfrac{{r}_\mathrm{EP}\sin\delta_\mathrm{PE}}{\dot{r}_\mathrm{EP}^2}a_\mathrm{P}.
\end{equation}
\begin{remark}
	The time-to-go estimate in \eqref{eq:tgoEP} is an underestimation in general, unless the pursuer is heading straight for the evader.
\end{remark}
Since the pursuer may execute arbitrary maneuvers, it is difficult to accurately compute the time it would require to capture the evader. However, if the evader and the pursuer are on the collision course, and the pursuer does not maneuver, then the time of the evader's capture by the pursuer can be accurately computed. This would also aid the defender to decide its own time of rendezvous with the evader or with the pursuer. Hence, the evader lures the pursuer by offering itself as a bait, and maneuvers accordingly to render the latter non-maneuvering. From the standpoint of the pursuer, this seems to be a favorable outcome and hence, it is likely that the pursuer would cease performing further maneuvers. In essence, the evader performs maneuvers to nullify its LOS rate with the pursuer as early as possible. Note that the pursuer becoming non-maneuvering is not a necessary condition to safeguard the evader but, as will be shown later, it aids the defender in reducing its maneuverability requirement. We present some intermediate results before proceeding with the formal design procedure.
\begin{lemma}\label{lem:LOSratedot}
	The dynamics of LOS rates of each pair of engagement has relative degree one with respect to the lateral accelerations of the corresponding agents in that engagement.
\end{lemma}
\begin{proof}
	On differentiating the LOS rate of the pursuer-evader engagement, \eqref{eq:lambdaEPdot}, with respect to time, one may obtain
	\begin{align}
		\dot{r}_\mathrm{EP}\dot{\lambda}_\mathrm{EP} + {r}_\mathrm{EP}\ddot{\lambda}_\mathrm{EP} = v_\mathrm{P} \cos\delta_\mathrm{PE}\dot{\delta}_\mathrm{PE} - v_\mathrm{E}\cos\delta_\mathrm{EP}\dot{\delta}_\mathrm{EP},
	\end{align}
	whose further simplification, using \eqref{eq:gammadot}, results in
	\begin{align}
		\ddot{\lambda}_\mathrm{EP} = & \dfrac{v_\mathrm{P}\cos\delta_\mathrm{PE}}{{r}_\mathrm{EP}}\left(\dfrac{a_\mathrm{P}}{v_\mathrm{P}}-\dot{\lambda}_\mathrm{EP}\right) - \dfrac{v_\mathrm{E}\cos\delta_\mathrm{EP}}{{r}_\mathrm{EP}}\left(\dfrac{a_\mathrm{E}}{v_\mathrm{E}}-\dot{\lambda}_\mathrm{EP}\right)  - \dfrac{\dot{r}_\mathrm{EP}}{{r}_\mathrm{EP}}\dot{\lambda}_\mathrm{EP}\nonumber \\
		=& -\left(\dot{r}_\mathrm{EP}+v_\mathrm{P}\cos\delta_\mathrm{PE}-v_\mathrm{E}\cos\delta_\mathrm{EP}\right)\dfrac{\dot{\lambda}_\mathrm{EP}}{{r}_\mathrm{EP}}  -\dfrac{\cos\delta_\mathrm{EP}}{{r}_\mathrm{EP}}a_\mathrm{E}+\dfrac{\cos\delta_\mathrm{PE}}{{r}_\mathrm{EP}}a_\mathrm{P}\nonumber\\
		=&-\dfrac{2\dot{r}_\mathrm{EP}\dot{\lambda}_\mathrm{EP}}{{r}_\mathrm{EP}}-\dfrac{\cos\delta_\mathrm{EP}}{{r}_\mathrm{EP}}a_\mathrm{E}+\dfrac{\cos\delta_\mathrm{PE}}{{r}_\mathrm{EP}}a_\mathrm{P}.\label{eq:lambdaEPddot}
	\end{align}
	Similarly, differentiating \eqref{eq:lambdaDPdot} and \eqref{eq:lambdaDEdot} with respect to time, and substituting $\dot{\gamma}_i$ from \eqref{eq:gammadot} yields
	\begin{equation}\label{eq:lambdaDPddot}
		\ddot{\lambda}_\mathrm{DP} = -\dfrac{2\dot{r}_\mathrm{DP}\dot{\lambda}_\mathrm{DP}}{r_\mathrm{DP}}-\dfrac{\cos\delta_\mathrm{DP}}{r_\mathrm{DP}}a_\mathrm{D}+\dfrac{\cos\delta_\mathrm{PD}}{r_\mathrm{DP}}a_\mathrm{P},
	\end{equation}
	and
	\begin{equation}\label{eq:lambdaDEddot}
		\ddot{\lambda}_\mathrm{DE} = -\dfrac{2\dot{r}_\mathrm{DE}\dot{\lambda}_\mathrm{DE}}{r_\mathrm{DE}}-\dfrac{\cos\delta_\mathrm{DE}}{r_\mathrm{DE}}a_\mathrm{D}+\dfrac{\cos\delta_\mathrm{ED}}{r_\mathrm{DE}}a_\mathrm{E}.
	\end{equation}
	From \eqref{eq:lambdaEPddot}--\eqref{eq:lambdaDEddot}, it follows that the dynamics of respective LOS rates possess relative degree one with respect to the lateral accelerations of the corresponding agents.
\end{proof}
Observations from \Cref{lem:LOSratedot} help in designing suitable guidance command for the evader. Though the pursuer's maneuver may be unknown to the guidance system of the evader-defender team, its upper bound can be known, $|a_\mathrm{P}|\leq a_\mathrm{P}^\mathrm{max}$. To compensate for the effect of this uncertainty, a nonlinear finite-time disturbance observer \cite{7265050} can be used to estimate the pursuer's maneuver when this information is unknown to the evader-defender team. In this regard, the error in estimation of the pursuer's maneuver can be written as 
\begin{equation}
	\tilde{e} = a_\mathrm{P}-\hat{a}_\mathrm{P},
\end{equation} 
which remains bounded, $|\tilde{e}|\leq \tilde{e}_\mathrm{max}$ \cite{7265050}. Consider \eqref{eq:lambdaEPddot}, in which the term $a_\mathrm{P} = d$ (say) may be treated as a disturbance. As $r_\mathrm{EP}\neq 0$ throughout the engagement, if $a_\mathrm{P}$ is bounded and differentiable, then so are $d$ and $\dot{d}$. The nonlinear disturbance observer to estimate $a_\mathrm{P}$ can be described by
\begin{align*}
	\dot{z}_0 =& -\dfrac{2\dot{r}_\mathrm{EP}\dot{\lambda}_\mathrm{EP}}{{r}_\mathrm{EP}}-\dfrac{\cos\delta_\mathrm{EP}}{{r}_\mathrm{EP}}a_\mathrm{E}+\dfrac{\cos\delta_\mathrm{PE}}{{r}_\mathrm{EP}}v_0,\\
	v_0 =& -G_2 L^{1/3}|z_0-\dot{\lambda}_\mathrm{EP}|^{2/3}\sign{(z_0-\dot{\lambda}_\mathrm{EP})} - H_2(z_0-\dot{\lambda}_\mathrm{EP}) + z_1,\\
	\dot{z}_1=&v_1 = -G_1 L^{1/2}|z_1-v_0|^{1/2}\sign{(z_1-v_0)} - H_2(z_1-v_0) + z_2,\\
	\dot{z}_2 =& -G_0 L \sign(z_2-v_1) - H_0(z_2-v_1),
\end{align*}
where $G_2>G_1>G_0>0$, $H_2>H_1>H_0>0$, $L$ is the Lipschitz constant such that $|\dot{d}|\leq L$, and $z_1$ gives an estimate of the pursuer's maneuver, $\hat{a}_\mathrm{P}$, which is used by the evader-defender team in their respective control laws (guidance strategies). As the estimation error, $\tilde{e}$ dies out, $\hat{a}_\mathrm{P}\to a_\mathrm{P}$ within a finite-time. We assume perfect state information (range, LOS angles, etc.) in the design, however, these states can be estimated/observed in practice using a state observer. The pursuer's maneuver is thus estimated using the nonlinear finite-time disturbance observer and the estimated value is used in our proposed design.

\begin{remark}
	It has been shown in \cite{doi:10.2514/1.49515,doi:10.2514/1.51765} that the pursuer's maneuver can be identified within a finite-time if it uses a linear class of guidance strategy. In such cases, $\hat{a}_\mathrm{P}={a}_\mathrm{P}$ right from the beginning of the engagement. When only bound on the pursuer's maneuverability, $|a_\mathrm{P}|<a_\mathrm{P}^\mathrm{max}$, is known, then, designing suitable controller gains (for example, gains larger than $\sup_{t\geq 0}a_\mathrm{P}^\mathrm{max}$) ensures that $\tilde{e}\to 0$. Thus, our consideration is more general, encompassing these possibilities.
\end{remark}
The evader is tasked with nullifying its LOS rate with respect to the pursuer. Consider the LOS rate of pursuer-evader engagement as the sliding manifold, such that
\begin{equation}\label{eq:s1}
	\sigma_1 = \dot{\lambda}_\mathrm{EP}.
\end{equation}
\begin{theorem}\label{thm:evader}
	For the three-agent engagement whose kinematics is governed by \eqref{eq:engdyn}, the fixed-time convergent guidance strategy for the evader, which nullifies the LOS rate for the pursuer-evader engagement is given by
	\begin{equation}\label{eq:aE}
		a_\mathrm{E} = -\dfrac{2\dot{r}_\mathrm{EP}\dot{\lambda}_\mathrm{EP}}{\cos\delta_\mathrm{EP}}+  \dfrac{\cos\delta_\mathrm{PE}}{\cos\delta_\mathrm{EP}}\hat{a}_\mathrm{P}+ \dfrac{r_\mathrm{EP}}{\cos\delta_\mathrm{EP}}\left[\left(\zeta_1 |\sigma_1|^{\alpha_1} + \xi_1 |\sigma_1|^{\beta_1}\right)^{\kappa_1}+\sec\delta_\mathrm{EP}\,\epsilon_1\right] \sign(\sigma_1),
	\end{equation}
	where $\zeta_1,\xi_1, \alpha_1,\beta_1,\kappa_1>0$ are the design parameters satisfying the constraints $\alpha_1\kappa_1<1$, $\beta_1\kappa_1>1$, and $\epsilon_1>\sup_{t\geq 0}\tilde{e}_\mathrm{max}$.
\end{theorem}
\begin{proof}
	Consider the sliding manifold, \eqref{eq:s1}, as the error variable. Differentiating \eqref{eq:s1} with respect to time, and substituting from \eqref{eq:lambdaEPddot} results in
	\begin{equation}\label{eq:s1dot}
		\dot{\sigma}_1 = \ddot{\lambda}_\mathrm{EP} = -\dfrac{2\dot{r}_\mathrm{EP}\dot{\lambda}_\mathrm{EP}}{{r}_\mathrm{EP}}-\dfrac{\cos\delta_\mathrm{EP}}{{r}_\mathrm{EP}}a_\mathrm{E}+\dfrac{\cos\delta_\mathrm{PE}}{{r}_\mathrm{EP}}a_\mathrm{P}.
	\end{equation}
	To derive the evader's control law, $a_\mathrm{E}$, consider a Lyapunov function candidate, $\mathcal{V}_1 = |\sigma_1|$, differentiation of which with respect to time yields
	\begin{align}\label{eq:V1dotstep1}
		\dot{\mathcal{V}}_1 = \sign(\sigma_1)\dot{\sigma}_1 = \sign(\sigma_1)\left[-\dfrac{2\dot{r}_\mathrm{EP}\dot{\lambda}_\mathrm{EP}}{{r}_\mathrm{EP}}-\dfrac{\cos\delta_\mathrm{EP}}{{r}_\mathrm{EP}}a_\mathrm{E}+\dfrac{\cos\delta_\mathrm{PE}}{{r}_\mathrm{EP}}a_\mathrm{P}\right].
	\end{align}
	If the evader's guidance strategy is chosen as \eqref{eq:aE}, then \eqref{eq:V1dotstep1} can be simplified as
	\begin{align}
		\dot{\mathcal{V}}_1 =& -\sign(\sigma_1)\left[\left\{\left(\zeta_1 |\sigma_1|^{\alpha_1} + \xi_1 |\sigma_1|^{\beta_1}\right)^{\kappa_1} +\sec\delta_\mathrm{EP}\,\epsilon_1 \right\}\sign(\sigma_1) -\dfrac{\cos\delta_\mathrm{PE}}{\cos\delta_\mathrm{EP}}\left(a_\mathrm{P}-\hat{a}_\mathrm{P}\right) \right] \nonumber\\
		=& -\left(\zeta_1 |\sigma_1|^{\alpha_1} + \xi_1 |\sigma_1|^{\beta_1}\right)^{\kappa_1} -\sec\delta_\mathrm{EP} \left(\epsilon_1-\sign(\sigma_1)\cos\delta_\mathrm{PE}\,\tilde{e}\right)\nonumber\\
		\leq & -\left(\zeta_1 |\sigma_1|^{\alpha_1} + \xi_1 |\sigma_1|^{\beta_1}\right)^{\kappa_1} - \left(\epsilon_1-\tilde{e}_\mathrm{max}\right) \nonumber\\
		\leq & -\left(\zeta_1 |\sigma_1|^{\alpha_1} + \xi_1 |\sigma_1|^{\beta_1}\right)^{\kappa_1} <0,\;\forall\,\sigma_1\neq 0. \label{eq:V1dot}
	\end{align}
	It follows from \eqref{eq:V1dot} that sliding mode is enforced on $\sigma_1$. Thus, $\mathcal{V}_1$, and hence $\sigma_1$, in \eqref{eq:s1} goes to zero in a fixed-time, $t_{c_1}\leq\dfrac{1}{\zeta_1^{\kappa_1}(1-\alpha_1\kappa_1)}+\dfrac{1}{\xi_1^{\kappa_1}(\beta_1\kappa_1-1)}$, for some design parameters $\zeta_1,\xi_1, \alpha_1,\beta_1,\kappa_1>0$  satisfying the constraints $\alpha_1\kappa_1<1$, $\beta_1\kappa_1>1$, and $\epsilon_1>\sup_{t\geq 0}\tilde{e}_\mathrm{max}$ irrespective of initial value of $\dot{\lambda}_\mathrm{EP}$ \cite{6104367}. This concludes the proof.
\end{proof}
\begin{remark}
	By a proper choice of the design parameters, nullification of the LOS rate for the pursuer-evader engagement can be guaranteed for any initial value of $\lambda_{\mathrm{EP}}$, thereby increasing the chances for the pursuer to become a non-maneuvering adversary within a fixed-time.
\end{remark}
In essence, the guidance strategy \eqref{eq:aE} can be seen as a variant of proportional-navigation guidance.
% As the pursuer-evader engagement proceeds, these agents home onto each other, making $\dot{r}_\mathrm{EP}<0$. 
The first term of \eqref{eq:aE} is a proportional-navigation term with navigation gain of 2, with the effect of nonlinearity captured through $\dfrac{\dot{r}_\mathrm{EP}}{\cos\delta_\mathrm{EP}}$. The second term captures the effect of the pursuer's maneuver. As soon as $\tilde{e}\to 0$, the evader's guidance strategy compensates for the pursuer's maneuver. The third term represents a corrective action that forces $\sigma_1$ to zero within a fixed-time, $t_{c_1}$. Effectively, $a_\mathrm{E}\propto \dot{\lambda}_\mathrm{EP}$, which tends to zero in the endgame. As a special case, when the pursuer uses a linear class of guidance strategies (such as a variant of proportional-navigation), it will becoming non-maneuvering after convergence of $\sigma_1$. For the defender, intercepting such an adversary requires lesser control effort than a maneuvering one.

Having designed the guidance strategy for the evader, that of the defender can now be delineated depending on its required stance during a particular mission. However, irrespective of the stance the defender takes, we propose two different cooperative schemes for the evader-defender team, depending on the nature of communication between them. We say that the evader is \emph{indirectly} cooperating with the defender when the former does not continuously share its own engagement information with the latter. The evader aids in capturing the pursuer by aiming to align itself on the collision course, thereby serving itself as a bait for the pursuer. The evader can also choose to actively cooperate with the defender by sharing some of its necessary engagement parameters during the course of engagement. We say that the evader is \emph{directly} cooperating with the defender when the former shares its own engagement information with the latter, in addition to luring the pursuer on the collision path, to render it non-maneuvering.

It is worth noting that even if the pursuer does not stop maneuvering despite the evader's attempt to maintain $\dot{\lambda}_\mathrm{EP}=0$, the defender can still capture the pursuer. However, the defender may need to execute relatively high lateral acceleration when the pursuer is maneuvering compared to when it is rendered non-maneuvering due to the pursuer's motion.

\section{Aggressive Stance of the Defender}\label{sec:defenderLawAgg}
Consider the scenario of defending an aerial target from an attacker wherein the target deploys an interceptor to safeguard it from the attacker. In this scenario, the aerial target is the evader, the attacker is the pursuer, and the interceptor is the defender. The pursuer is required to be intercepted by the defender to protect the evader in such scenarios. Thus, the defender may need to adopt an aggressive stance in such cases. The defender is said to be in aggressive stance when it rendezvous with the pursuer to neutralize it before the latter can come in the vicinity of the evader.

To this day, most of the impact time constrained guidance strategies (for example, see \cite{7376242,Cho2015,SINHA2021106824,doi:10.2514/1.G005367,doi:10.2514/1.G005180} and references therein) are based on proportional-navigation guidance and its close variants. These guidance strategies were shown to be quite effective against stationary adversaries. However, in general, the pursuer may perform certain maneuvers, the information regarding which might be difficult to obtain a priori. As a result, the performance of such guidance strategies is likely to degrade against maneuvering adversaries. Besides proportional-navigation based strategies, deviated pursuit guidance \cite{shneydor1998missile}, which was originally developed for a mobile adversary has also drawn some attention of late. Unlike most of the existing strategies, deviated pursuit guidance offers an exact expression of time-to-go for a non-maneuvering target, which aids in interception of a constant velocity adversary in a time constrained setting \cite{doi:10.2514/1.G004284}. When the pursuer is lured on the collision path, it is likely to become a non-maneuvering adversary. For such adversaries, an exact time-to-go in a closed-form can be obtained, thereby making it easier for the defender to intercept the pursuer at some desired impact time. For a constant deviation angle of the defender guided by deviated pursuit, the time to intercept the pursuer can be exactly evaluated using the time-to-go for the defender-pursuer engagement, and is given by \cite{shneydor1998missile,doi:10.2514/1.G004284}
\begin{equation}\label{eq:tgoDP} 
	t_\mathrm{go}^\mathrm{DP} = \dfrac{r_\mathrm{DP}\sec\delta_\mathrm{DP}}{v_\mathrm{D}^2 - v_\mathrm{P}^2}\left[v_\mathrm{D}+v_\mathrm{P}\cos\left(\delta_\mathrm{PD}+\delta_\mathrm{DP}\right)\right] = \dfrac{r_\mathrm{DP}\left(v_{r_\mathrm{DP}}+2v_\mathrm{D}\cos\delta_\mathrm{DP}-v_{\lambda_\mathrm{DP}}\tan\delta_\mathrm{DP}\right)}{v_\mathrm{D}^2 - v_\mathrm{P}^2},
\end{equation}
whose dynamics is expressed as
\begin{equation}\label{eq:tgoDPdot}
	\dot{t}_\mathrm{go}^\mathrm{DP} = -1 + \dfrac{r_\mathrm{DP}^2\dot{\lambda}_\mathrm{DP}^2 \sec^2\delta_\mathrm{DP}}{v_\mathrm{D}^2 - v_\mathrm{P}^2} - \dfrac{r_\mathrm{DP}^2\dot{\lambda}_\mathrm{DP}\sec^2\delta_\mathrm{DP}}{v_\mathrm{D}\left(v_\mathrm{D}^2 - v_\mathrm{P}^2\right)} a_\mathrm{D} -\dfrac{r_\mathrm{DP}\sin\left(\delta_\mathrm{PD}+\delta_\mathrm{DP}\right)\sec\delta_\mathrm{DP}}{v_\mathrm{D}^2 - v_\mathrm{P}^2}a_\mathrm{P}.
\end{equation}
Note from \eqref{eq:tgoDP} that $t_\mathrm{go}^\mathrm{DP} =0\iff r_\mathrm{DP}=0$ is true irrespective of the pursuer maneuver, $a_\mathrm{P}$, since neither $\sec\delta_\mathrm{DP}$ nor $v_\mathrm{D}+v_\mathrm{P}\cos\left(\delta_\mathrm{PD}+\delta_\mathrm{DP}\right)$ is equal to zero for $v_\mathrm{D}>v_\mathrm{P}$. This means that even if the pursuer maneuvers, the defender can still use \eqref{eq:tgoDP} as an estimate for time-to-go. However, in such a case, the time-to-go expression for defender-pursuer engagement, \eqref{eq:tgoDP}, is not exact. From results in \cite{doi:10.2514/1.G004284}, it is evident that a successful interception of the adversary can be guaranteed in such engagement scenarios for $v_\mathrm{D}>v_\mathrm{P}$. We now present two different cooperative guidance techniques that can be used by the evader-defender team depending on the availability of communication resources.

\subsection{Indirect Cooperation}
In indirect cooperation, the evader does not share information about its engagement with the defender. This scheme is useful in the scenarios when there are constraints on the communication resources. The time-to-go of the pursuer-evader engagement, $t_\mathrm{go}^\mathrm{EP}$, can be estimated at the beginning of the engagement. Using this information, a suitable impact time can be fed to the guidance system of the defender against the pursuer. For the defender, a desired impact time, $T_f$, is feasible if it is chosen sufficiently smaller than $t_\mathrm{go}^\mathrm{EP}(t=0)$ so that it can capture the pursuer before the latter can intercept the evader. However, if the pursuer is lured on the collision path to become non-maneuvering, $t_{\mathrm{go}}^\mathrm{DP}$, given by \eqref{eq:tgoDP}, becomes exact, which makes it easier for the defender to capture the pursuer at a desired impact time, using a control law/guidance strategy based on deviated pursuit.

We define the sliding manifold representing impact time error for the engagement between the defender and the pursuer as
\begin{equation}\label{eq:s2}
	\sigma_2 = t_{\mathrm{go}}^\mathrm{DP} - (T_f - t),
\end{equation}
where $T_f$ denotes the \emph{desired} impact time, that is, the desired time at which the pursuer should be captured.
\begin{theorem}\label{thm:aD0}
	For a three-agent engagement, whose kinematics is governed by \eqref{eq:engdyn}, if the evader-defender team uses indirect cooperation, and the aggressive strategy for the defender is designed as
	\begin{align}
		a_{\mathrm{D}} =&~ v_{\mathrm{D}}\dot{\lambda}_{\mathrm{DP}} - \dfrac{v_{\mathrm{D}}\sin\left(\delta_{\mathrm{PD}}+\delta_{\mathrm{DP}}\right)}{v_{\lambda_{\mathrm{DP}}}\sec\delta_{\mathrm{DP}}}\hat{a}_\mathrm{P} \nonumber\\
		&+ \dfrac{v_{\mathrm{D}}\left(v_\mathrm{D}^2-v_\mathrm{P}^2\right)\cos^2\delta_{\mathrm{DP}}}{r_{\mathrm{DP}}v_{\lambda_{\mathrm{DP}}}}\left[\left(\zeta_{2} |\sigma_{2}|^{\alpha_{2}} + \xi_{2} |\sigma_{2}|^{\beta_{2}}\right)^{\kappa_{2}}+\sec\delta_{\mathrm{DP}}\,\epsilon_{2}\right]\sign(\sigma_{2}),\label{eq:aDindirectAgg}
	\end{align}
	where the parameters $\zeta_{2},\xi_{2}, \alpha_{2},\beta_{2},\kappa_{2}>0$ satisfy constraints $\alpha_{2}\kappa_{2}<1$, $\beta_{2}\kappa_{2}>1$, and $\epsilon_{2}>\sup_{t\geq 0}\left\{\dfrac{r_{\mathrm{DP}}}{v_\mathrm{D}^2-v_\mathrm{P}^2} \tilde{e}_\mathrm{max}\right\}$, then the defender converges to the deviated pursuit course in a fixed-time,
	\begin{equation}\label{eq:tc2}
		t_{c_{2}}\leq\dfrac{1}{\zeta_{2}^{\kappa_{2}}(1-\alpha_{2}\kappa_{2})}+\dfrac{1}{\xi_{2}^{\kappa_{2}}(\beta_{2}\kappa_{2}-1)}<\min\left\{t_{\mathrm{go}}^\mathrm{DP},t_\mathrm{go}^\mathrm{EP}\right\},
	\end{equation}
	and captures the pursuer in the desired impact time, $T_f$.
\end{theorem}
\begin{proof}
	Differentiating \eqref{eq:s2} with respect to time, as $T_f$ is a constant, one obtains using \eqref{eq:tgoDPdot}, $\dot{\sigma}_{2} = \dot{t}_{\mathrm{go}}^\mathrm{DP} + 1$, that is,	
	\begin{align}
		\dot{\sigma}_{2} = \dfrac{r_{\mathrm{DP}}^2\dot{\lambda}_{\mathrm{DP}}^2 \sec^2\delta_{\mathrm{DP}}}{v_\mathrm{D}^2-v_\mathrm{P}^2} - \dfrac{r_{\mathrm{DP}}^2\dot{\lambda}_{\mathrm{DP}}\sec^2\delta_{\mathrm{DP}}}{v_{\mathrm{D}}\left(v_\mathrm{D}^2-v_\mathrm{P}^2\right)} a_{\mathrm{D}} -\dfrac{r_{\mathrm{DP}}\sin\left(\delta_{\mathrm{PD}}+\delta_{\mathrm{DP}}\right)\sec\delta_{\mathrm{DP}}}{v_\mathrm{D}^2-v_\mathrm{P}^2}a_\mathrm{P}.
	\end{align}
	In a similar manner used to derive the evader's strategy, we consider a Lyapunov function candidate $\mathcal{V}_{2} = |\sigma_{2}|$, differentiating which with respect to time yields
	\begin{align}\label{eq:V2dotstep1}
		\dot{\mathcal{V}}_{2} =& \sign(\sigma_{2})\dot{\sigma}_{2} \nonumber\\
		=& \sign(\sigma_{2})\left[\dfrac{r_{\mathrm{DP}}^2\dot{\lambda}_{\mathrm{DP}}^2 \sec^2\delta_{\mathrm{DP}}}{v_\mathrm{D}^2-v_\mathrm{P}^2} - \dfrac{r_{\mathrm{DP}}^2\dot{\lambda}_{\mathrm{DP}}\sec^2\delta_{\mathrm{DP}}}{v_{\mathrm{D}}\left(v_\mathrm{D}^2-v_\mathrm{P}^2\right)} a_{\mathrm{D}}-\dfrac{r_{\mathrm{DP}}\sin\left(\delta_{\mathrm{PD}}+\delta_{\mathrm{DP}}\right)\sec\delta_{\mathrm{DP}}}{v_\mathrm{D}^2-v_\mathrm{P}^2}a_\mathrm{P}\right].
	\end{align}
	As before, the estimated value of the pursuer's maneuver is used to design the control law for the defender. If $a_{\mathrm{D}}$ is chosen as the one given in \eqref{eq:aDindirectAgg}, then \eqref{eq:V2dotstep1} can be simplified as
	\begin{align}
		\dot{\mathcal{V}}_{2} =& -\sign(\sigma_{2})\left[\dfrac{r_{\mathrm{DP}}\sin\left(\delta_{\mathrm{PD}}+\delta_{\mathrm{DP}}\right)}{\left(v_\mathrm{D}^2-v_\mathrm{P}^2\right) \cos\delta_{\mathrm{D}}}\left(a_\mathrm{P}-\hat{a}_\mathrm{P}\right)+\left\{\left(\zeta_{2} |\sigma_{2}|^{\alpha_{2}} + \xi_{2} |\sigma_{2}|^{\beta_{2}}\right)^{\kappa_{2}}\right.\right.\nonumber\\
		&\left.\left.+\sec\delta_{\mathrm{DP}}\,\epsilon_{2}\right\}\sign(\sigma_{2}) \right] \nonumber\\ 
		=& -\left(\zeta_{2} |\sigma_{2}|^{\alpha_{2}} + \xi_{2} |\sigma_{2}|^{\beta_{2}}\right)^{\kappa_{2}} -\sec\delta_{\mathrm{DP}} \left(\epsilon_{2}-\dfrac{\sign(\sigma_{2})r_{\mathrm{DP}}\sin\left(\delta_{\mathrm{PD}}+\delta_{\mathrm{DP}}\right)}{v_\mathrm{D}^2-v_\mathrm{P}^2}\tilde{e}\right)\nonumber\\
		\leq &-\left(\zeta_{2} |\sigma_{2}|^{\alpha_{2}} + \xi_{2} |\sigma_{2}|^{\beta_{2}}\right)^{\kappa_{2}} - \left(\epsilon_{2}-\dfrac{r_{\mathrm{DP}}}{v_\mathrm{D}^2-v_\mathrm{P}^2}\tilde{e}_\mathrm{max}\right) \nonumber\\
		\leq & -\left(\zeta_{2} |\sigma_{2}|^{\alpha_{2}} + \xi_{2} |\sigma_{2}|^{\beta_{2}}\right)^{\kappa_{2}} <0,\;\forall\,\sigma_{2}\neq 0,\label{eq:V2dot}
	\end{align}
	from which it is immediate that sliding mode is enforced on $\sigma_{2}$ within a time, \eqref{eq:tc2}, which can be made arbitrarily small by a proper choice of the design parameters, regardless of initial error in impact time of the defender \cite{6104367}. This leads to an interception of the pursuer by the defender at the desired impact time, $T_f$.
\end{proof}	
The defender's strategy is based on deviated pursuit, as observed from the first term of \eqref{eq:aDindirectAgg}, with second and third terms accounting for the pursuer's maneuver and impact time error correction, respectively. With the proposed dynamics of $\sigma_{2}$, the defender's time-to-go converges to its desired value within a fixed-time irrespective of the initial error. Since the initial time-to-go values of the adversaries depend on the engagement geometry between them, it can be concluded that by using the law, \eqref{eq:aDindirectAgg}, the defender arrives at the requisite collision path in a fixed-time irrespective of its initial engagement geometry with the pursuer. As soon as the estimation error, $\tilde{e}$, becomes zero, the defender's guidance system compensates for the pursuer's maneuver, leading the second term in \eqref{eq:aDindirectAgg} to zero. The LOS rate, $\dot{\lambda}_{\mathrm{DP}}$, becomes zero when the defender is on the collision course with the pursuer, which can happen only when the defender is on a suitable path that would lead to interception of the pursuer at a desired time. Hence, the numerator of the third term of \eqref{eq:aDindirectAgg} goes to zero faster than its denominator, making the third term zero, once sliding mode is enforced on \eqref{eq:s2}. This prevents any singularity in the expression of the defender's control law. 

With a proper choice of parameters, $\tilde{e}$, $\sigma_1$, and $\sigma_{2}$ can be made to vanish early in the engagement. This makes the defender follow a deviated pursuit trajectory toward the pursuer with a fixed deviation angle, $\delta_{\mathrm{DP}}$, leading to an interception of the pursuer at a desired impact time. The nature of evolution of the deviation angle, $\delta_{\mathrm{DP}}$, can be verified by observing its dynamics when $\sigma_{2}$ vanishes, that is, for $t\geq t_{c_{2}}$, one gets 
\begin{equation}
	\dot{\delta}_{\mathrm{DP}} = \dfrac{\left(v_\mathrm{D}^2-v_\mathrm{P}^2\right)\cos^2\delta_{\mathrm{D}}}{r_{\mathrm{DP}}v_{\lambda_{\mathrm{DP}}}}\left[\left(\zeta_{2} |\sigma_{2}|^{\alpha_{2}} + \xi_{2} |\sigma_{2}|^{\beta_{2}}\right)^{\kappa_{2}} +\sec\delta_{\mathrm{DP}}\,\epsilon_{2}\right]\sign(\sigma_{2}) = 0.
\end{equation}

Essentially, indirect cooperation is similar to classical impact time guidance between two agents \cite{1597196,7376242,Cho2015,9000526}. Here, we have used deviated pursuit to design the defender's strategy. However, any other strategy that ensures interception of a maneuvering adversary at a desired impact time can also be used in this design.

\subsection{Direct Cooperation}
This scheme is useful when the agents have sufficient communication capabilities. In direct cooperation, the defender aims to capture the pursuer at a certain time instant which is $t_\mathrm{margin}$ sec before the time instant when the pursuer would have captured the evader, where $t_\mathrm{margin}>0$ is a fixed time-margin. Therefore, exact information of initial $t_\mathrm{go}^\mathrm{EP}$ is not needed, and the defender can adapt as the engagement proceeds. In this cooperative scheme, the guidance strategy for the defender is slightly modified when compared to the case of indirect cooperation. 

The above condition can be translated mathematically to derive the guidance strategy of the defender. Consider the sliding manifold representing the error between the estimated time-to-go of the pursuer-evader engagement and that of the defender-pursuer one, that is
\begin{equation}\label{eq:s2direct}
	\sigma_{3} = t_{\mathrm{go}}^\mathrm{DP} - (t_\mathrm{go}^\mathrm{EP} - t_\mathrm{margin}),
\end{equation}
where $t_\mathrm{margin}>0$ is the constant time-margin which we want to keep between the expected interception of the two pairs of adversaries (pursuer-evader and defender-pursuer pairs).
%, and must be chosen so that the evader avoids falling within the lethal radius of the defenders or that of the pursuer. Therefore, we have
%\begin{equation*}
%	t_\mathrm{margin}>\max\left\{\dfrac{r_{\mathrm{D}_i}^\mathrm{L}}{\left(v_{\mathrm{D}_i}+v_\mathrm{P}\right)},~\dfrac{r_{\mathrm{D}_i}^\mathrm{L}}{\left(v_{\mathrm{D}_i}+v_\mathrm{E}\right)},~\dfrac{r_\mathrm{P}^\mathrm{L}}{\left(v_\mathrm{E}+v_\mathrm{P}\right)}\right\},
%\end{equation*} 
%where $r_{\mathrm{D}_i}^\mathrm{L}$ and $r_\mathrm{P}^\mathrm{L}$ are the lethal radii of the $i$\textsuperscript{th} defender and the pursuer, respectively. Note that with a zero lethal radius, $t_\mathrm{margin}>0$.
\begin{theorem}
	For a three-agent engagement, whose kinematics is governed by \eqref{eq:engdyn}, if the evader-defender team uses direct cooperation, and the aggressive strategy for the defender is designed as
	\begin{align}\label{eq:aDdirectAgg}
		a_{\mathrm{D}} =& v_{\mathrm{D}}\dot{\lambda}_{\mathrm{DP}} - \dfrac{v_{\lambda_\mathrm{EP}}^2 v_{\mathrm{D}}\left(v_\mathrm{D}^2-v_\mathrm{P}^2\right)}{r_{\mathrm{DP}}\sec^2\delta_{\mathrm{DP}}v_{r_\mathrm{EP}}^2 v_{\lambda_{\mathrm{DP}}}} - \dfrac{r_\mathrm{EP}\sin\delta_\mathrm{EP}v_{\mathrm{D}}\left(v_\mathrm{D}^2-v_\mathrm{P}^2\right)}{r_{\mathrm{DP}}\sec^2\delta_{\mathrm{DP}}v_{r_\mathrm{EP}}^2 v_{\lambda_{\mathrm{DP}}}}a_\mathrm{E}\nonumber\\
		& + \dfrac{v_{\mathrm{D}}\left[ r_\mathrm{EP}\left(v_\mathrm{D}^2-v_\mathrm{P}^2\right)\sin\delta_\mathrm{PE}\cos\delta_{\mathrm{DP}}-v_{r_\mathrm{EP}}^2r_{\mathrm{DP}}\sin\left(\delta_{\mathrm{PD}}+\delta_{\mathrm{DP}}\right)\right]}{r_{\mathrm{DP}}\sec\delta_{\mathrm{DP}}v_{\lambda_{\mathrm{DP}}}v_{r_\mathrm{EP}}^2}\hat{a}_\mathrm{P}\nonumber\\
		&+\dfrac{v_{\mathrm{D}}\left(v_\mathrm{D}^2-v_\mathrm{P}^2\right)}{r_{\mathrm{DP}}\sec^2\delta_{\mathrm{DP}}v_{\lambda_{\mathrm{DP}}}}\left[\left(\zeta_{3} |\sigma_{3}|^{\alpha_{3}} + \xi_{3} |\sigma_{3}|^{\beta_{3}}\right)^{\kappa_{3}} +\sec\delta_{\mathrm{DP}}\,\epsilon_{3}\right]\sign(\sigma_{3}),
	\end{align}
	where the parameters $\zeta_{3},\xi_{3},\alpha_{3},\beta_{3},\kappa_{3}>0$ satisfy constraints $\alpha_{3}\kappa_{3}<1$, $\beta_{3}\kappa_{3}>1$, and\newline $\epsilon_{3}>\sup_{t\geq 0}\left\{\dfrac{r_\mathrm{EP}\left(v_\mathrm{D}^2-v_\mathrm{P}^2\right)+r_{\mathrm{DP}}v_{r_\mathrm{EP}}^2}{\left(v_\mathrm{D}^2-v_\mathrm{P}^2\right)  v_{r_\mathrm{EP}}^2}\tilde{e}_\mathrm{max}\right\}$, then the defender converges to the deviated pursuit course in a fixed-time,
	\begin{equation}\label{eq:tc20}
		t_{c_{3}}\leq\dfrac{1}{\zeta_{3}^{\kappa_{3}}(1-\alpha_{3}\kappa_{3})}+\dfrac{1}{\xi_{3}^{\kappa_{3}}(\beta_{3}\kappa_{3}-1)}<\min\left\{t_{\mathrm{go}}^\mathrm{DP},t_\mathrm{go}^\mathrm{EP}- t_\mathrm{margin}\right\},
	\end{equation}
	and captures the pursuer in a time-margin of $t_\mathrm{margin}$ s.
\end{theorem}
\begin{proof}
	Differentiating \eqref{eq:s2direct} with respect to time, and using \eqref{eq:tgoEPdot} and \eqref{eq:tgoDPdot}, one may obtain
	\begin{align}
		\dot{\sigma}_{3} =& \left(\dfrac{v_{\lambda_{\mathrm{DP}}}^2 }{\left(v_\mathrm{D}^2-v_\mathrm{P}^2\right)\cos^2\delta_{\mathrm{DP}}} - \dfrac{v_{\lambda_\mathrm{EP}}^2}{v_{r_\mathrm{EP}}^2}\right) - \dfrac{{r}_\mathrm{EP}\sin\delta_\mathrm{EP}}{v_{r_\mathrm{EP}}^2}a_\mathrm{E} - \dfrac{r_{\mathrm{DP}}v_{\lambda_{\mathrm{DP}}}\sec^2\delta_{\mathrm{DP}}}{v_{\mathrm{D}}\left(v_\mathrm{D}^2-v_\mathrm{P}^2\right)} a_{\mathrm{D}} \nonumber\\
		&+ \left(\dfrac{{r}_\mathrm{EP}\sin\delta_\mathrm{PE}}{v_{r_\mathrm{EP}}^2} - \dfrac{r_{\mathrm{DP}}\sin\left(\delta_{\mathrm{PD}}+\delta_{\mathrm{DP}}\right)}{\left(v_\mathrm{D}^2-v_\mathrm{P}^2\right)\cos\delta_{\mathrm{DP}}}\right)a_\mathrm{P},\label{eq:s2directdot}
	\end{align}
	which includes the information of the evader's guidance strategy, $a_\mathrm{E}$, obtained previously. To derive the guidance strategy for the defender in this cooperative scheme, consider the Lyapunov function candidate, $\mathcal{V}_{3}=|\sigma_{3}|$, whose time differentiation now yields
	\begin{align}\label{eq:V2dotstep1direct}
		\dot{\mathcal{V}}_{3} =& \sign(\sigma_{3})\left[\left(\dfrac{v_{\lambda_{\mathrm{DP}}}^2 }{\left(v_\mathrm{D}^2-v_\mathrm{P}^2\right)\cos^2\delta_{\mathrm{DP}}} - \dfrac{v_{\lambda_\mathrm{EP}}^2}{v_{r_\mathrm{EP}}^2}\right) - \dfrac{{r}_\mathrm{EP}\sin\delta_\mathrm{EP}}{v_{r_\mathrm{EP}}^2}a_\mathrm{E} - \dfrac{r_{\mathrm{DP}}v_{\lambda_{\mathrm{DP}}}\sec^2\delta_{\mathrm{DP}}}{v_{\mathrm{D}}\left(v_\mathrm{D}^2-v_\mathrm{P}^2\right)} a_{\mathrm{D}} \right.\nonumber\\
		&\left.+ \left(\dfrac{{r}_\mathrm{EP}\sin\delta_\mathrm{PE}}{v_{r_\mathrm{EP}}^2} - \dfrac{r_{\mathrm{DP}}\sin\left(\delta_{\mathrm{PD}}+\delta_{\mathrm{DP}}\right)}{\left(v_\mathrm{D}^2-v_\mathrm{P}^2\right)\cos\delta_{\mathrm{DP}}}\right)a_\mathrm{P}\right].
	\end{align}
	If the defender's guidance command is chosen as the one given in \eqref{eq:aDdirectAgg}, then  $\dot{\mathcal{V}}_3$ in \eqref{eq:V2dotstep1direct} simplifies to
	\begin{align}
		\dot{\mathcal{V}}_3 =& -\sign(\sigma_{3})\left(\left[\left(\zeta_{3} |\sigma_{3}|^{\alpha_{3}} + \xi_{3} |\sigma_{3}|^{\beta_{3}}\right)^{\kappa_{3}} +\sec\delta_{\mathrm{DP}}\,\epsilon_{3}\right]\sign(\sigma_{3})\right.\nonumber\\
		&\left.+\dfrac{r_\mathrm{EP}\left(v_\mathrm{D}^2-v_\mathrm{P}^2\right)\sin\delta_\mathrm{PE}\cos\delta_{\mathrm{DP}}-r_{\mathrm{DP}}v_{r_\mathrm{EP}}^2\sin\left(\delta_{\mathrm{PD}}+\delta_{\mathrm{DP}}\right)}{\left(v_\mathrm{D}^2-v_\mathrm{P}^2\right) \cos\delta_{\mathrm{DP}} v_{r_\mathrm{EP}}^2}\left(a_\mathrm{P}-\hat{a}_\mathrm{P}\right)\right)\nonumber\\
		=& -\left(\zeta_{3} |\sigma_{3}|^{\alpha_{3}} + \xi_{3} |\sigma_{3}|^{\beta_{3}}\right)^{\kappa_{3}} -\sec\delta_{\mathrm{DP}}\,\epsilon_{3}\nonumber\displaybreak\\
		&-\dfrac{r_\mathrm{EP}\left(v_\mathrm{D}^2-v_\mathrm{P}^2\right)\sin\delta_\mathrm{PE}\cos\delta_{\mathrm{DP}}-r_{\mathrm{DP}}v_{r_\mathrm{EP}}^2\sin\left(\delta_{\mathrm{PD}}+\delta_{\mathrm{DP}}\right)}{\left(v_\mathrm{D}^2-v_\mathrm{P}^2\right) \cos\delta_{\mathrm{DP}} v_{r_\mathrm{EP}}^2}\tilde{e}\,\sign(\sigma_3)\nonumber\\
		\leq & -\left(\zeta_{3} |\sigma_{3}|^{\alpha_{3}} + \xi_{3} |\sigma_{3}|^{\beta_{3}}\right)^{\kappa_{3}} \nonumber \\
		& -\sec\delta_{\mathrm{DP}}\left(\epsilon_{3}-\dfrac{r_\mathrm{EP}\left(v_\mathrm{D}^2-v_\mathrm{P}^2\right)\sin\delta_\mathrm{PE}\cos\delta_{\mathrm{DP}}-r_{\mathrm{DP}}v_{r_\mathrm{EP}}^2\sin\left(\delta_{\mathrm{PD}}+\delta_{\mathrm{DP}}\right)}{\left(v_\mathrm{D}^2-v_\mathrm{P}^2\right)  v_{r_\mathrm{EP}}^2}\tilde{e}\right)\nonumber\\
		\leq & -\left(\zeta_{3} |\sigma_{3}|^{\alpha_{3}} + \xi_{3} |\sigma_{3}|^{\beta_{3}}\right)^{\kappa_{3}} -\left(\epsilon_{3}-\dfrac{r_\mathrm{EP}\left(v_\mathrm{D}^2-v_\mathrm{P}^2\right)+r_{\mathrm{DP}}v_{r_\mathrm{EP}}^2}{\left(v_\mathrm{D}^2-v_\mathrm{P}^2\right)  v_{r_\mathrm{EP}}^2}\tilde{e}_\mathrm{max}\right)\nonumber\\
		\leq & -\left(\zeta_{3} |\sigma_{3}|^{\alpha_{3}} + \xi_{3} |\sigma_{3}|^{\beta_{3}}\right)^{\kappa_{3}} <0\quad\forall\;\sigma_{3}\neq 0,\label{eq:V2dotdirect}
	\end{align}
	inferring that sliding mode is enforced on $\sigma_3$ in a time, $t_{c_{3}}$, regardless of initial value of $\sigma_{3}$ \cite{6104367}, leading to an interception of the pursuer in $t_\mathrm{margin}$. This concludes the proof.
\end{proof}	
In direct cooperation, the defender's guidance strategy uses the information of the pursuer-evader engagement, which is evident from various coupled terms in the control law \eqref{eq:aDdirectAgg} pertaining to the pursuer-evader engagement. As the error in \eqref{eq:s2direct} vanishes, the term
\begin{align*}
	&\dfrac{v_{\mathrm{D}}\left(v_\mathrm{D}^2-v_\mathrm{P}^2\right)}{r_{\mathrm{DP}}\sec^2\delta_{\mathrm{DP}}v_{\lambda_{\mathrm{DP}}}}\left[\left(\zeta_{3} |\sigma_{3}|^{\alpha_{3}} + \xi_{3} |\sigma_{3}|^{\beta_{3}}\right)^{\kappa_{3}} +\sec\delta_{\mathrm{DP}}\,\epsilon_{3}\right]\sign(\sigma_{3})%\\
	%=& v_{\mathrm{D}}\dot{\delta}_{\mathrm{DP}}\left[\left(\zeta_{3} |\sigma_{3}|^{\alpha_{3}} + \xi_{3} |\sigma_{3}|^{\beta_{3}}\right)^{\kappa_{3}} +\sec\delta_{\mathrm{DP}}\,\epsilon_{3}\right]\sign(\sigma_{3})
\end{align*} 
becomes zero. With the evader and the nonlinear finite-time disturbance observer achieving their objectives, rest of the terms in the expression of $a_\mathrm{D}$ in \eqref{eq:aDdirectAgg} vanish, except the first term, which corresponds to the deviated pursuit law. After $\sigma_{3}=0$, $t_\mathrm{margin}= t_\mathrm{go}^\mathrm{EP} - t_{\mathrm{go}}^\mathrm{DP}$ is maintained by the defender to capture the pursuer within the desired time-margin. 
\begin{remark}
	By maintaining a constant time-margin (as a designer's choice) between the estimated time-to-go of the pursuer-evader engagement and that of the defender-pursuer one, the defender is always able to capture the pursuer, before the latter can capture the evader, but it requires greater interaction between the evader and the defender.
\end{remark}

\section{Defensive Stance of the Defender}\label{sec:defenderLawDef}
Consider an asset guarding scenario wherein a mobile high-value asset is being targeted by an attacker. To protect itself from the threat, the asset calls for a protective vehicle as a backup to shield it from the incoming threat. This situation may arise in cases such as surveillance, patrolling, and secure transportation. In this scenario, the asset is the evader, the attacker is the pursuer and the protective vehicle is the defender.  Thus, the defender needs to act defensive in this case. The defender is said to be in defensive stance when it is required to rendezvous with the evader in a lesser time than what the pursuer would take to do the same.

Similar to the previous design, the defender can either use indirect cooperation or direct cooperation to achieve its objective, which is to rendezvous with the evader before the pursuer can capture the evader. In order to prevent the evader from being captured by the pursuer, the defender's strategy should be designed to be an anticipatory one. 
%Rather than pointing directly to the evader, the defender employs deviated pursuit law (as used in the previous design), which is a variant of ordinary pursuit law. 
With the deviated pursuit control law, the defender always points somewhat ahead of the defender-evader LOS, $\lambda_\mathrm{DE}$, and anticipates to arrive at a location where the evader will shortly be, thereby reducing the chances of missing the evader. The defender now aims to maintain a constant deviation angle, $\delta_\mathrm{DE}$, to the defender-evader LOS.
%Unlike the ordinary pursuit law, wherein the defender would directly follow the maneuvering evader's position, the defender now aims to maintain a constant deviation angle, $\delta_\mathrm{DE}$, to the defender-evader LOS.

The time-to-go for the defender-evader engagement can be expressed as
\begin{equation}\label{eq:tgoDE} 
	t_\mathrm{go}^\mathrm{DE} = \dfrac{r_\mathrm{DE}\sec\delta_\mathrm{DE}}{v_\mathrm{D}^2 - v_\mathrm{E}^2}\left[v_\mathrm{D}+v_\mathrm{E}\cos\left(\delta_\mathrm{ED}+\delta_\mathrm{DE}\right)\right]= \dfrac{r_\mathrm{DE}\left(v_{r_\mathrm{DE}}+2v_\mathrm{D}\cos\delta_\mathrm{DE}-v_{\lambda_\mathrm{DE}}\tan\delta_\mathrm{DE}\right)}{v_\mathrm{D}^2 - v_\mathrm{E}^2},
\end{equation} 
whose dynamics is given by
\begin{equation}\label{eq:tgoDEdot}
	\dot{t}_\mathrm{go}^\mathrm{DE} = -1 + \dfrac{r_\mathrm{DE}^2\dot{\lambda}_\mathrm{DE}^2 \sec^2\delta_\mathrm{DE}}{v_\mathrm{D}^2 - v_\mathrm{E}^2} - \dfrac{r_\mathrm{DE}^2\dot{\lambda}_\mathrm{DE}\sec^2\delta_\mathrm{DE}}{v_\mathrm{D}\left(v_\mathrm{D}^2 - v_\mathrm{E}^2\right)} a_\mathrm{D} -\dfrac{r_\mathrm{DE}\sin\left(\delta_\mathrm{ED}+\delta_\mathrm{DE}\right)\sec\delta_\mathrm{DE}}{v_\mathrm{D}^2 - v_\mathrm{E}^2}a_\mathrm{E},
\end{equation}
where $v_\mathrm{D}> v_\mathrm{E}$. The expression in \eqref{eq:tgoDE} is an exact value when the evader does not maneuver \cite{shneydor1998missile}. Hence, the pursuer-evader pair becoming non-maneuvering early would help the defender in accurately fixing its own time of rendezvous with the evader. From \eqref{eq:tgoDE}, it is also worth noting that $t_\mathrm{go}^\mathrm{DE} =0\iff r_\mathrm{DE}=0$, which is similar to the previous scenario. This means that even if the evader maneuvers, the defender can still use \eqref{eq:tgoDE} as an estimate for deciding its time of rendezvous with the evader. However, in such a case, \eqref{eq:tgoDE}, is not exact. These observations are similar to the previous case when the defender was required to rendezvous with the pursuer. Based on a similar premise, we proceed to design the defender's defensive strategy using various cooperative modes.

\subsection{Indirect Cooperation}
At the beginning of the engagement, the pursuer's time to capture the evader can be estimated by the evader, and the evader shares this information with the defender. Using this information, the defender decides at the onset of the engagement a suitable time of rendezvous with the evader, $T_f$, which is sufficiently lower than the initial value of $t_\mathrm{go}^\mathrm{EP}$. Note that if the pursuer becomes non-maneuvering and stays on the collision course with the evader, then the estimates of $t_\mathrm{go}^\mathrm{DE}$ and $t_\mathrm{go}^\mathrm{EP}$ become accurate. However, the pursuer becoming non-maneuvering is not a stringent requirement for the evader to be protected. The following theorem summarizes the defender's strategy when it uses this cooperation with a defensive stance.
\begin{theorem}
	For a three-agent engagement whose kinematics is governed by \eqref{eq:engdyn}, if the defender effects a defensive control law, using indirect cooperation, given by
	\begin{align}
		a_\mathrm{D} = &v_\mathrm{D}\dot{\lambda}_\mathrm{DE} + \dfrac{v_\mathrm{D}\left(v_\mathrm{D}^2 - v_\mathrm{E}^2\right)\cos^2\delta_\mathrm{DE}}{r_\mathrm{DE}v_{\lambda_\mathrm{DE}}}\left\{\left(\zeta_4 |\sigma_4|^{\alpha_4} + \xi_4 |\sigma_4|^{\beta_4}\right)^{\kappa_4}+ \sec\delta_\mathrm{DE}\,\epsilon_4\right\} \sign(\sigma_4) ,\label{eq:aDindirectDef}
	\end{align}
	where the sliding manifold, $\sigma_4 = t_\mathrm{go}^\mathrm{DE} - \left(T_f - t\right)$, and the parameters $\zeta_4,\xi_4, \alpha_4,\beta_4,\kappa_4>0$ satisfy constraints $\alpha_4\kappa_4<1$, $\beta_4\kappa_4>1$, and $\epsilon_4>\sup_{t\geq 0}\dfrac{r_\mathrm{DE}}{v_\mathrm{D}^2 - v_\mathrm{E}^2}a_\mathrm{E}^\mathrm{max}$, then the defender converges on the deviated pursuit trajectory in a fixed-time, 
	\begin{equation}
		t_{c_4}\leq\dfrac{1}{\zeta_4(1-\alpha_4\kappa_4)} + \dfrac{1}{\xi_4(\beta_4\kappa_4-1)}<\min\left\{t_{\mathrm{go}}^\mathrm{DE},t_\mathrm{go}^\mathrm{EP}\right\},
	\end{equation}
	eventually leading it to the evader at the chosen time of rendezvous, $T_f$.
\end{theorem}
\begin{proof}
	Taking time derivative of $\sigma_4$ yields
	\begin{align}
		\dot{\sigma}_4 =& 1+\dot{t}_\mathrm{go}^\mathrm{DE}  = \dfrac{r_\mathrm{DE}^2\dot{\lambda}_\mathrm{DE}^2 \sec^2\delta_\mathrm{DE}}{v_\mathrm{D}^2 - v_\mathrm{E}^2} - \dfrac{r_\mathrm{DE}^2\dot{\lambda}_\mathrm{DE}\sec^2\delta_\mathrm{DE}}{v_\mathrm{D}\left(v_\mathrm{D}^2 - v_\mathrm{E}^2\right)} a_\mathrm{D} -\dfrac{r_\mathrm{DE}\sin\left(\delta_\mathrm{ED}+\delta_\mathrm{DE}\right)\sec\delta_\mathrm{DE}}{v_\mathrm{D}^2 - v_\mathrm{E}^2}a_\mathrm{E}.\label{eq:s4dot}
	\end{align}
	In a manner similar to the one used previously, we construct a Lyapunov function candidate $\mathcal{V}_4 = |\sigma_4|$, differentiating which with respect to time yields $\dot{\mathcal{V}}_4 = \sign(\sigma_4)\dot{\sigma}_4 = \sign(\sigma_4)\left(1+\dot{t}_\mathrm{go}^\mathrm{DE}\right)$.
	On substituting the defender's control law, \eqref{eq:aDindirectDef}, in $\dot{\mathcal{V}}_4$, and performing simplifications using \eqref{eq:s4dot}, one may obtain
	\begin{align}
		\dot{\mathcal{V}}_4 =&  -\sec\delta_\mathrm{DE}\left[\epsilon_4-\dfrac{\sign(\sigma_4)r_\mathrm{DE}\sin\left(\delta_\mathrm{DE}+\delta_\mathrm{ED}\right)}{v_\mathrm{D}^2 - v_\mathrm{E}^2}a_\mathrm{E}\right]-\left(\zeta_4 |\sigma_4|^{\alpha_4} + \xi_4 |\sigma_4|^{\beta_4}\right)^{\kappa_4}\nonumber\\
		\leq & -\left(\zeta_4 |\sigma_4|^{\alpha_4} + \xi_4 |\sigma_4|^{\beta_4}\right)^{\kappa_4} - \left[\epsilon_4-\dfrac{r_\mathrm{DE}}{v_\mathrm{D}^2 - v_\mathrm{E}^2}a_\mathrm{E}^\mathrm{max}\right]\nonumber\\
		\leq & -\left(\zeta_4 |\sigma_4|^{\alpha_4} + \xi_4 |\sigma_4|^{\beta_4}\right)^{\kappa_4} < 0,~\forall~ \sigma_4 \neq 0,
	\end{align}
	inferring that the defender converges on the deviated pursuit trajectory in a fixed-time, $t_{c_4}<T_f$, that will eventually lead it to the evader in the desired time, $T_f$, regardless of the initial value of $t_\mathrm{go}^\mathrm{DE}$ \cite{6104367}. Stated differently, the defender will rendezvous with the evader in the desired time, $T_f$, irrespective of the initial defender-evader engagement geometry. This concludes the proof.
\end{proof}
The defender's control law, \eqref{eq:aDindirectDef}, consists of a deviated pursuit term, $v_\mathrm{D}\dot{\lambda}_\mathrm{DE}$, and a term to alter its trajectory towards the time-constrained deviated pursuit course. The defender's deviation angle, $\delta_\mathrm{DE}$, varies initially to place the defender on the deviated pursuit course but gets fixed up once sliding mode is enforced on $\sigma_4$. This essentially means that the defender \emph{looks ahead} toward the anticipated position of the evader in the future with a fixed deviation angle, $\delta_\mathrm{DE}$, once sliding mode is enforced. The time derivative of $\delta_\mathrm{DE}$, that is,
\begin{align}
	\dot{\delta}_\mathrm{DE} =& \dfrac{\left(v_\mathrm{D}^2 - v_\mathrm{E}^2\right)\cos^2\delta_\mathrm{DE}}{r_\mathrm{DE}v_{\lambda_\mathrm{DE}}}\left\{\left(\zeta_4 |\sigma_4|^{\alpha_4} + \xi_4 |\sigma_4|^{\beta_4}\right)^{\kappa_4} +\sec\delta_\mathrm{DE}\epsilon_4\right\} \sign(\sigma_4),\nonumber
\end{align}
becomes zero as soon as $\sigma_4=0$, thereby ensuring $\delta_\mathrm{DE}$ attains a constant value when the defender is on the deviated pursuit course towards the evader.

\subsection{Direct Cooperation}
While the defender uses direct cooperation with a defensive stance, it aims to rendezvous with the evader at a certain time instant which is ${t}_\mathrm{margin}$ sec before the pursuer would have captured the evader. Thus, by maintaining a fixed time-margin between $t_\mathrm{go}^\mathrm{EP}$ and $t_\mathrm{go}^\mathrm{DE}$, the defender is always able to rendezvous with the evader by being on the deviated pursuit course and maintaining a positive constant time-margin, ${t}_\mathrm{margin}>0$, before the pursuer can capture the evader. This is summarized in the following theorem.
\begin{theorem}
	For a three-agent engagement whose kinematics is governed by \eqref{eq:engdyn}, if the defender effects a defensive control law, using direct cooperation, given by
	\begin{align}
		a_\mathrm{D} =& v_\mathrm{D}\dot{\lambda}_{\mathrm{DE}} - \dfrac{v_{\lambda_\mathrm{EP}}^2 v_\mathrm{D}\left(v_\mathrm{D}^2 - v_\mathrm{E}^2\right)}{r_\mathrm{DE}v_{r_\mathrm{EP}}^2 v_{\lambda_\mathrm{DE}}\sec^2\delta_\mathrm{DE}} - \dfrac{r_\mathrm{EP}v_\mathrm{D}\left(v_\mathrm{D}^2 - v_\mathrm{E}^2\right)\sin\delta_\mathrm{EP}}{r_\mathrm{DE}v_{r_\mathrm{EP}}^2 v_{\lambda_\mathrm{DE}}\sec^2\delta_\mathrm{DE}}a_\mathrm{E} \nonumber\\
		&+ \dfrac{v_{\mathrm{D}}\left[ r_\mathrm{EP}\left(v_\mathrm{D}^2 - v_\mathrm{E}^2\right)\sin\delta_\mathrm{PE}\cos\delta_{\mathrm{DE}}-v_{r_\mathrm{EP}}^2r_{\mathrm{DE}}\sin\left(\delta_{\mathrm{ED}}+\delta_{\mathrm{DE}}\right)\right]}{r_{\mathrm{DE}}v_{\lambda_{\mathrm{DE}}}v_{r_\mathrm{EP}}^2\sec\delta_{\mathrm{DE}}}\hat{a}_\mathrm{P}\nonumber\\
		&+\dfrac{v_\mathrm{D}\left(v_\mathrm{D}^2 - v_\mathrm{E}^2\right)}{r_\mathrm{DE}v_{\lambda_\mathrm{DE}}\sec^2\delta_\mathrm{DE}} \left\{\left(\zeta_5 |\sigma_5|^{\alpha_5} + \xi_5 |\sigma_5|^{\beta_5}\right)^{\kappa_5} +\sec\delta_\mathrm{DE}\epsilon_5\right\}\sign(\sigma_5),\label{eq:aDdirectDef}
	\end{align}
	where the sliding manifold, $\sigma_5 = t_\mathrm{go}^\mathrm{DE} - t_\mathrm{go}^\mathrm{EP} + {t}_\mathrm{margin}$, $\epsilon_5>\sup_{t\geq 0}\dfrac{r_\mathrm{EP}\left(v_\mathrm{D}^2 - v_\mathrm{E}^2\right)+r_{\mathrm{DE}}v_{r_\mathrm{EP}}^2}{v_{r_\mathrm{EP}}^2\left(v_\mathrm{D}^2 - v_\mathrm{E}^2\right)}\tilde{e}_\mathrm{max}$, and the parameters $\zeta_5,\xi_5,\alpha_5,\beta_5,\kappa_5>0$ satisfy  $\alpha_5\kappa_5<1$, $\beta_5\kappa_5>1$, then the defender converges to the deviated pursuit trajectory in a fixed-time, 
	\begin{equation}
		t_{c_5} \leq \dfrac{1}{\zeta_5^{\kappa_5}(1-\alpha_5\kappa_5)}+\dfrac{1}{\xi_5^{\kappa_5}(\beta_5\kappa_5-1)}<\min\left\{t_{\mathrm{go}}^\mathrm{DE},t_\mathrm{go}^\mathrm{EP}- t_\mathrm{margin}\right\},
	\end{equation} 
	eventually leading it to the evader ${t}_\mathrm{margin}$ sec before the pursuer can capture the evader. 
\end{theorem}

\begin{proof}
	Differentiating $\sigma_5$ with respect to time, and using the results in \eqref{eq:tgoEPdot} and \eqref{eq:tgoDEdot}, one may obtain
	\begin{align}
		\dot{\sigma}_5 =& \left(\dfrac{v_{\lambda_\mathrm{DE}}^2\sec^2\delta_\mathrm{DE} }{v_\mathrm{D}^2 - v_\mathrm{E}^2} - \dfrac{v_{\lambda_\mathrm{EP}}^2}{v_{r_\mathrm{EP}}^2}\right) - \dfrac{{r}_\mathrm{EP}\sin\delta_\mathrm{EP}}{v_{r_\mathrm{EP}}^2}a_\mathrm{E} - \dfrac{{r}_\mathrm{DE}v_{\lambda_\mathrm{DE}}\sec^2\delta_\mathrm{DE}}{v_\mathrm{D}\left(v_\mathrm{D}^2 - v_\mathrm{E}^2\right)} a_\mathrm{D}\nonumber\\
		&   + \left(\dfrac{{r}_\mathrm{EP}\sin\delta_\mathrm{PE}}{v_{r_\mathrm{EP}}^2} - \dfrac{r_\mathrm{DE}\sin\left(\delta_\mathrm{DE}+\delta_\mathrm{ED}\right)}{\left(v_\mathrm{D}^2 - v_\mathrm{E}^2\right)\cos\delta_\mathrm{DE}}\right)a_\mathrm{P},\label{eq:s5directdot}
	\end{align}
	which includes the information of the evader's engagement with the pursuer, and the evader's control law. Note that the evader's control law also includes the information about the estimate of the pursuer's maneuver. To design the feedback control law for the defender, we consider a Lyapunov function candidate, $\mathcal{V}_5=\vert\sigma_5\vert$, whose time differentiation results in $\dot{\mathcal{V}}_5 =  \sign(\sigma_5) \dot{\sigma}_5$. On substituting the defender's control law, \eqref{eq:aDdirectDef}, in $\dot{\mathcal{V}}_5$ and using \eqref{eq:s5directdot}, one may obtain
	\begin{align}
		\dot{\mathcal{V}}_5 =&-\left(\zeta_5 |\sigma_5|^{\alpha_5} + \xi_5 |\sigma_5|^{\beta_5}\right)^{\kappa_5} - \sec\delta_\mathrm{DE}\epsilon_5 \nonumber\\
		&- \sign(\sigma_5)\left(\dfrac{{r}_\mathrm{EP}\sin\delta_\mathrm{PE}}{v_{r_\mathrm{EP}}^2} - \dfrac{r_\mathrm{DE}\sin\left(\delta_\mathrm{DE}+\delta_\mathrm{DE}\right)}{\left(v_\mathrm{D}^2 - v_\mathrm{E}^2\right)\cos\delta_\mathrm{DE}}\right)\left(a_\mathrm{P}-\hat{a}_\mathrm{P}\right) \nonumber\\
		\leq & -\left(\zeta_5 |\sigma_5|^{\alpha_5} + \xi_5 |\sigma_5|^{\beta_5}\right)^{\kappa_5} - \left[\epsilon_5-\dfrac{r_\mathrm{EP}\left(v_\mathrm{D}^2 - v_\mathrm{E}^2\right)+r_{\mathrm{DE}}v_{r_\mathrm{EP}}^2}{v_{r_\mathrm{EP}}^2\left(v_\mathrm{D}^2 - v_\mathrm{E}^2\right)}\tilde{e}_\mathrm{max}\right]\nonumber\\
		\leq & -\left(\zeta_5 |\sigma_5|^{\alpha_5} + \xi_5 |\sigma_5|^{\beta_5}\right)^{\kappa_5} < 0~\forall~\sigma_5\neq 0,
	\end{align}
	which establishes the fixed-time convergence of the manifold, $\sigma_5$, in $t_{c_5}$. This essentially implies that the defender is able to rendezvous with the evader ${t}_\mathrm{margin}$ sec before the pursuer would have captured the evader regardless of the defender-evader initial engagement geometry. This concludes the proof.
\end{proof}
In the expression of the defender's defensive control law, \eqref{eq:aDdirectDef}, the first term corresponds to the deviated pursuit law while the second term represents the information from various engagements between the agents. The third and the fourth terms in \eqref{eq:aDdirectDef} contain the information about the evader's and the pursuer's strategies, respectively. The last term in \eqref{eq:aDdirectDef} is the fixed-time convergent sliding mode error correction term responsible for enforcing sliding mode on the manifold, $\sigma_5$. After sliding mode is enforced on $\sigma_5$, the last term becomes zero. If the evader achieves its objective, then the terms containing $a_\mathrm{E}$ and $v_{\lambda_\mathrm{EP}}$ vanish, thus leaving the deviated pursuit term (the first term in \eqref{eq:aDdirectDef}) along with the term containing the information about the pursuer's strategy. If, however, the pursuer is rendered non-maneuvering and the defender attains the deviated pursuit trajectory, then the control law, \eqref{eq:aDdirectDef}, reduces to zero in the endgame.

\begin{figure}[h!]
	\centering
	\includegraphics[width=\linewidth]{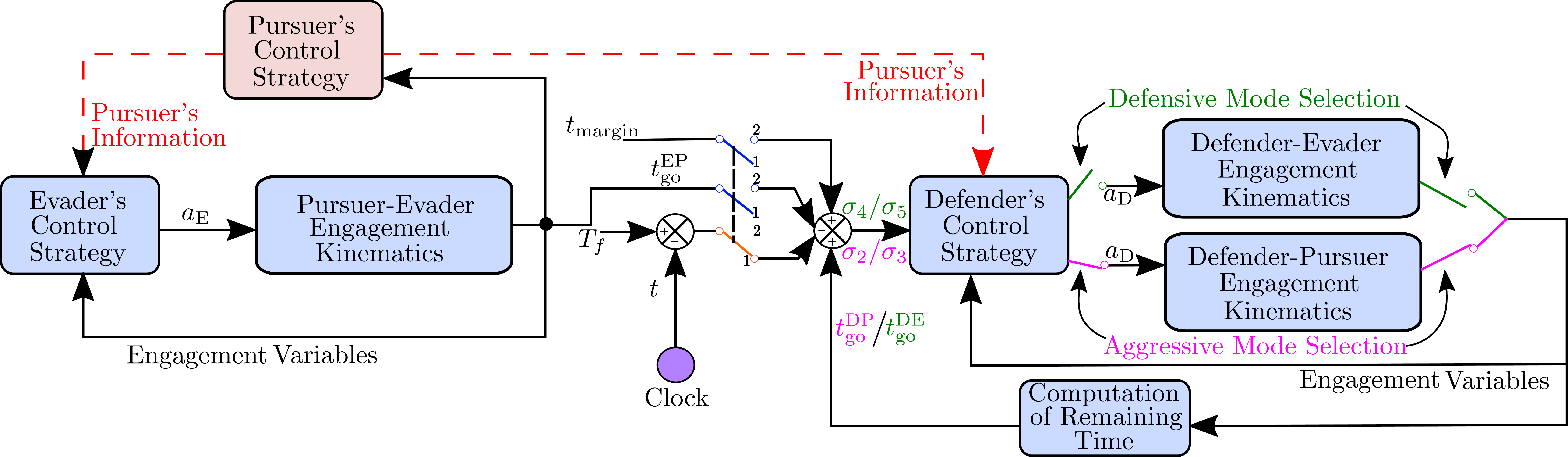}
	\caption{A schematic diagram of the proposed cooperative schemes.}
	\label{fig:schematic}
\end{figure}

A schematic of the proposed cooperation scheme has been illustrated in \Cref{fig:schematic}. The mode of cooperation must be decided before the three-agent engagement begins, which remains unchanged throughout the course of engagement. The red dashed line represent the availability of the pursuer's information. In the absence of this information, the pursuer's information is not fed to the guidance systems of the evader and the defender, and the red dashed line can be ignored.

When the coupled switches are at position `$1$', the orange switch is closed, while the blue switches are open. In this state, the evader-defender team are in indirect cooperation, in which the evader's current engagement information is unavailable to the defender. Under this scheme, the guidance strategies of the evader and the defender are independent of each other. The guidance system of the evader utilizes the information of LOS rate, $\dot{\lambda}_\mathrm{EP}$, to generate suitable guidance command, while the defender's guidance system requires the user to provide the impact time (less than interception time for evader-pursuer engagement) prior to homing. The current time is referenced from a clock, which, together with the time of interception, generates desired time-to-go.

On the contrary, if the switches are at position `$2$', the blue switches are closed, and the orange switch is open. This corresponds to direct cooperation, in which the the evader shares its engagement information with the defender. The evader's strategy remains essentially the same as that in the indirect cooperation scheme. In this scheme, explicit information of the time of interception is not required to be fed to the defender's guidance system. In addition to the time-to-go of the pursuer-evader, and the defender-pursuer engagements, a time-margin, $t_\mathrm{margin}$, is needed by the defender's guidance system to generate the cooperative control signals. This time-margin is crucial to the direct cooperative scheme, as the defender can alter its trajectory to maintain this constant time-margin in $t_\mathrm{go}^\mathrm{EP} - t_\mathrm{go}^\mathrm{DP}$ with the pursuer, and the pursuer's time-to-go, $t_\mathrm{go}^\mathrm{EP}$, need not be accurately known initially. However, in indirect cooperation, such erroneous estimate might lead to failure of mission.

Either of the two cooperative schemes can be used by the defender in a particular mission depending on the stance it takes. In \Cref{fig:schematic}, the connection through green coupled switches correspond to defender's defensive stance wherein it shields the evader from the pursuer while those via magenta coupled switches means that the defender is required to neutralize the pursuer by taking an aggressive stance. In addition to the mode of cooperation (indirect/direct), the defender's stance (aggressive/defensive) must also be decided prior to the three-agent engagement and remains unaltered throughout the engagement.

\section{Simulations}\label{sec:simulations}
The efficacy of the proposed time-constrained cooperative approach, to safeguard an evader from a pursuer using a defender, is illustrated in this section via numerical simulations. The pursuer is assumed to use a strategy that is a variant of the proportional-navigation guidance, given as $a_\mathrm{P}=-N\dot{r}_\mathrm{EP}\dot{\lambda}_\mathrm{EP}$, with the navigation constant chosen as $5$. This strategy is advantageous to the pursuer since by traveling along a trajectory with a constant bearing angle to the evader, it can be sure of eventually capturing the evader. The pursuer, then, does not chase or follow the evader, but it maneuvers in a way that will place it in the same location as the evader some time in the future. However, the proposed designs work equally well for any arbitrary maneuver that the pursuer executes. The controller parameters are selected as $\kappa_1=\kappa_{2}=\kappa_{3}=\kappa_4=\kappa_5=1$, $\beta_1=\beta_{2}=\beta_{3}=\beta_4=\beta_5=2$, $\alpha_1=0.2$, $\alpha_{2}=\alpha_{3}=\alpha_4=\alpha_5=0.3$, $\zeta_{2}=\zeta_{3}=\xi_1=0.005$, $\zeta_1=\zeta_4=\zeta_5=0.05$, $\xi_{2}=\xi_{3}=0.5$, $\xi_4=\xi_5=25$, while $\epsilon_1,\epsilon_{2},\epsilon_{3},\epsilon_4,\epsilon_5>0$. The observer parameters are chosen as $G_0=0.01$, $G_1=0.05$, $G_2=1.3$, $H_0=0.005$, $H_1=3.25$, $H_2=3.3$, and $L=0.5$. The desired time-to-go in each case is denoted by $t_\mathrm{go}^\mathrm{d}$.%=T_f-t$.

\subsection{Aggressive Mode}
When the defender takes an aggressive stance, the speeds of the evader, the pursuer, and the defender are taken to be $100$, $300$, and $400$ m/s, respectively. The maximum acceleration that a vehicle can apply is always limited in practice, hence, the lateral accelerations of the evader and the defender are limited to $5$ g and $40$ g, respectively, in simulations to reflect the agents' different maneuver capabilities. The launch position of the adversaries are represented by a circle marker ($\bullet$), while the interception/rendezvous is signified by a cross ($\times$), in the trajectory plots. The results are illustrated when the clock starts with the evader launching the defender. The initial distance between the evader and the pursuer, and the defender and the pursuer, are therefore identical, and is $10$ km. Similarly, the LOS between the pursuer-evader and the defender-pursuer pairs are also same at the time of launch, as are the heading angles of the evader and the defender.

\begin{figure}[h!]
	\centering
	\begin{subfigure}[t]{.5\textwidth}
		\centering
		\includegraphics[width=\textwidth]{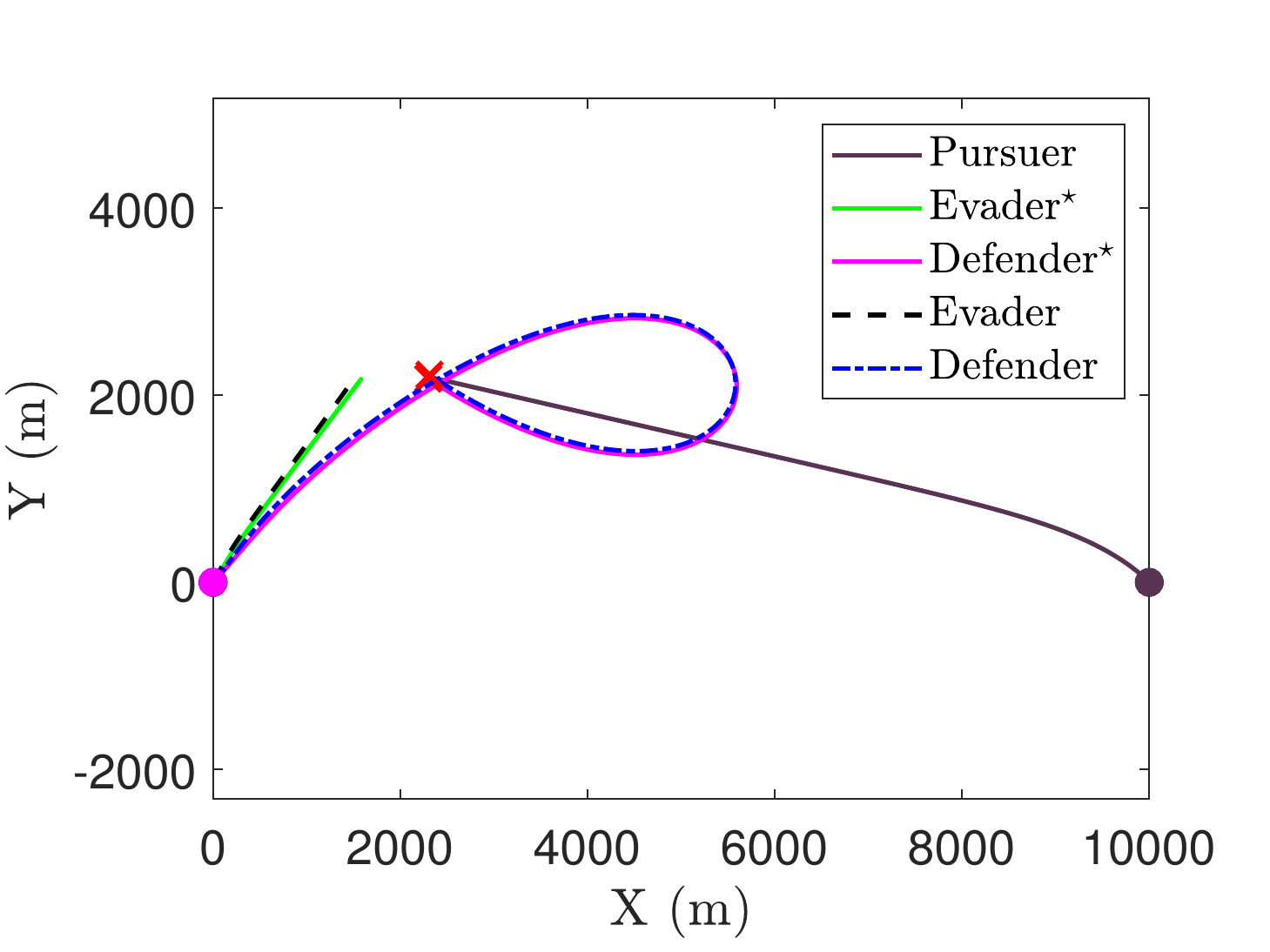}
		\caption{Trajectories.}
		\label{fig:ex27traj}
	\end{subfigure}%
	\begin{subfigure}[t]{.5\textwidth}
		\centering
		\includegraphics[width=\textwidth]{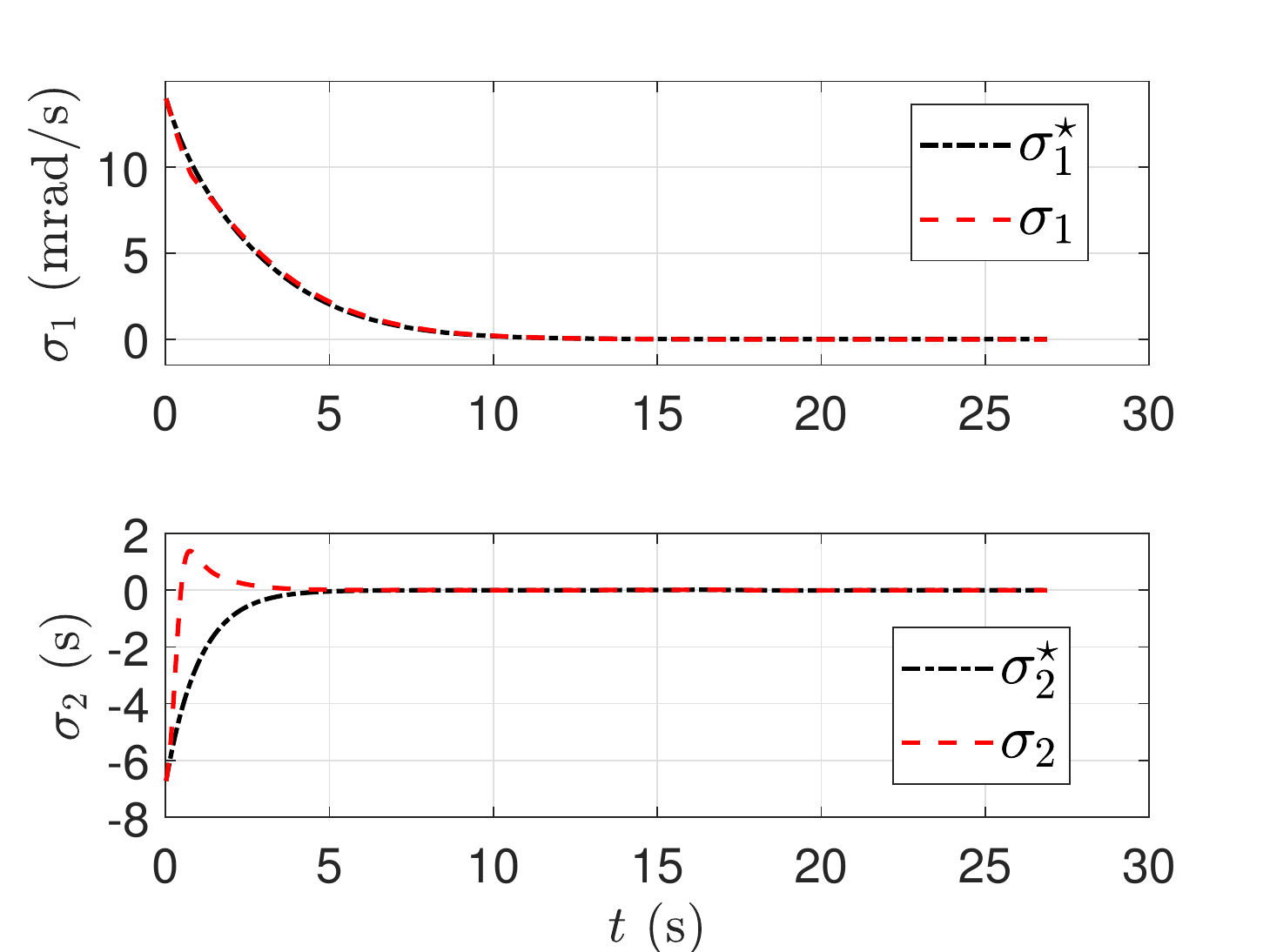}
		\caption{Sliding manifolds (error profiles).}
		\label{fig:ex27surface}
	\end{subfigure}
	\begin{subfigure}[t]{.5\textwidth}
		\centering
		\includegraphics[width=\textwidth]{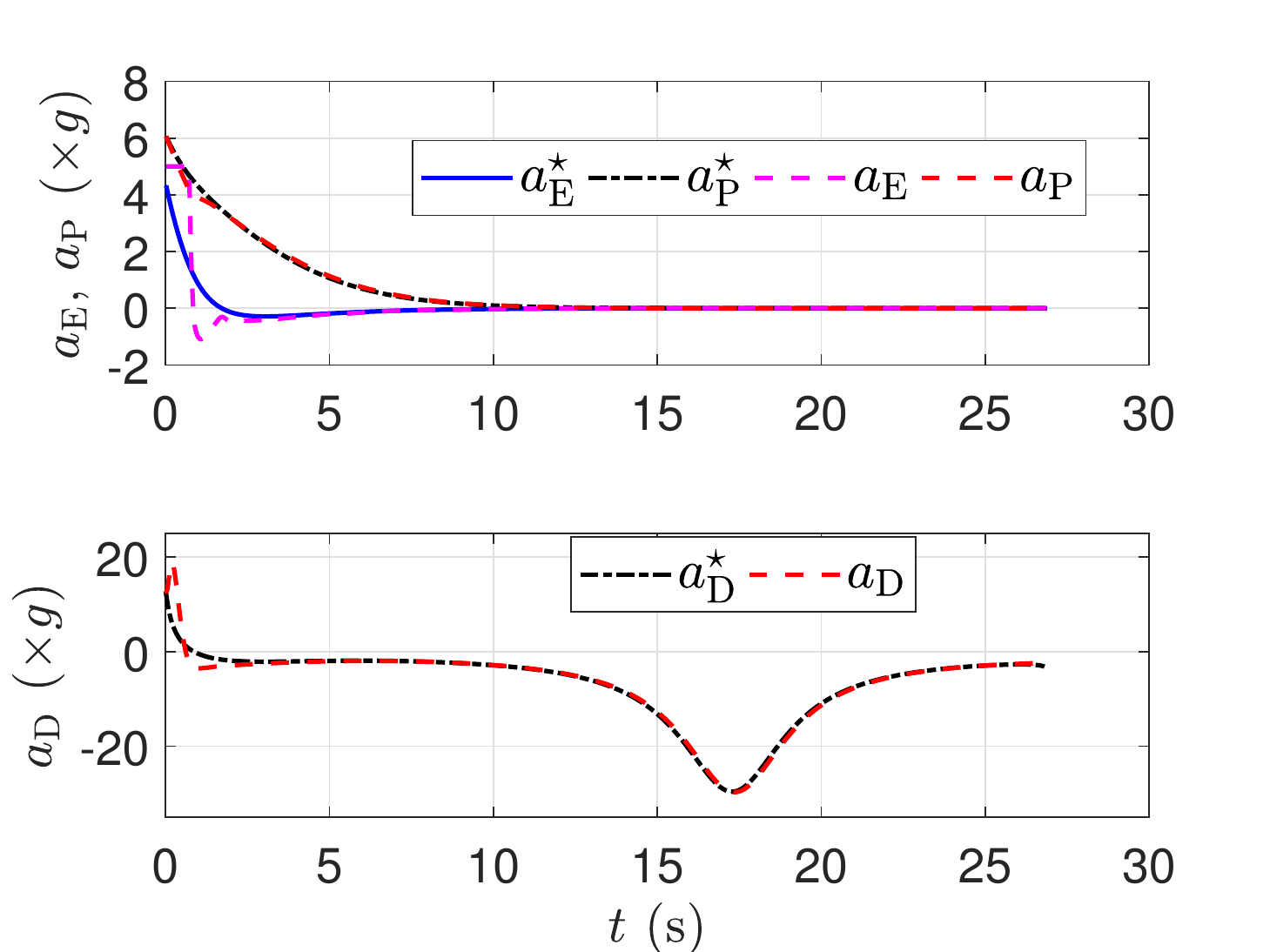}
		\caption{Lateral accelerations (steering controls).}
		\label{fig:ex27accn}
	\end{subfigure}%
	\begin{subfigure}[t]{.5\textwidth}
		\centering
		\includegraphics[width=\textwidth]{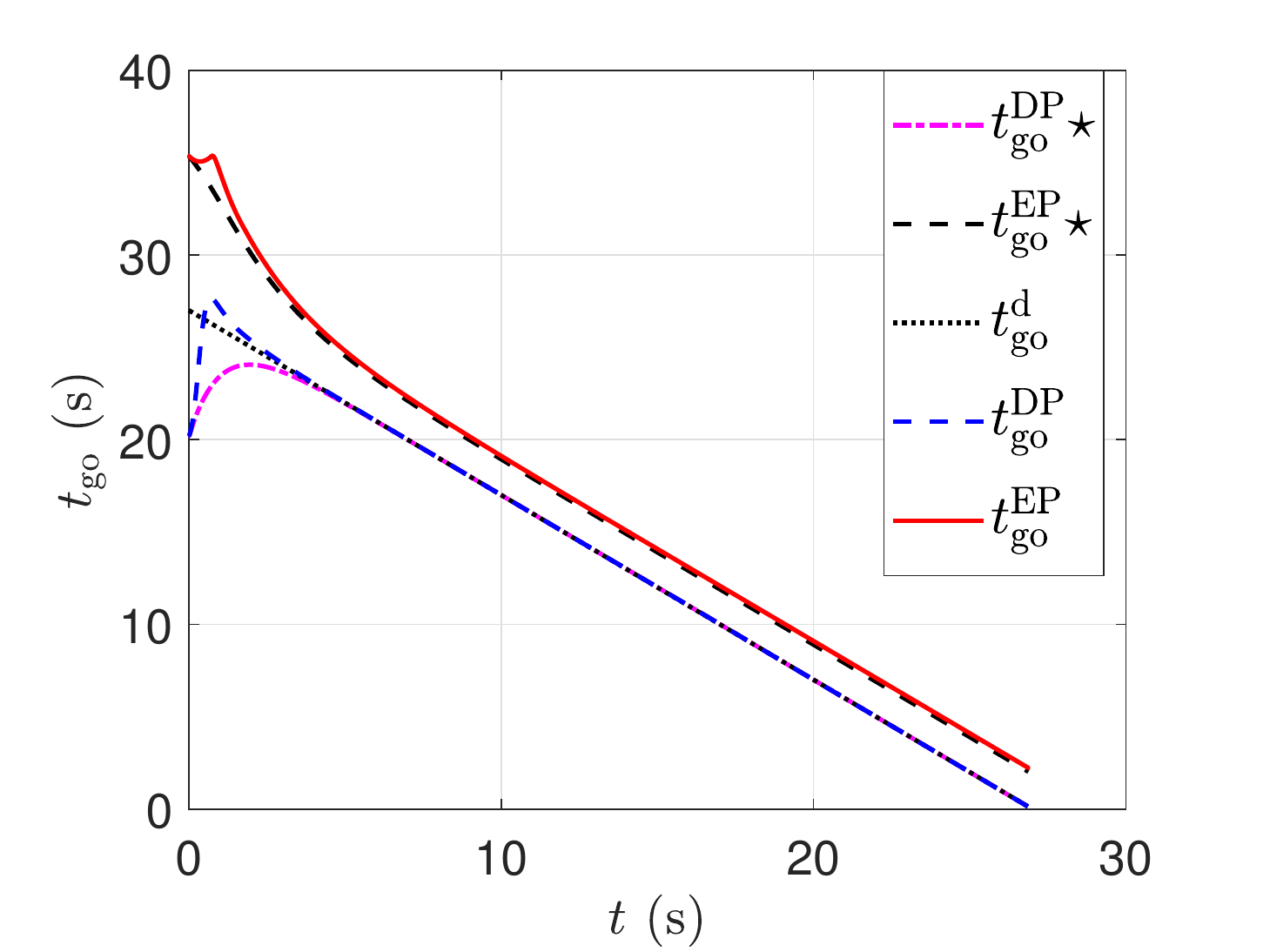}
		\caption{Time-to-go.}
		\label{fig:ex27tgo}
	\end{subfigure}
	\caption{Indirect cooperation with the defender's impact time set at $27$ s: A $\star$ indicates that $a_\mathrm{P}$ is known.}
	\label{fig:ex27s}
\end{figure}

\Cref{fig:ex27s} depicts an indirect cooperation scenario when the defender is required to intercept the pursuer at an impact time of $27$ s (chosen sufficiently less than $t_\mathrm{go}^\mathrm{EP}(0)\approxeq 36$ s). The initial LOS between the adversaries, $\lambda_\mathrm{EP}$, is $0^\circ$, and the flight path angles of the evader and the pursuer are $45^\circ$, and $135^\circ$, respectively. Results are also presented for the case when the pursuer's maneuver is known.  A $\star$ in the legends of \Cref{fig:ex27s} indicate the result when the pursuer's maneuver is identified using the results in \cite{doi:10.2514/1.49515,doi:10.2514/1.51765}. The trajectories of the adversaries are depicted in \Cref{fig:ex27traj}, which shows that successful defender-pursuer engagement is a tail-chase mode. This might come as an element of surprise for the pursuer, which may not be expecting a defender on its tail. The defender's trajectory, when $a_\mathrm{P}$ is estimated, differs initially from the one when $a_\mathrm{P}$ is completely known, but converges to the same trajectory once the estimated value resembles the actual one. This can also be observed from \Cref{fig:ex27surface}, where the defender's time-to-go error first decreases, then increases, resulting in a different initial time-to-go in the beginning (as shown in \Cref{fig:ex27tgo}). The trajectories of the agents, however, remain almost same in both scenarios. \Cref{fig:ex27surface} shows that with the proposed strategy, the LOS rate, $\dot{\lambda}_\mathrm{EP}$, becomes zero quite early in the engagement, luring the pursuer to become non-maneuvering. The defender does not have information about the pursuer-evader engagement, hence, the proposed strategy first becomes a deviated pursuit for a maneuvering target with a time-to-go estimate, \eqref{eq:tgoDP}. As soon as the pursuer becomes non-maneuvering, the estimate in \eqref{eq:tgoDP}, becomes exact, and the pursuer is guaranteed to be intercepted at the desired impact time. The profiles of the lateral accelerations (steering control signals), given by \eqref{eq:aE} and~\eqref{eq:aDindirectAgg}, are shown in \Cref{fig:ex27accn}, depicting that the accelerations/control efforts of the evader and the pursuer become zero soon after pursuer stops maneuvering. The demand in case of the defender is more in the beginning of the engagement, when $a_\mathrm{P}$ is estimated, but reduces to zero in the endgame, as discussed in the previous section. With the errors in \eqref{eq:s1} and ~\eqref{eq:s2} converging to zero, the time-to-go for the pursuer-evader engagement becomes more accurate, and that of the defender-pursuer engagement attains its desired value, as shown in \Cref{fig:ex27tgo}. In this case, the desired time-to-go for the defender is higher than its initial value. Hence, the defender adjusts its trajectory to capture the pursuer at the desired impact time of $27$ s.
\begin{figure}[h!]
	\centering
	\begin{subfigure}[t]{.5\textwidth}
		\centering
		\includegraphics[width=\textwidth]{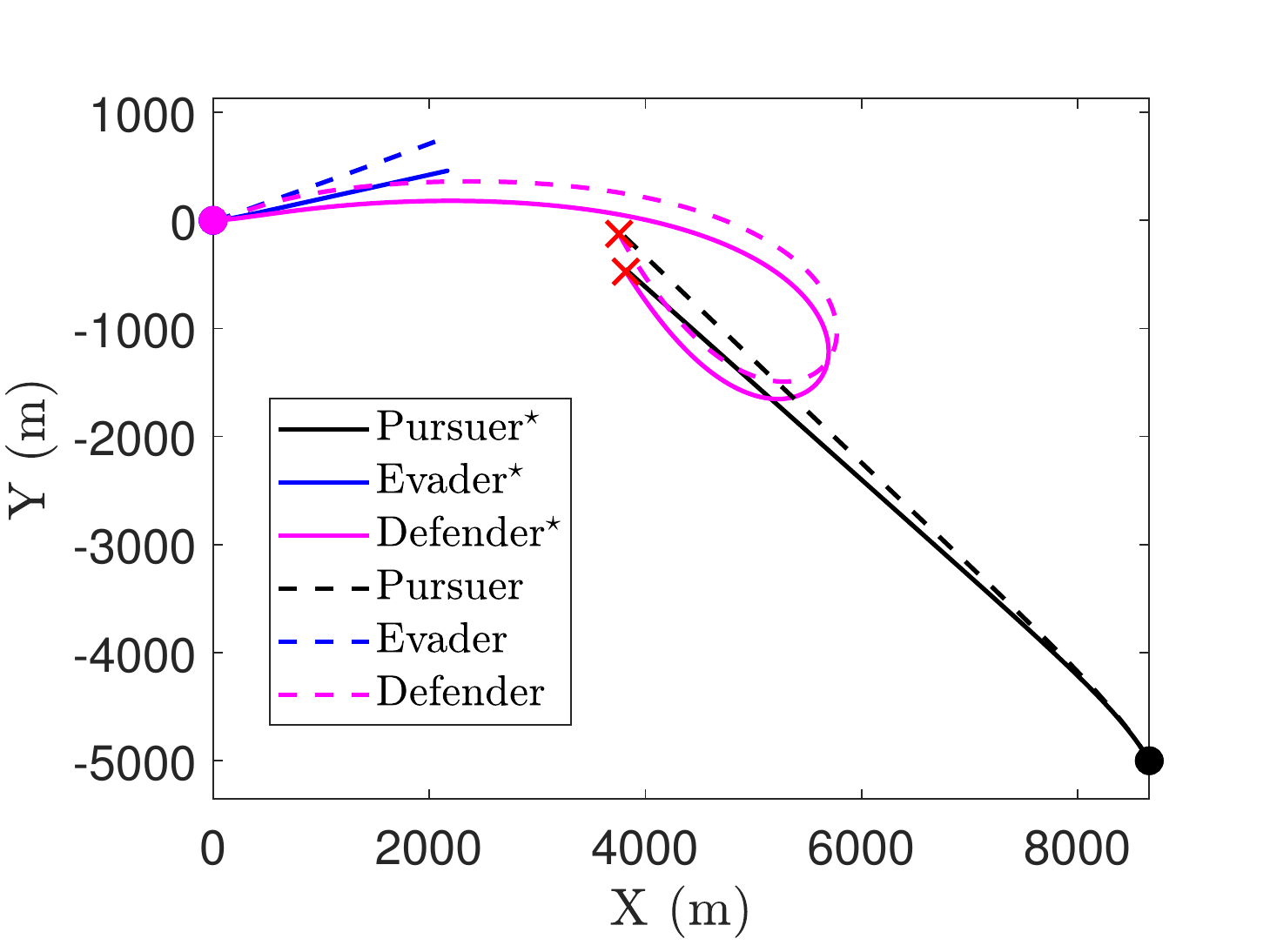}
		\caption{Trajectories.}
		\label{fig:in5traj}
	\end{subfigure}%
	\begin{subfigure}[t]{.5\textwidth}
		\centering
		\includegraphics[width=\textwidth]{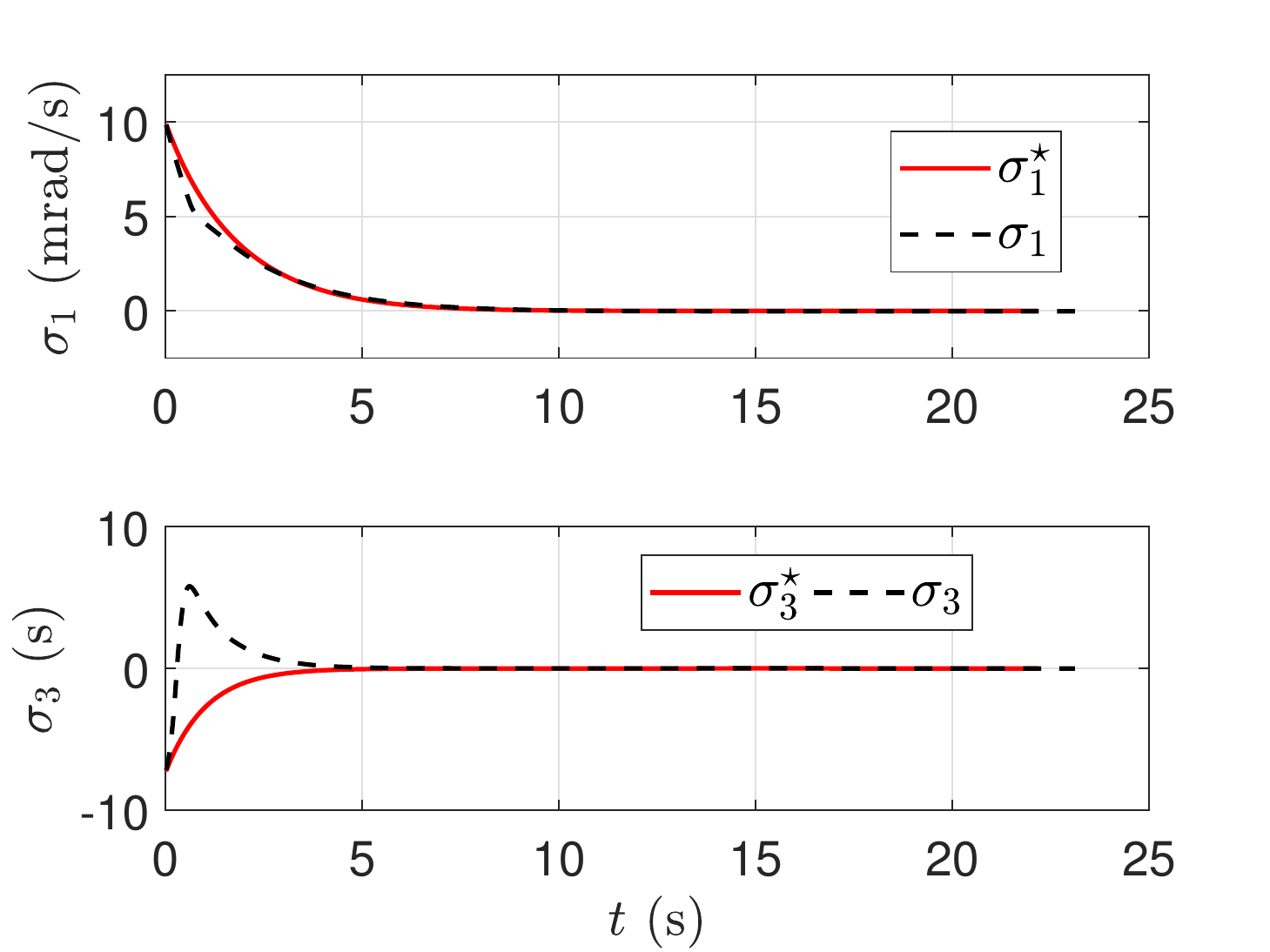}
		\caption{Sliding manifolds (error profiles).}
		\label{fig:in5surface}
	\end{subfigure}
	\begin{subfigure}[t]{.5\textwidth}
		\centering
		\includegraphics[width=\textwidth]{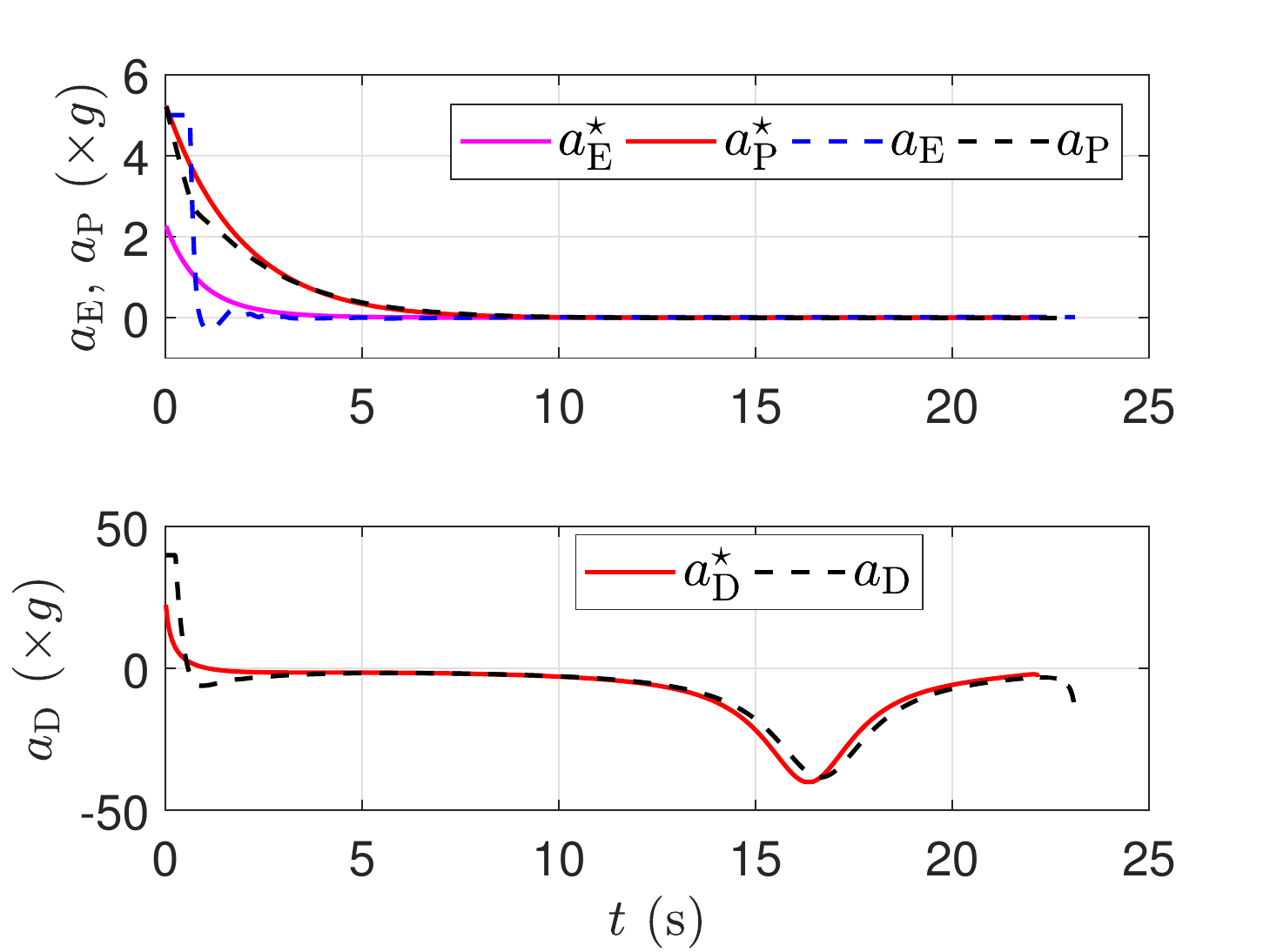}
		\caption{Lateral accelerations (steering controls).}
		\label{fig:in5accn}
	\end{subfigure}%
	\begin{subfigure}[t]{.5\textwidth}
		\centering
		\includegraphics[width=\textwidth]{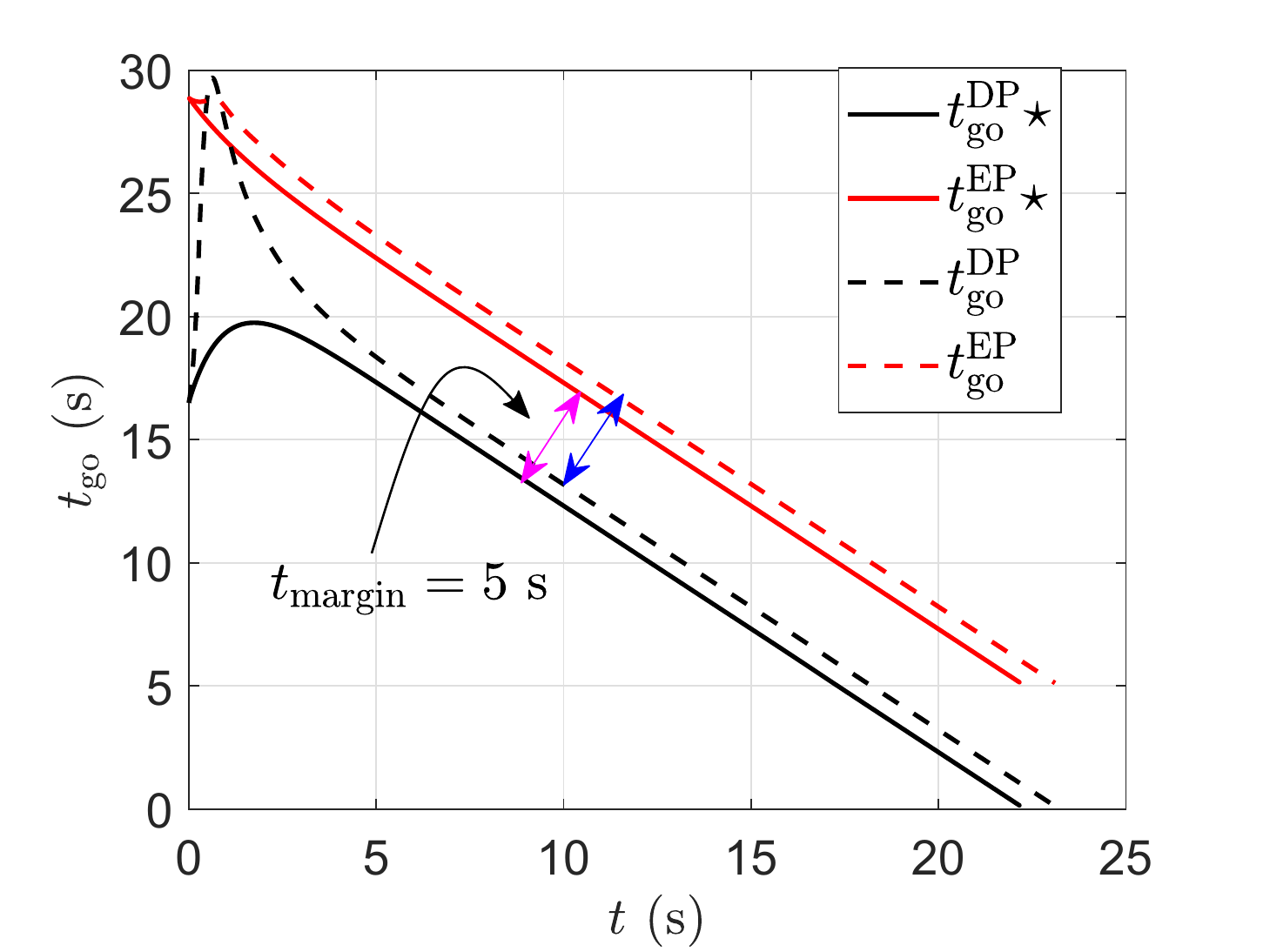}
		\caption{Time-to-go.}
		\label{fig:in5tgo}
	\end{subfigure}
	\caption{Direct cooperation with time-margin set at $5$ s: A $\star$ indicates that $a_\mathrm{P}$ is known.}
	\label{fig:in5s}
\end{figure}

A scenario when the evader-defender team uses direct cooperation is portrayed in \Cref{fig:in5s}. The initial $\lambda_\mathrm{EP}$ is taken as $-30^\circ$ with heading angles, $\gamma_\mathrm{E}=0^\circ$, and $\gamma_\mathrm{P}=120^\circ$. The time-margin, $t_\mathrm{margin}$, is chosen as $5$ s. Trajectories of the agents are shown in \Cref{fig:in5traj}, wherein the defender-pursuer engagement is still of a tail-chase nature. In this scenario, the trajectories of all the agents are slightly different when $a_\mathrm{P}$ is known. The slight change in the pursuer's trajectory can be attributed to the change in the evader's trajectory when $a_\mathrm{P}$ is known. Consequently, the defender also takes the requisite course to intercept the pursuer within the pre-specified time-margin. With a proper selection of design parameters, sliding mode is enforced on \eqref{eq:s1} and~\eqref{eq:s2direct} quickly, as evidenced by \Cref{fig:in5surface}, rendering the pursuer non-maneuvering. This is utilized by the defender, which now possesses the information of the pursuer-evader engagement, and adjusts its course to intercept the pursuer in a specified time-margin. The behaviors of the lateral accelerations, given by \eqref{eq:aE} and~\eqref{eq:aDdirectAgg}, are portrayed in \Cref{fig:in5accn}, from which it can be observed that the control demands are similar to those in the previous case of indirect cooperation. The change in the agents' trajectories, when $a_\mathrm{P}$ is known, can be attributed to the lateral acceleration profiles in the beginning of the engagement, as shown in \Cref{fig:in5accn}. This also reflects in the time-to-go of the adversaries, as shown in \Cref{fig:in5tgo}. Although the time-to-go profiles of the pursuer-evader and the defender-pursuer engagements take different values depending on whether $a_\mathrm{P}$ is known or estimated, a constant time-margin is maintained in both the cases once the errors converge to zero.
\begin{figure}[h!]
	\centering
	\begin{subfigure}[t]{.5\textwidth}
		\centering
		\includegraphics[width=\textwidth]{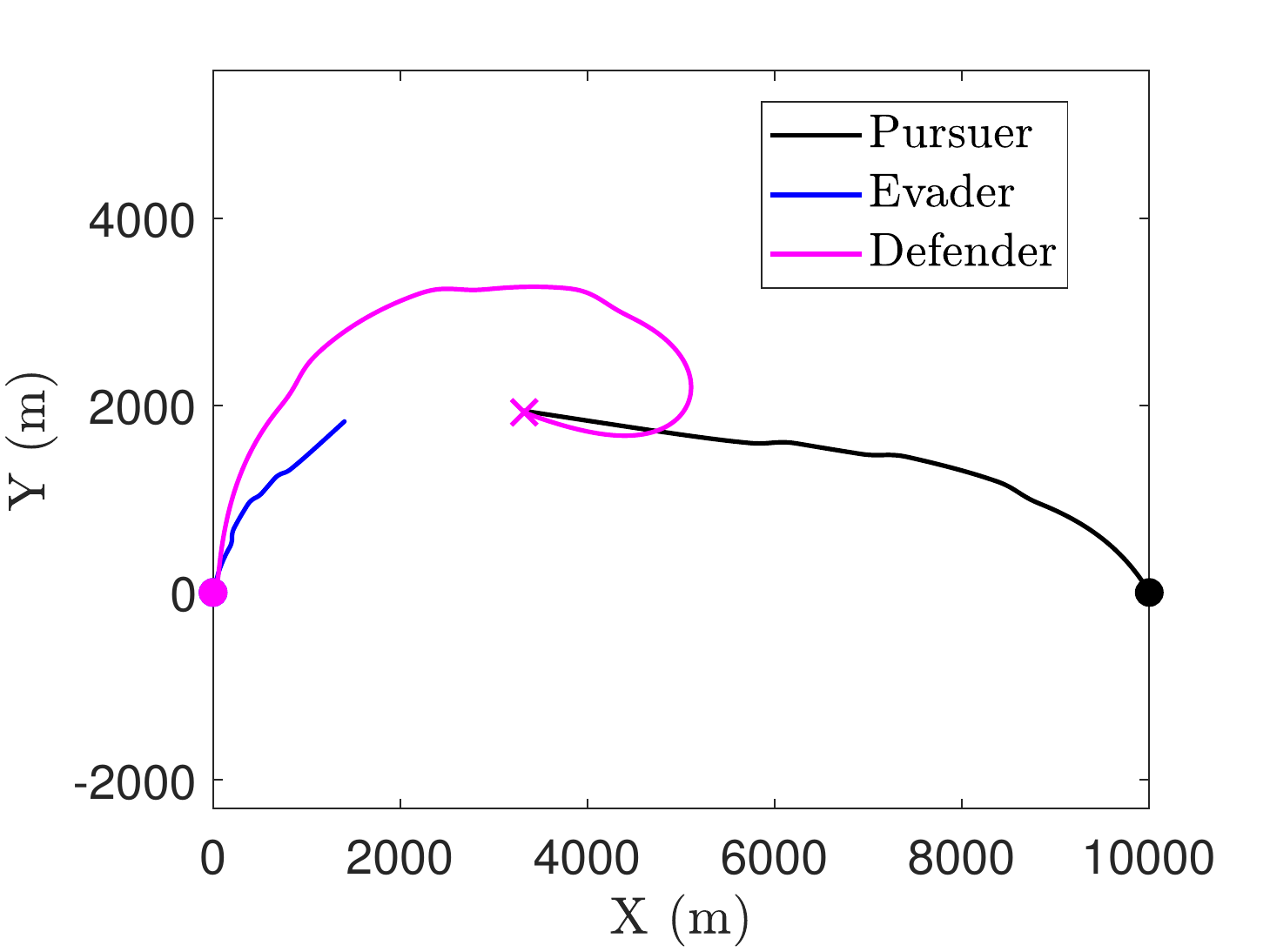}
		\caption{Trajectories.}
		\label{fig:in3traj}
	\end{subfigure}%
	\begin{subfigure}[t]{.5\textwidth}
		\centering
		\includegraphics[width=\textwidth]{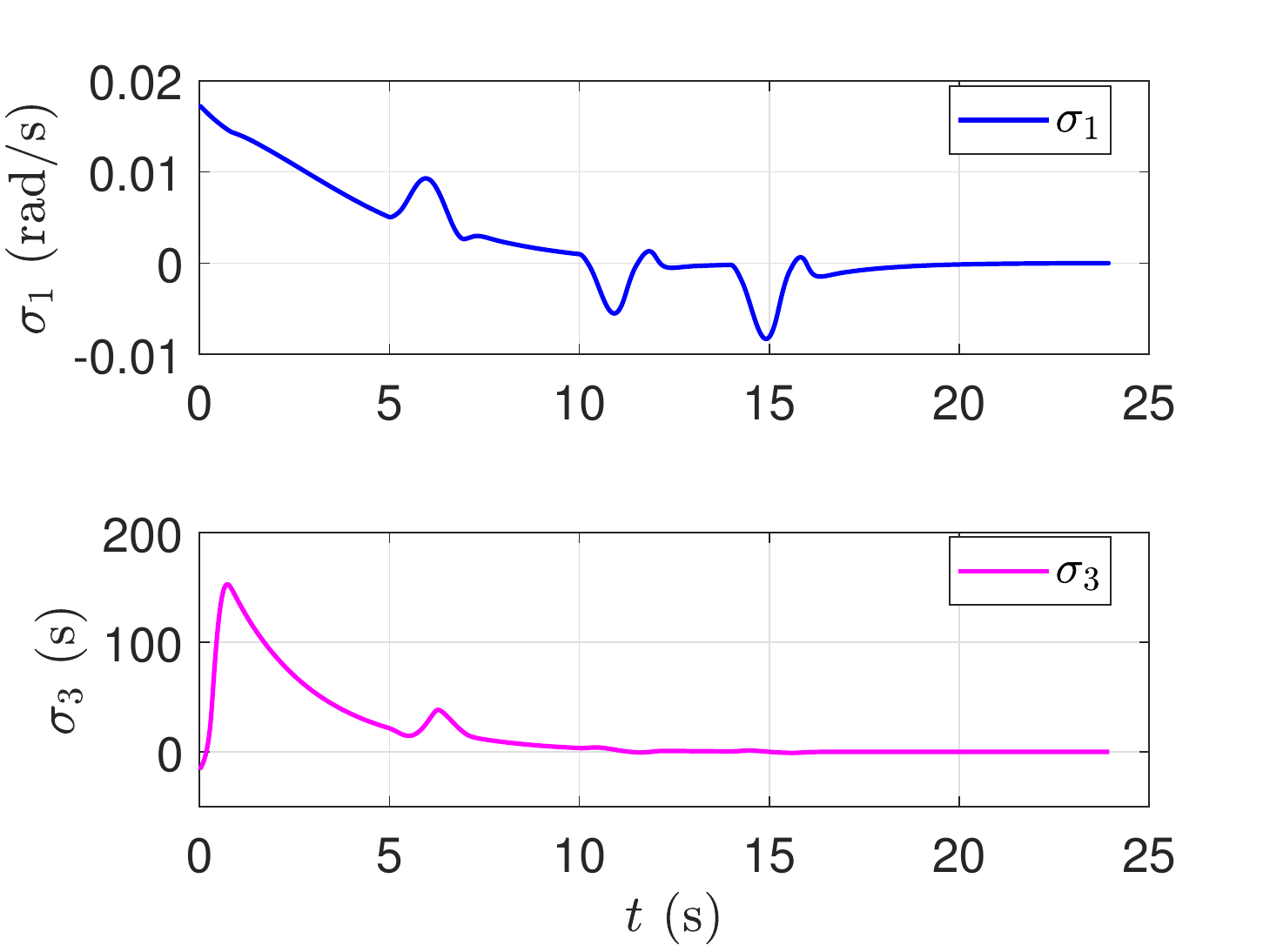}
		\caption{Sliding manifolds (error profiles).}
		\label{fig:in3surface}
	\end{subfigure}
	\begin{subfigure}[t]{.5\textwidth}
		\centering
		\includegraphics[width=\textwidth]{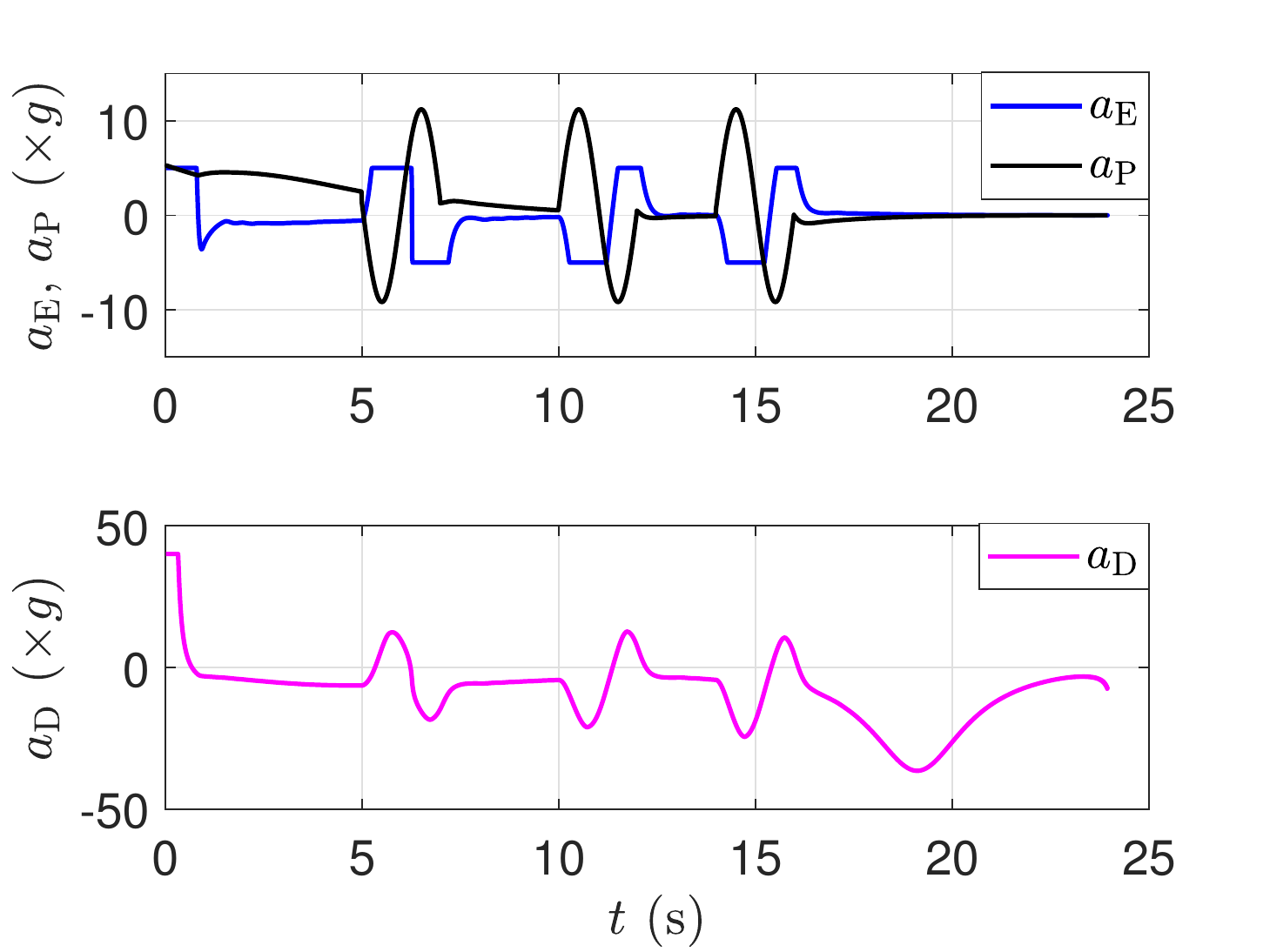}
		\caption{Lateral accelerations (steering controls).}
		\label{fig:in3accn}
	\end{subfigure}%
	\begin{subfigure}[t]{.5\textwidth}
		\centering
		\includegraphics[width=\textwidth]{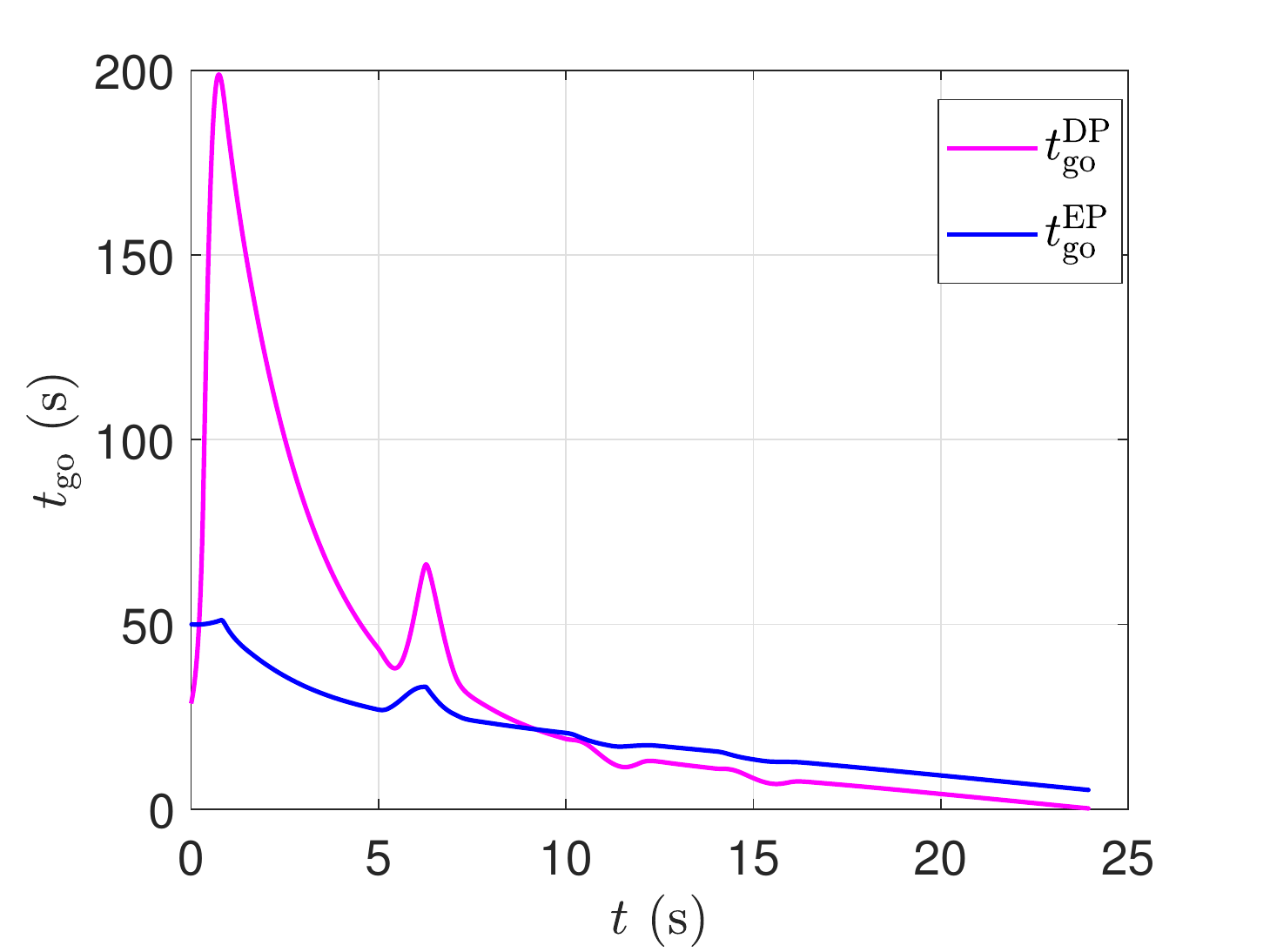}
		\caption{Time-to-go.}
		\label{fig:in3tgo}
	\end{subfigure}
	\caption{Direct cooperation with time-margin set at $3$ s.}
	\label{fig:in3s}
\end{figure}

In the three-agent engagement shown in \Cref{fig:in3s}, the initial LOS, $\lambda_\mathrm{EP}=0^\circ$, and the heading angles of the adversaries are $\gamma_\mathrm{E}=60^\circ$, and $\gamma_\mathrm{P}=120^\circ$, with the time-margin reduced to $3$ s. Suppose the pursuer is somehow alerted that the defender is on its tail, and makes sudden evasive maneuvers at arbitrary intervals. The evader responds to these sudden changes in the pursuer's maneuver and adjusts its trajectory, forcing a zero LOS rate with respect to the pursuer. The defender utilizes this opportunity, and makes necessary maneuvers to maintain the prescribed time-margin of $t_\mathrm{margin}$ sec. \Cref{fig:in3traj} depicts the trajectories of the adversaries. Even for a smaller time-margin and the pursuer's sudden maneuvers, the pursuer misses the evader by a radial separation of $1911.6$ m at the time of its capture by the defender. The error profiles are shown in \Cref{fig:in3surface}, with a similar behavior as in the previous cases. Control signals of the agents are depicted in \Cref{fig:in3accn}, showing the sudden maneuvers of the pursuer (taken as $10+100\sin(\pi t)$) at various time instants during the course of engagement, and the maneuvers executed by the evader-defender team in response to such behavior. Since $t_\mathrm{margin}$ is set at a low value, $r_\mathrm{EP}$ decreases as the engagement proceeds. From the pursuer's standpoint, remaining on the collision course with the evader is a favorable outcome, so it does not maneuver when $r_\mathrm{EP}$, and hence, $t_\mathrm{go}^\mathrm{EP}$ is decreasing. The defender finds the pursuer non-maneuvering, computes $t_\mathrm{go}^\mathrm{DP}$ accurately, and suitably adjusts its trajectory that will lead to the pursuer's capture in the prescribed time-margin. It can be observed from \Cref{fig:in3tgo} that the initial estimate of $t_\mathrm{go}^\mathrm{EP}$ is $50$ s, which drops to $27$ s in the endgame, and $t_\mathrm{go}^\mathrm{DP}$ changes in accordance with changes in $t_\mathrm{go}^\mathrm{EP}$ to maintain a constant $t_\mathrm{margin}$.
\begin{figure}[h!]
	\centering
	\begin{subfigure}[t]{.5\textwidth}
		\centering
		\includegraphics[width=\textwidth]{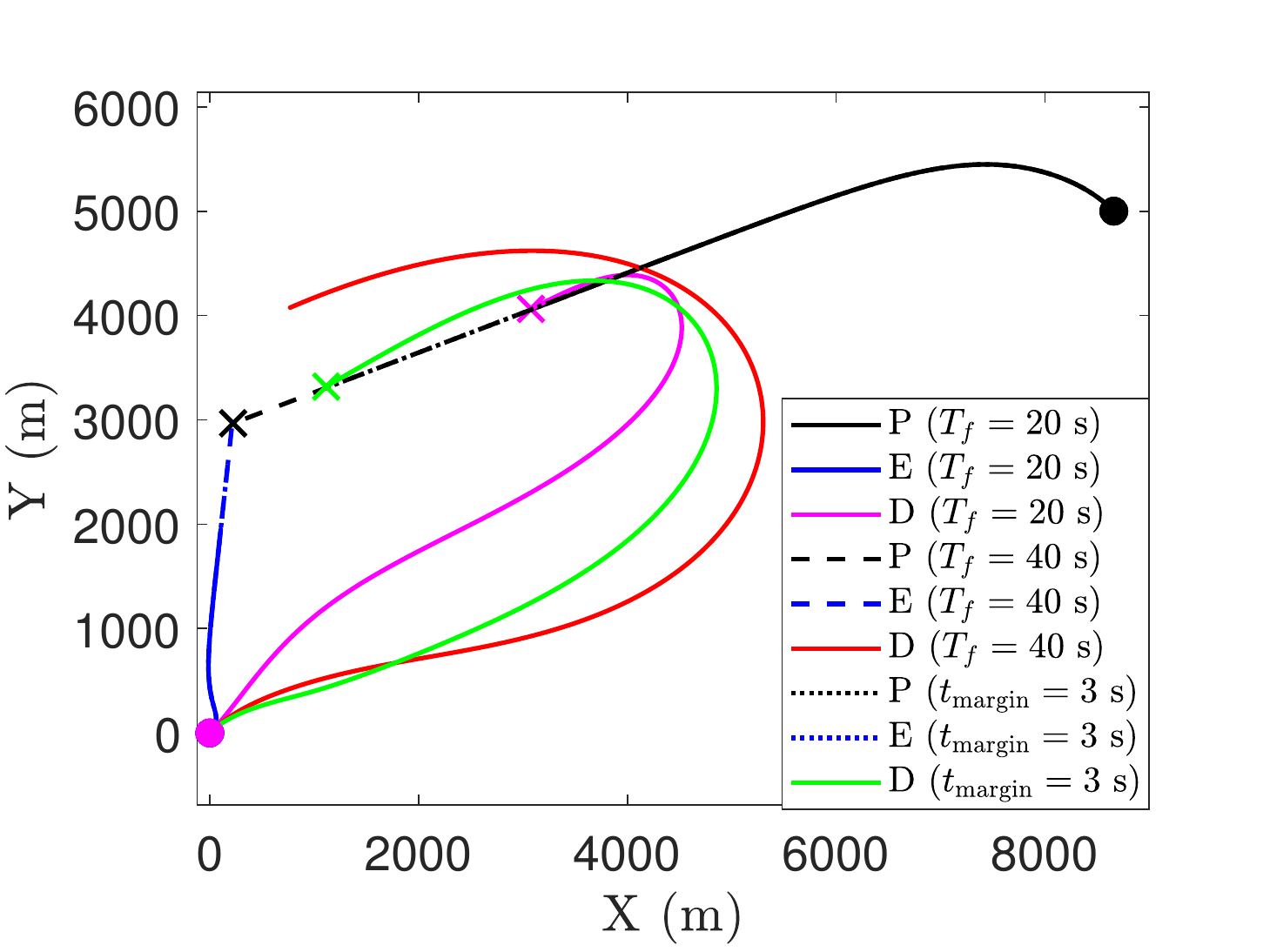}
		\caption{Trajectories.}
		\label{fig:ex20traj}
	\end{subfigure}%
	\begin{subfigure}[t]{.5\textwidth}
		\centering
		\includegraphics[width=\textwidth]{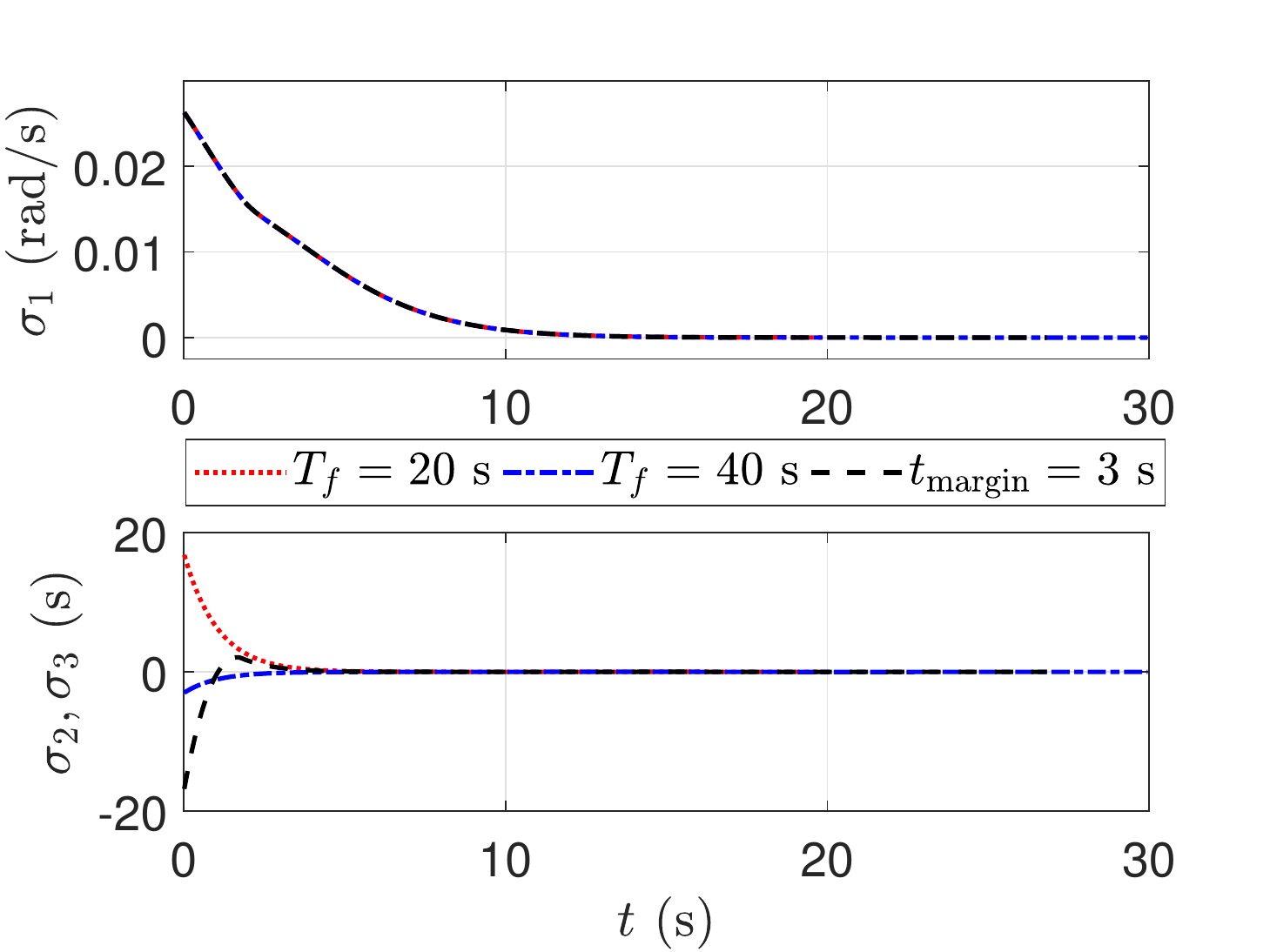}
		\caption{Sliding manifolds (error profiles).}
		\label{fig:ex20surface}
	\end{subfigure}
	\begin{subfigure}[t]{.5\textwidth}
		\centering
		\includegraphics[width=\textwidth]{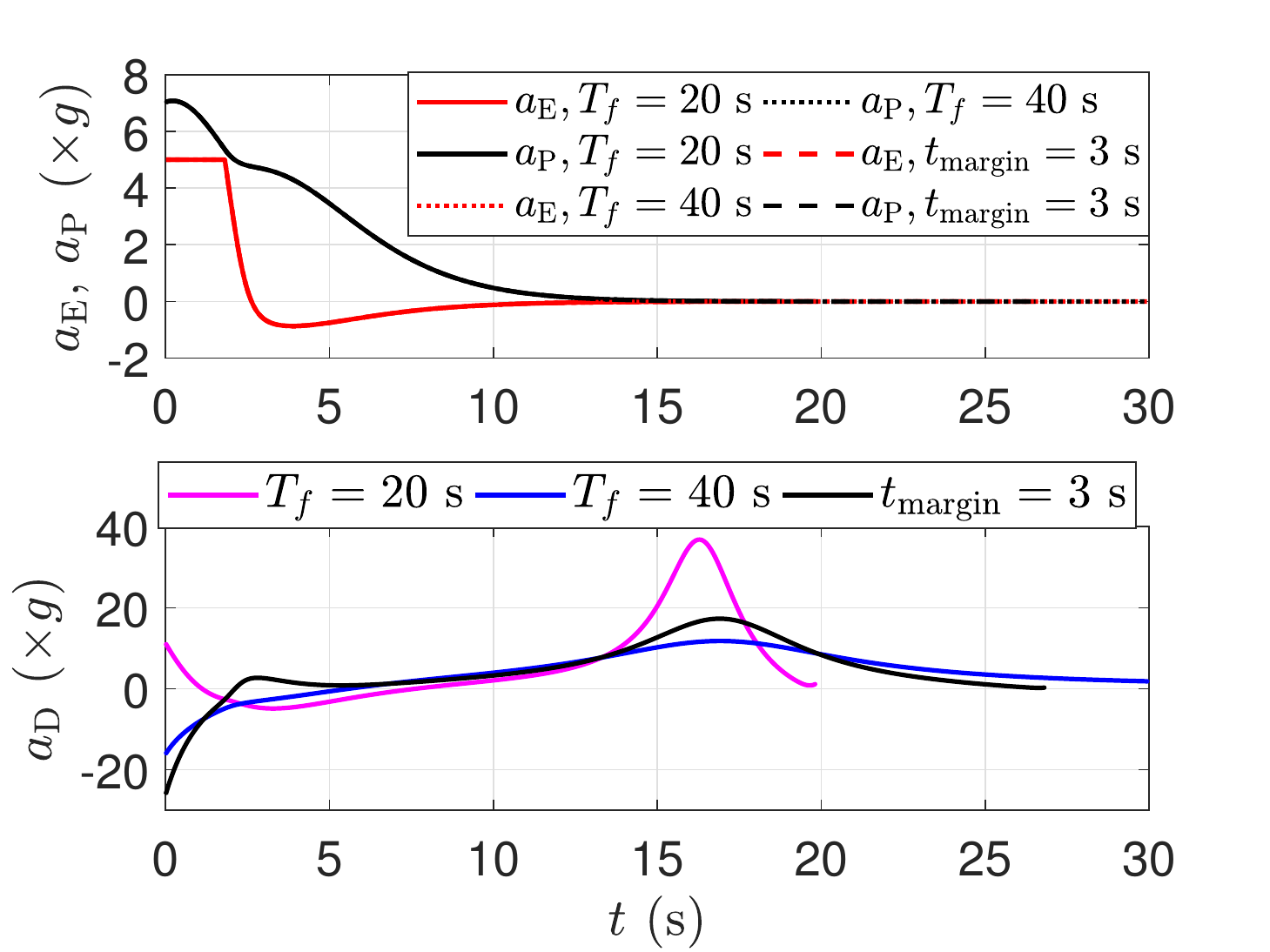}
		\caption{Lateral accelerations (steering controls).}
		\label{fig:ex20accn}
	\end{subfigure}%
	\begin{subfigure}[t]{.5\textwidth}
		\centering
		\includegraphics[width=\textwidth]{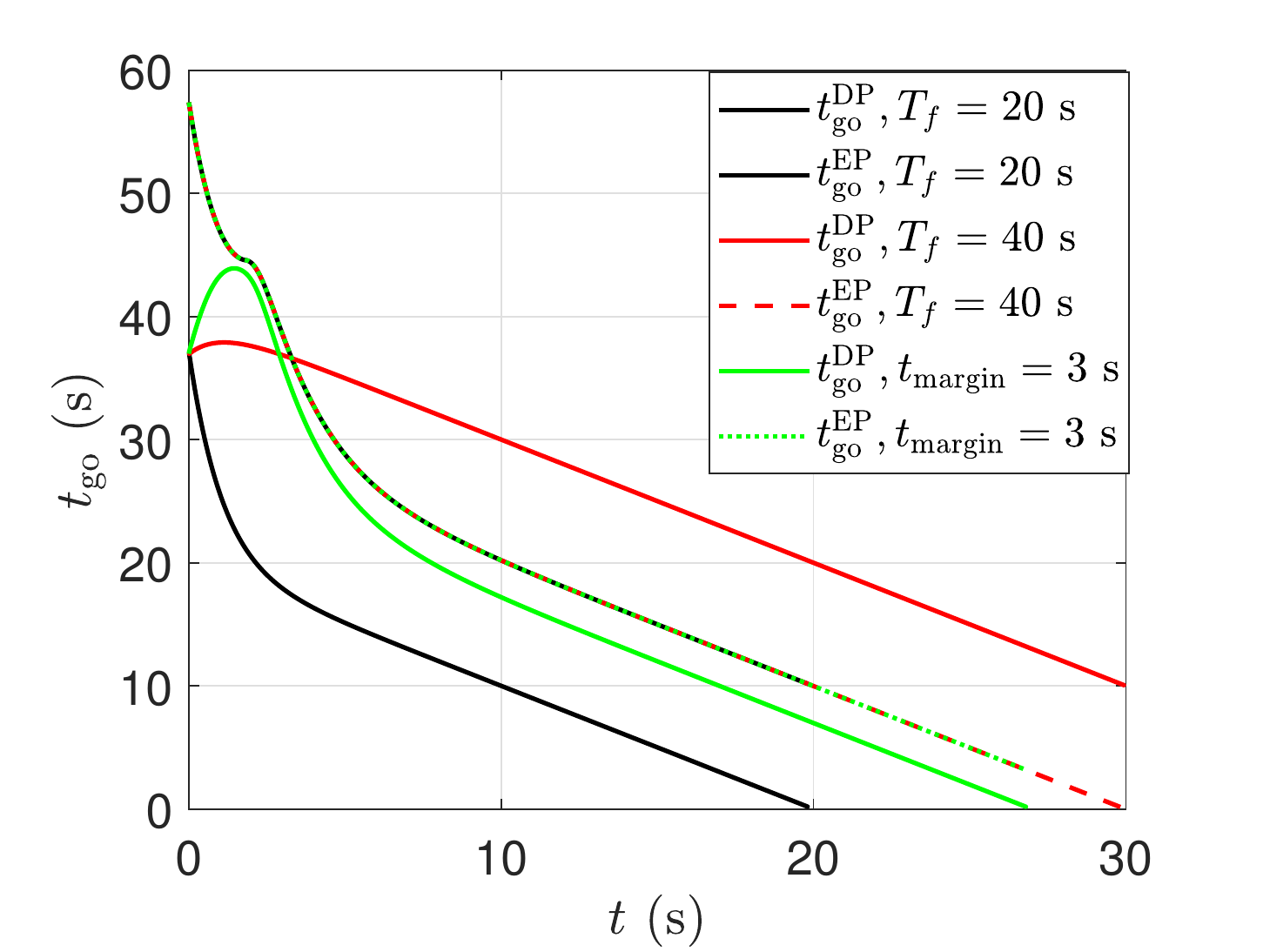}
		\caption{Time-to-go.}
		\label{fig:ex20tgo}
	\end{subfigure}
	\caption{Direct and indirect cooperation for various impact time requirements.}
	\label{fig:ex20s}
\end{figure}

\Cref{fig:ex20s} illustrates the case when the initial LOS, ${\lambda}_\mathrm{EP}$, is changed to $30^\circ$, keeping rest of the parameters same as that in \Cref{fig:ex27s}. In this configuration, the estimated time-to-go for the pursuer-evader engagement, $t_\mathrm{go}^\mathrm{EP}$, is $57.39$ s at the beginning of the engagement. If the evader-defender team uses indirect cooperation to neutralize the pursuer, a sufficiently small impact time has to be provided to the defender. This is necessary as $t_\mathrm{go}^\mathrm{EP}$ is subject to change in accordance with the pursuer's maneuver. For the defender's desired interception time set at $20$ s, the pursuer is neutralized, for which the trajectories of the adversaries are depicted in \Cref{fig:ex20traj}. The case when the pursuer intercepts the evader, leading to failure of the mission, is also shown in \Cref{fig:ex20traj}, where the defender's impact time is chosen as $40$ s. Although an impact time of $40$ s is less than the initial $t_\mathrm{go}^\mathrm{EP}$ of $57.39$ s, this estimate is susceptible to errors due to the pursuer's maneuver. It can be inferred that specifying an explicit desired impact time at the beginning is a tricky thing to do. \Cref{fig:ex20tgo} shows that  $t_\mathrm{go}^\mathrm{EP}=30$ s, which is accurately known only when the pursuer and the evader are on the collision course. The evader may therefore choose to launch the defender after the pursuer has become non-maneuvering, in case of indirect cooperation.

However, if the evader-defender team uses direct cooperation with a suitable time margin of $3$ s, the defender captures the pursuer in $27$ s, whose trajectories, and time-to-go are presented in \Cref{fig:ex20traj}, and \Cref{fig:ex20tgo}, respectively. Thus, even if $t_\mathrm{go}^\mathrm{EP}$ changes later in the engagement, the defender adjusts its trajectory to maintain the time margin. In both the cases, the profiles of error and lateral accelerations are similar. The error variables vanish quickly, as evidenced by \Cref{fig:ex20surface}, after which the demands of the lateral accelerations for all the agents reduce. It is worth noting that the accelerations of the evader and the pursuer remain same in all the three cases, as observed from \Cref{fig:ex20accn}, while that of the defender changes in accordance with the specified impact time, or time-margin. The lateral accelerations of the pursuer and the evader become zero when they are on the collision path, while that of the defender reduces to zero in the endgame. Effects of the pursuer's maneuver manifests in $t_\mathrm{go}^\mathrm{EP}$, which is less accurate until the pursuer is maneuvering, as illustrated in \Cref{fig:ex27tgo}. Note that with the proposed strategy, there is no restriction on the desired impact time to be greater than the initial time-to-go of the defender. Consequently, the defender performs necessary maneuvers to reduce its time-to-go to intercept the pursuer.

\begin{figure}[h!]
	\centering
	\begin{subfigure}[t]{.5\columnwidth}
		\centering
		\includegraphics[width=\linewidth]{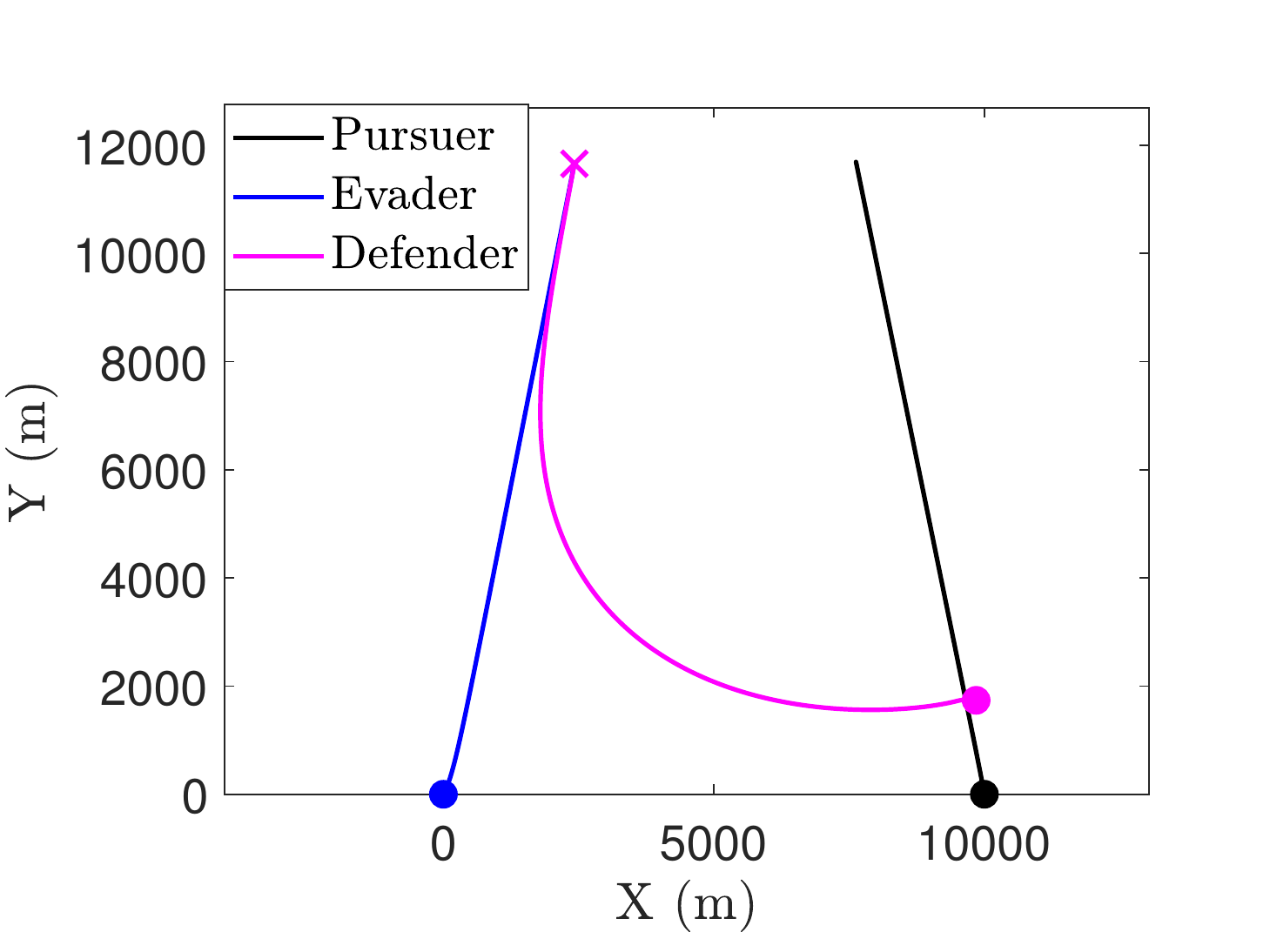}
		\caption{Trajectories.}
		\label{fig:ex40traj}
	\end{subfigure}%
	\begin{subfigure}[t]{.5\columnwidth}
		\centering
		\includegraphics[width=\linewidth]{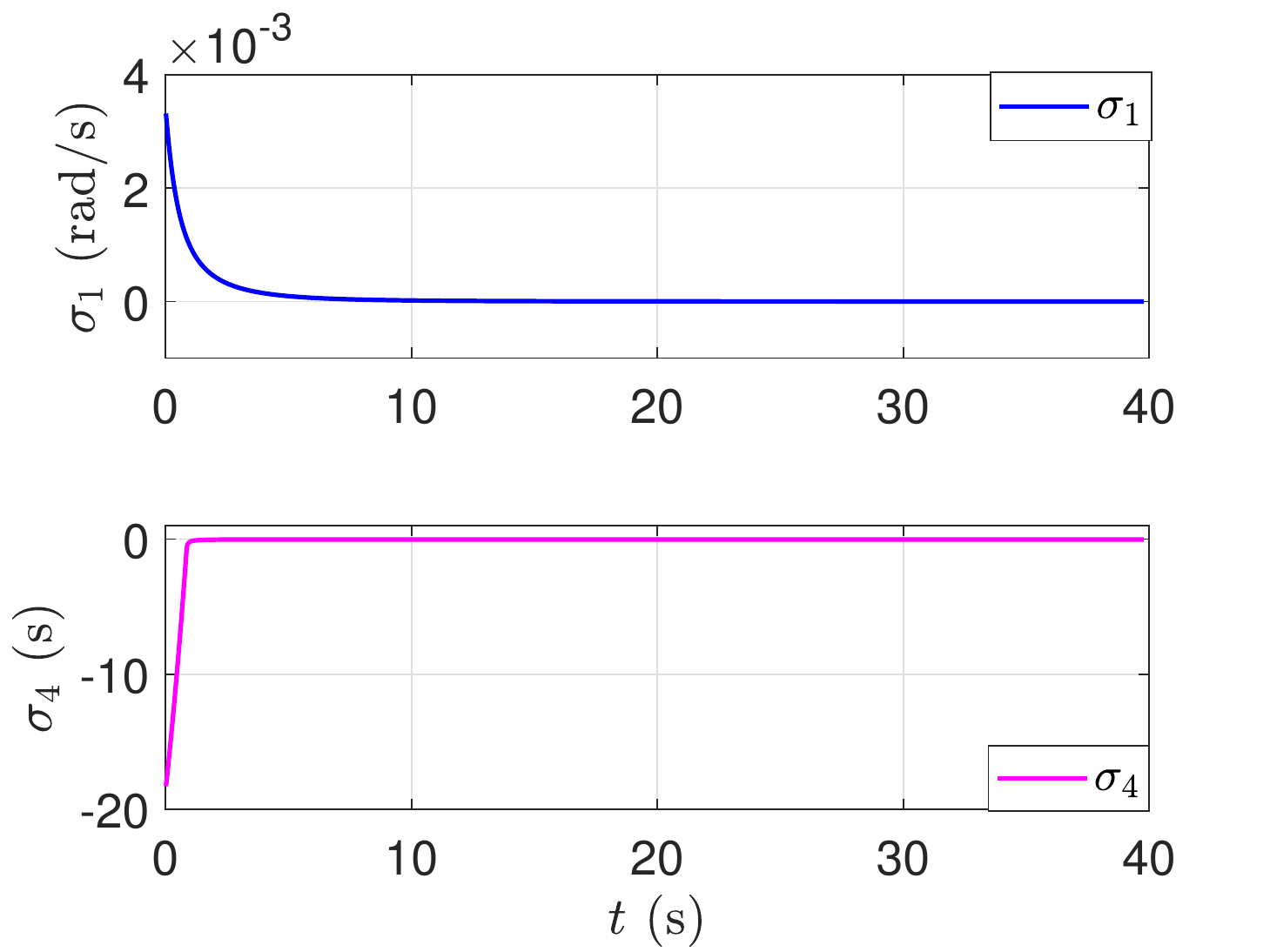}
		\caption{Sliding manifolds (error profiles).}
		\label{fig:ex40surface}
	\end{subfigure}
	\begin{subfigure}[t]{.5\columnwidth}
		\centering
		\includegraphics[width=\linewidth]{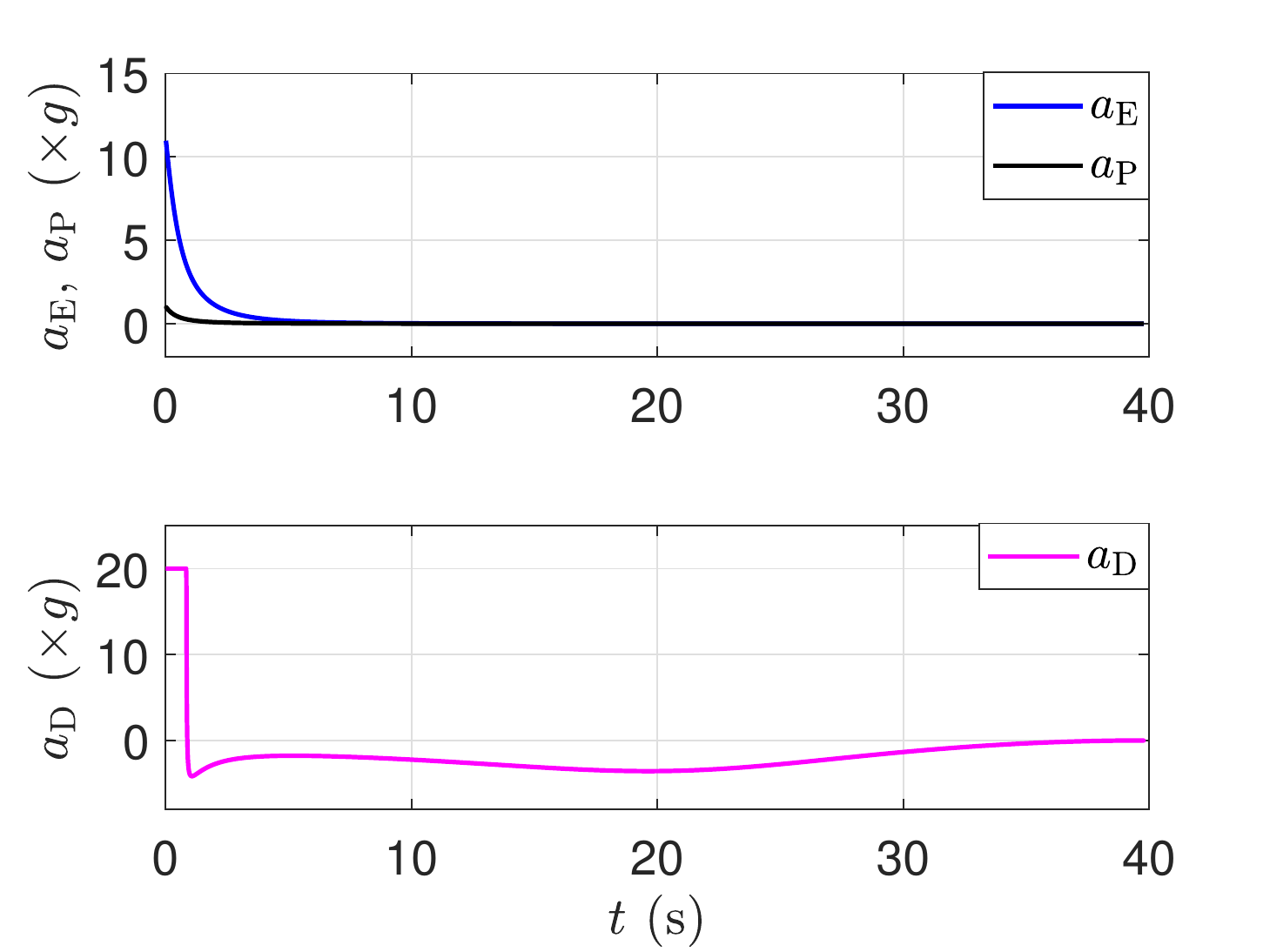}
		\caption{Lateral accelerations (steering controls).}
		\label{fig:ex40accn}
	\end{subfigure}%
	\begin{subfigure}[t]{.5\columnwidth}
		\centering
		\includegraphics[width=\linewidth]{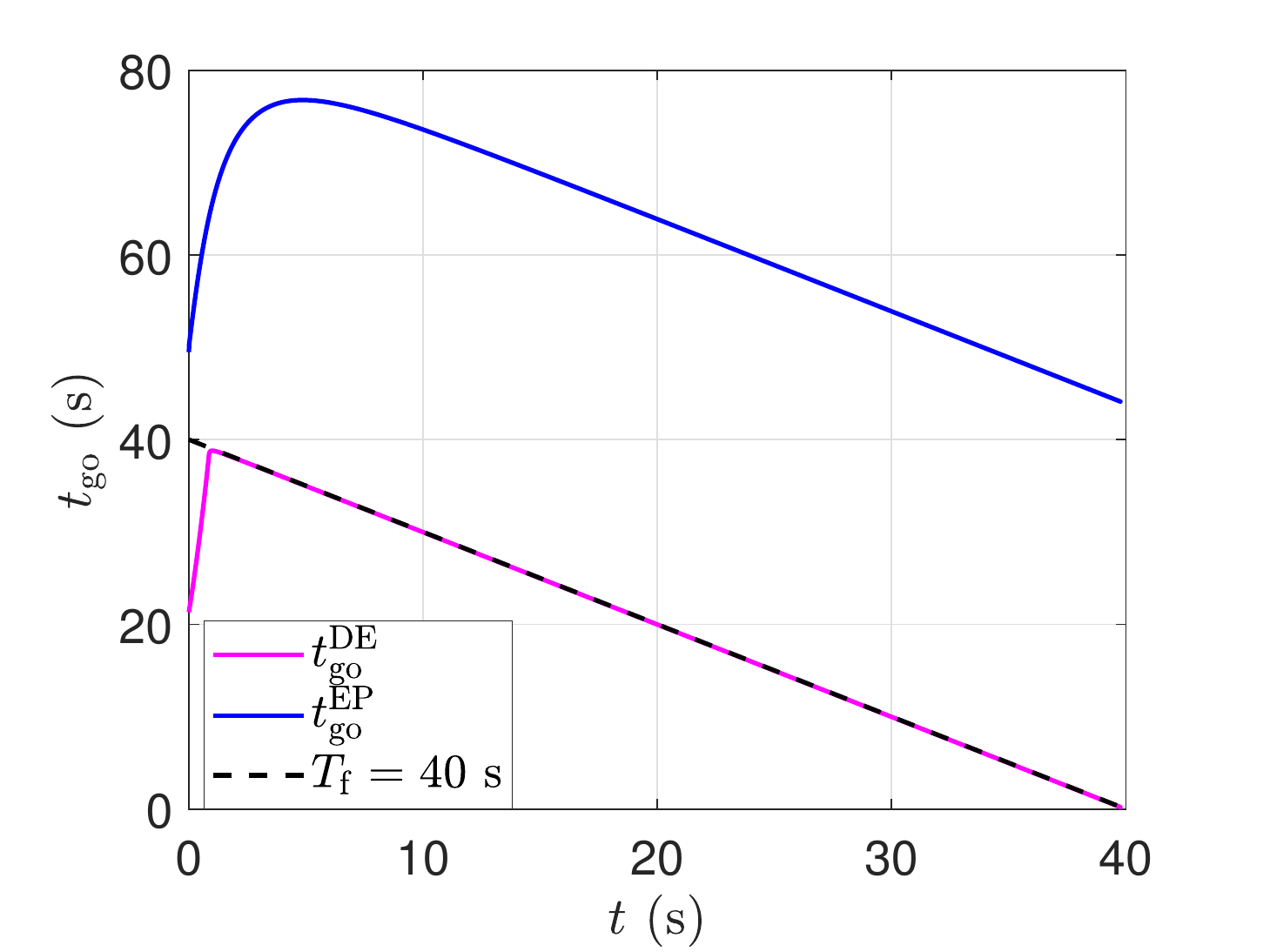}
		\caption{Time-to-go.}
		\label{fig:ex40tgo}
	\end{subfigure}
	\caption{The defender chooses to rendezvous with the evader at $40$ s.}
	\label{fig:ex40s}
\end{figure}
\begin{figure}[h!]
	\centering
	\begin{subfigure}[t]{.5\columnwidth}
		\centering
		\includegraphics[width=\linewidth]{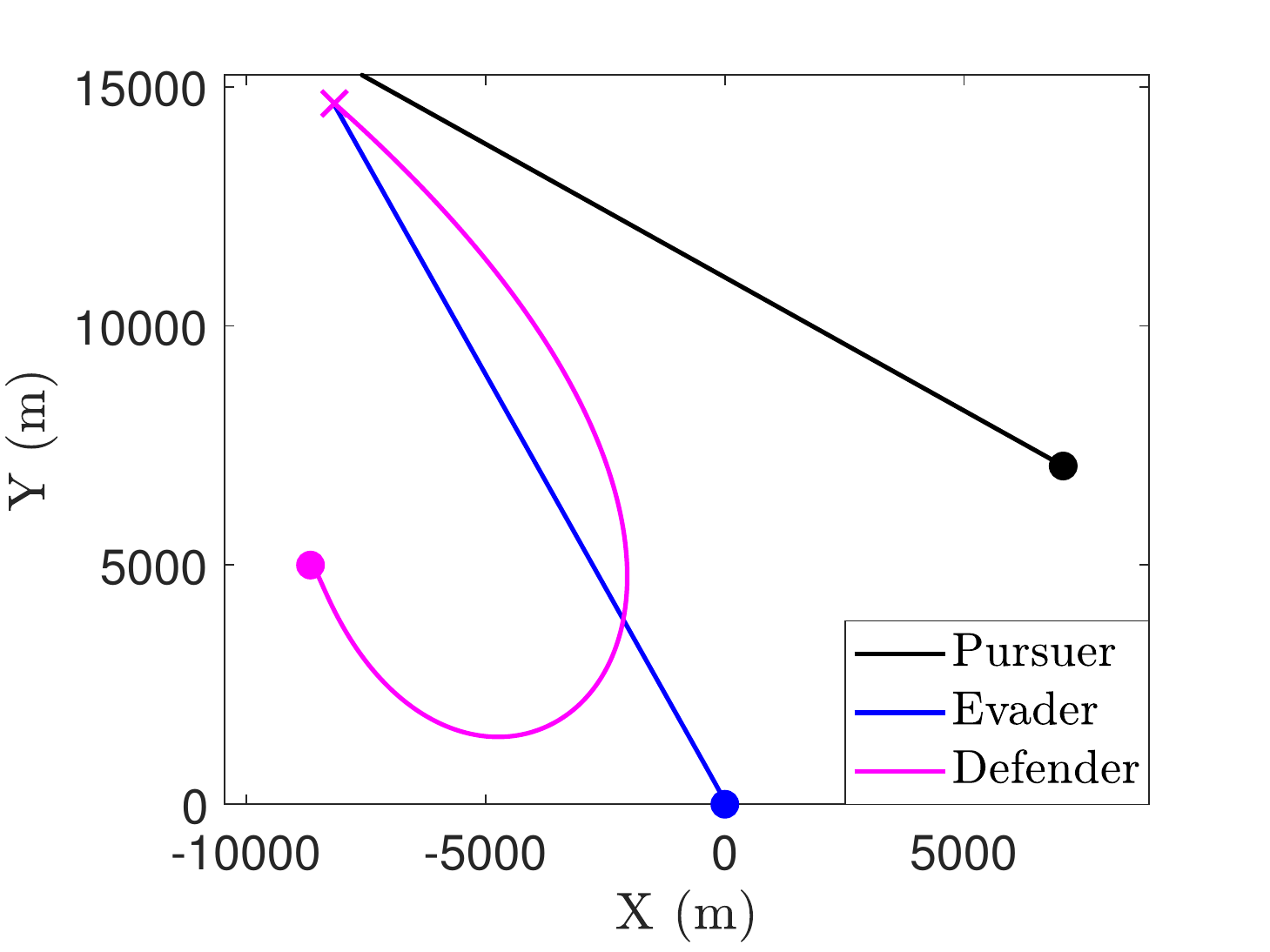}
		\caption{Trajectories.}
		\label{fig:in5Deftraj}
	\end{subfigure}%
	\begin{subfigure}[t]{.5\columnwidth}
		\centering
		\includegraphics[width=\linewidth]{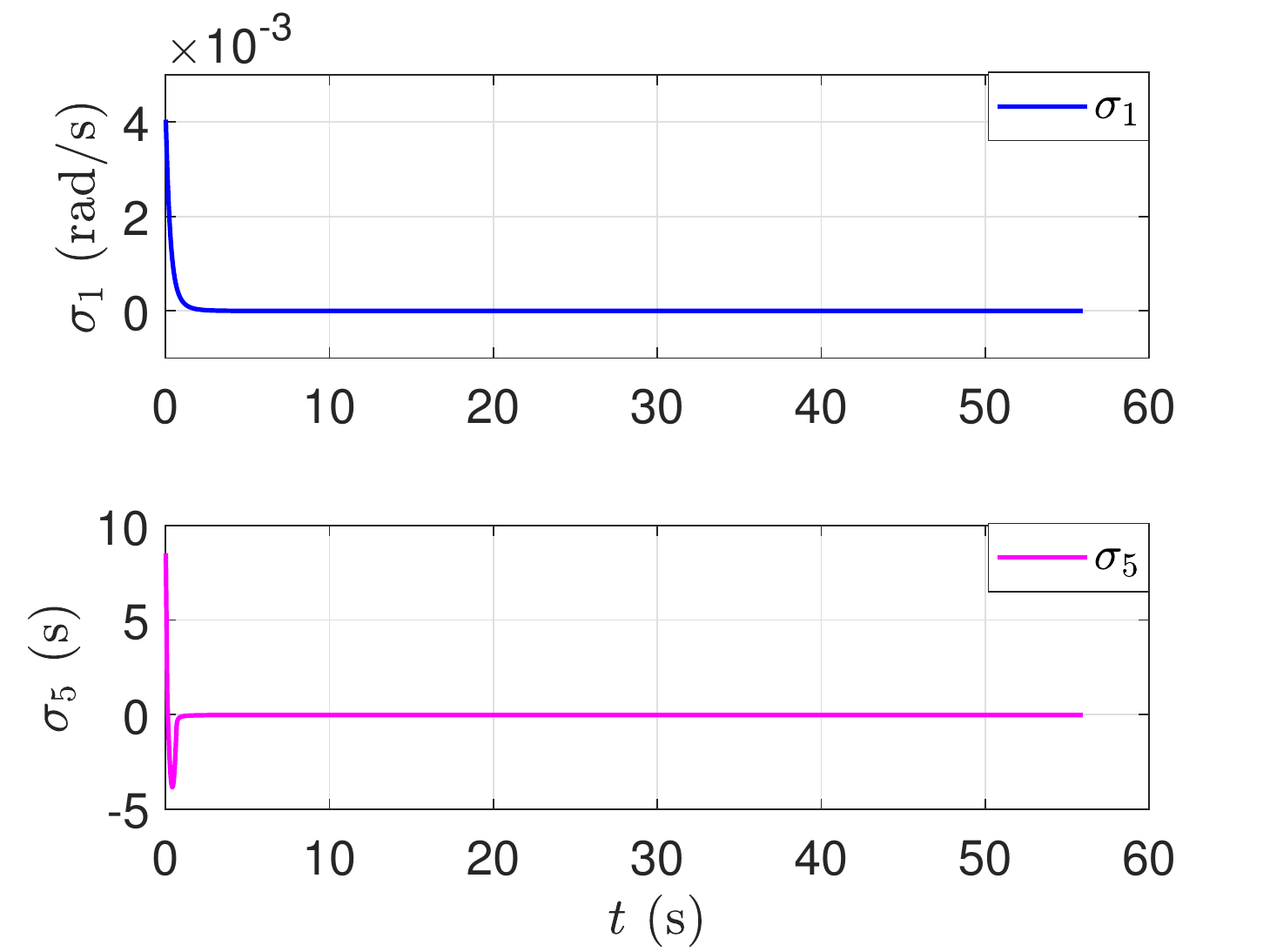}
		\caption{Sliding manifolds (error profiles).}
		\label{fig:in5Defsurface}
	\end{subfigure}
	\begin{subfigure}[t]{.5\columnwidth}
		\centering
		\includegraphics[width=\linewidth]{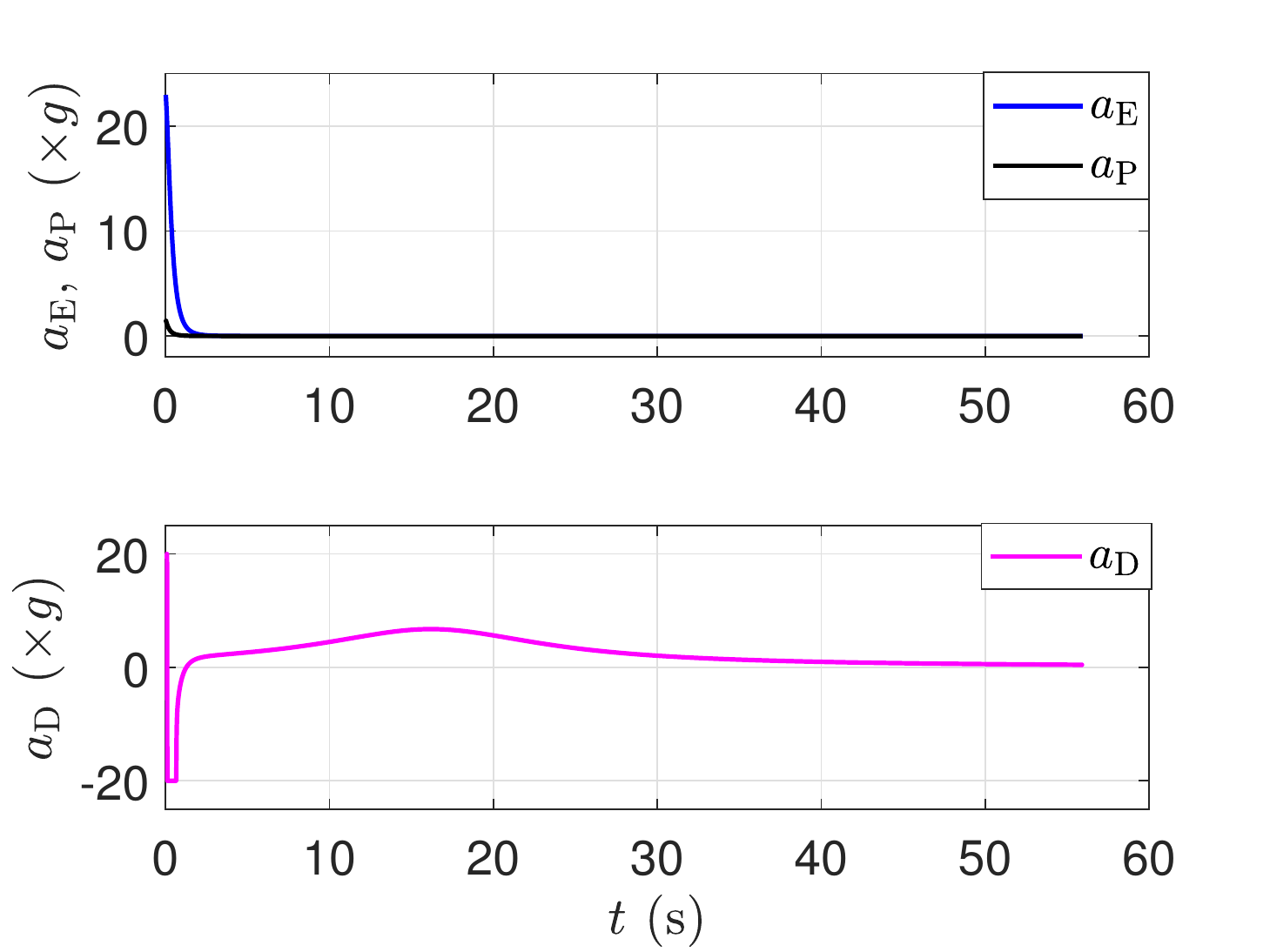}
		\caption{Lateral accelerations (steering controls).}
		\label{fig:in5Defaccn}
	\end{subfigure}%
	\begin{subfigure}[t]{.5\columnwidth}
		\centering
		\includegraphics[width=\linewidth]{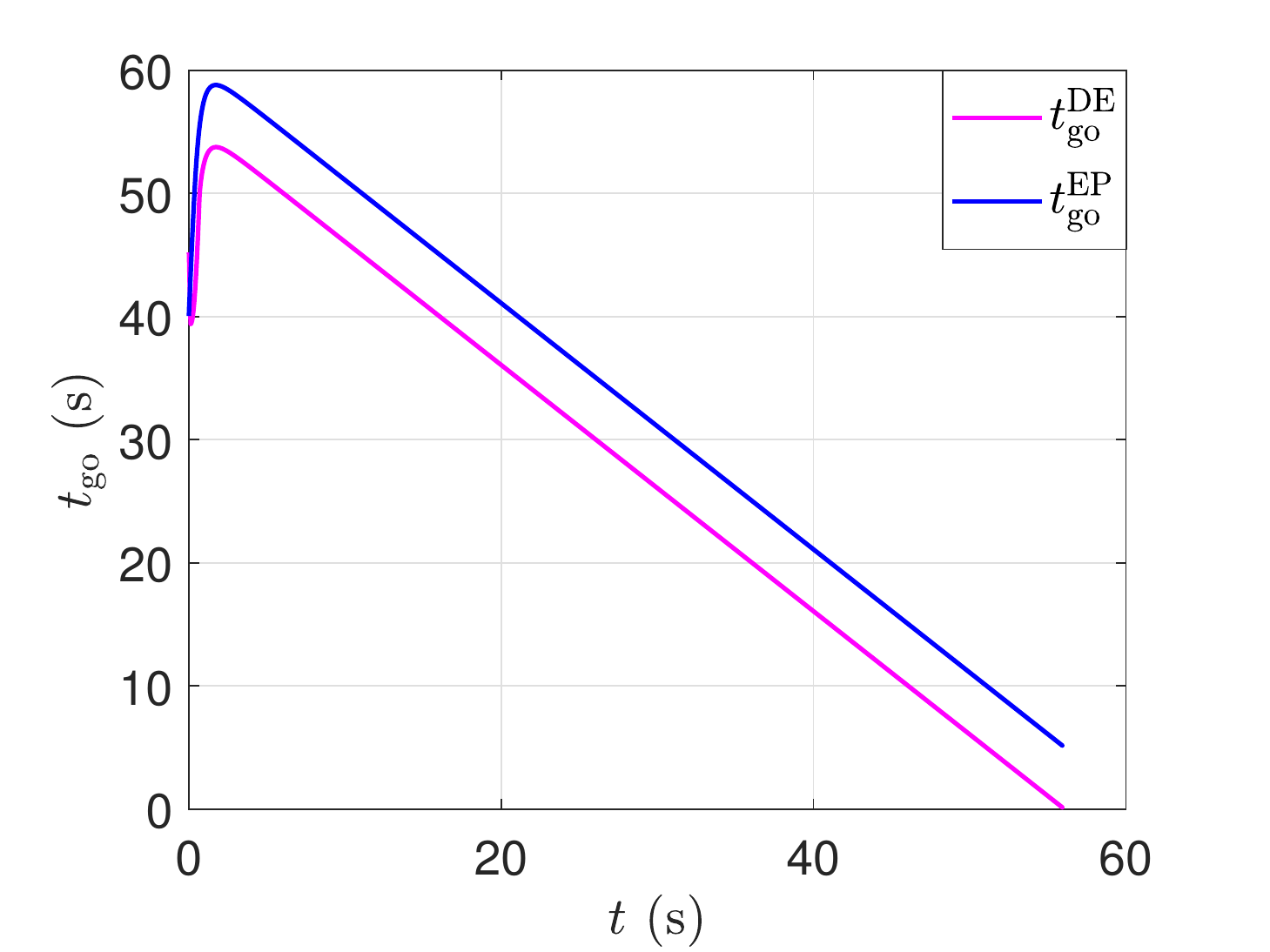}
		\caption{Time-to-go.}
		\label{fig:in5Deftgo}
	\end{subfigure}
	\caption{The defender rendezvous with the evader in a time-margin of $5$ s.}
	\label{fig:in5sDef}
\end{figure}
\begin{figure}[h!]
	\centering
	\begin{subfigure}[t]{.5\columnwidth}
		\centering
		\includegraphics[width=\linewidth]{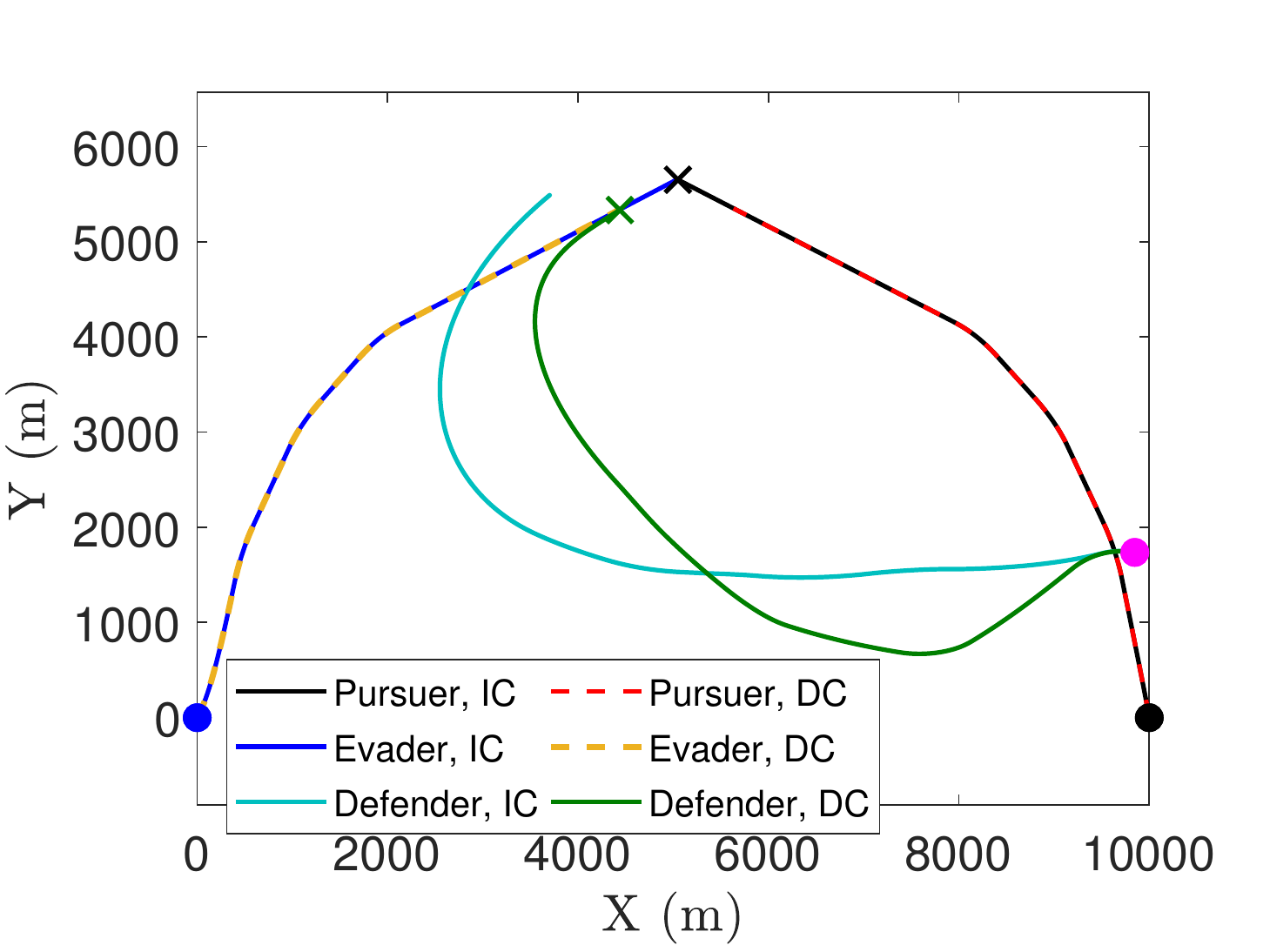}
		\caption{Trajectories.}
		\label{fig:inex2traj}
	\end{subfigure}%
	\begin{subfigure}[t]{.5\columnwidth}
		\centering
		\includegraphics[width=\linewidth]{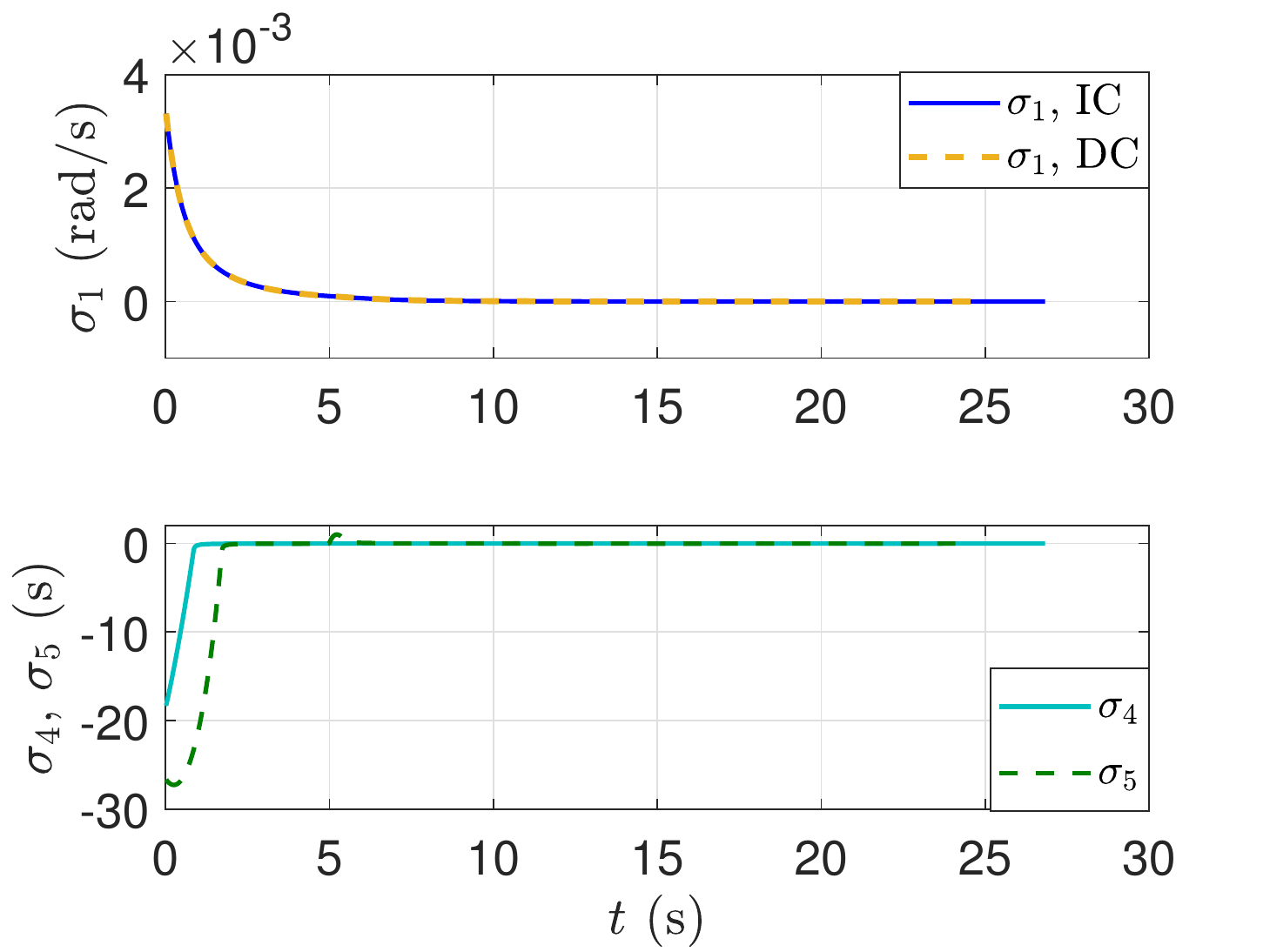}
		\caption{Sliding manifolds (error profiles).}
		\label{fig:inex2surface}
	\end{subfigure}
	\begin{subfigure}[t]{.5\columnwidth}
		\centering
		\includegraphics[width=\linewidth]{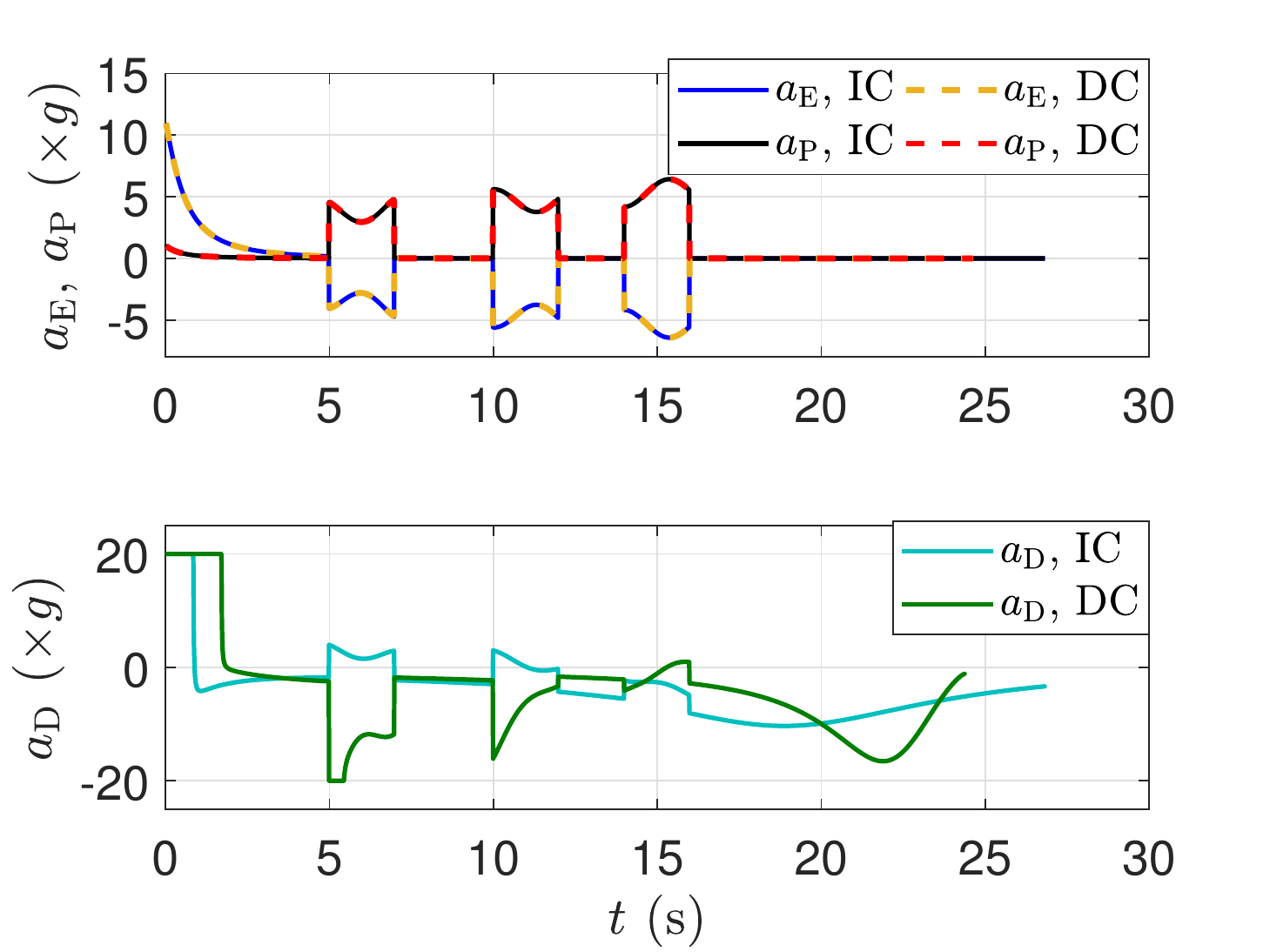}
		\caption{Lateral accelerations (steering controls).}
		\label{fig:inex2accn}
	\end{subfigure}%
	\begin{subfigure}[t]{.5\columnwidth}
		\centering
		\includegraphics[width=\linewidth]{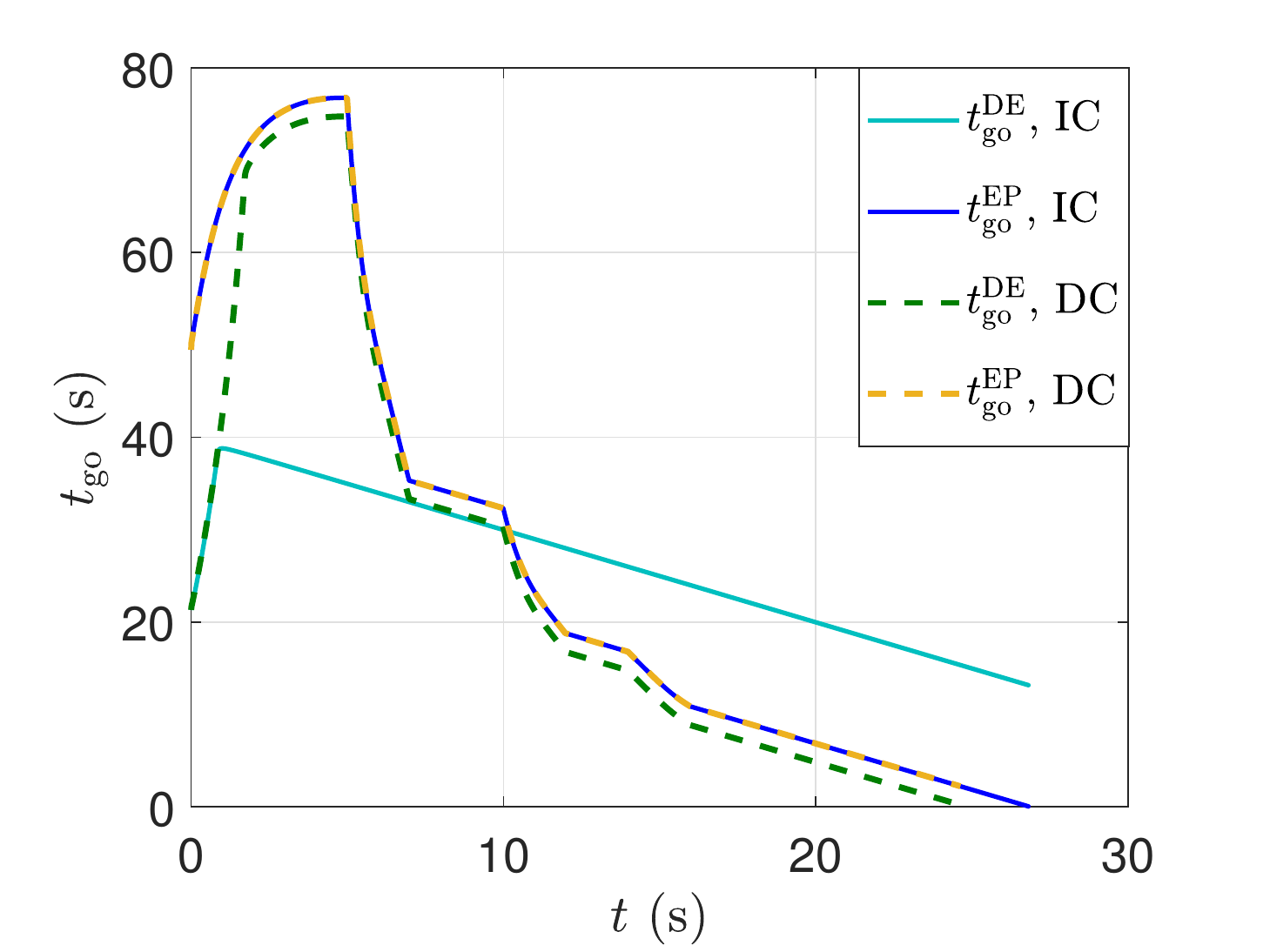}
		\caption{Time-to-go.}
		\label{fig:inex2tgo}
	\end{subfigure}
	\caption{Indirect and direct cooperation between the evader and the defender for various time requirements.}
	\label{fig:inex2s}
\end{figure}

\subsection{Defensive Mode}
The efficacy of the proposed cooperative mechanisms is now illustrated for the case when the defender takes a defensive stance. We assume that the pursuer and the evader move with a speed of $300$ m/s, while the defender has a speed of $400$ m/s. For simplicity, we further assume that $a_\mathrm{P}$ has been identified using the results in \cite{doi:10.2514/1.49515,doi:10.2514/1.51765}. Each agent's steering control (lateral acceleration) is bounded, and capped at $20$ g. In the plots that follow, the evader and the defender are separated at the start unlike in the previous case. The pursuer and the defender are radially separated from the evader by $10$ km. 

In \Cref{fig:ex40s}, the initial LOS angle between the pursuer and the evader, $\lambda_{\mathrm{EP}}$, is $0^\circ$, while that between the defender and the evader, $\lambda_{\mathrm{DE}}$, is $-170^\circ$. Initially, the pursuer's heading is $100^\circ$, while those of the defender and the evader are $170^\circ$ and $60^\circ$, respectively. Under the action of defender's control law, \eqref{eq:aDindirectDef}, the defender is able to rendezvous with the evader using indirect cooperation (IC) at a pre-specified time of capture (set at $40$ s). From the trajectory plot in \Cref{fig:ex40traj} and the error profiles (sliding manifolds) in \Cref{fig:ex40surface}, it is evident that early nullification of $\sigma_4$ makes the pursuer and the evader non-maneuvering. Hence, they appear as moving on a straight line. This is favorable for the pursuer as it can capture the evader with lesser effort. The defender uses this to its advantage by exploiting its higher maneuvering capability, attains the requisite time-constrained deviated pursuit course when $\sigma_4$ becomes zero, to rendezvous with the evader at the desired time. Due to the evader and the pursuer becoming non-maneuvering, their acceleration requirements also reduce to zero (as shown in \Cref{fig:ex40accn}), while that of the defender is large initially to steer itself on the requisite course. Since the time required by the pursuer to capture the evader is estimated to be near $50$ s in the beginning, the defender attempts to rendezvous with the evader in $40$ s, which is sufficiently lower than that value, as seen from \Cref{fig:ex40tgo}, and alters its trajectory accordingly.

\Cref{fig:in5sDef} depicts the case when the defender and the evader use direct cooperation (DC) with a time-margin of ${t}_\mathrm{margin}=5$ s, similar to the earlier case of aggressive defender. The initial values of $\lambda_{\mathrm{EP}}$ and $\lambda_{\mathrm{DE}}$ are $45^\circ$ and $-30^\circ$, respectively, while the agents' headings are $\gamma_\mathrm{P} = 150^\circ, \gamma_\mathrm{E}=100^\circ, \gamma_\mathrm{D}=-50^\circ$. In this case, the defender does not need the time of rendezvous with the evader to be specified beforehand. Rather, it shapes its remaining time profile in a way to maintain a constant time-margin of $5$ s between its own time of rendezvous with the evader and the time at which the pursuer would have captured the evader. Under the action of defender's control law, \eqref{eq:aDdirectDef}, the sliding mode is enforced on $\sigma_5$ in a fixed-time. This also reduces the effort of the agents thereafter, and they attain their requisite courses.

In \Cref{fig:inex2s}, we revisit the scenario depicted in and \Cref{fig:ex20s} and \Cref{fig:ex40s}. In spite of the evader trying to nullify the LOS rate between itself and the pursuer, the latter executes an additional maneuver (taken as $a_\mathrm{P}=30+1.5t-10\sin(0.75\pi t)$) at arbitrary time instants to confuse the evader and the defender. In response to this, the evader also executes counter-maneuvers to keep the LOS rate at zero. However, by executing maneuvers at arbitrary instants, the time needed by the pursuer to capture the evader rapidly varies, and is difficult for the defender to decide its own time of arrival at the evader. The initial $t_\mathrm{go}^\mathrm{EP}$ is nearly $50$ s, and if the defender uses indirect cooperation, it fails to protect the evader because the pursuer is able to capture the evader in nearly $27$ s (owing to its arbitrary maneuvers). However, if the defender uses direct cooperation with a time-margin of ${t}_\mathrm{margin}=2$ s, then the defender can shape its own time profile in accordance with that of the pursuer, and is always able to rendezvous with the evader that much time before the pursuer can do so. The choice to use a particular cooperation, thus, depends on the availability of communication resources during implementation, and the stance the defender is required to take.

\section{Conclusions}\label{sec:conclusion}
In this work, a novel active protection approach was proposed to safeguard the evader from an attacking pursuer. The evader deployed a defender, usually a protective vehicle with similar capabilities as that of the pursuer, which actively cooperated with the evader. The complete guidance design was presented in a nonlinear framework, that allowed the proposed strategies to remain applicable even for initial conditions far away from the collision course. The strategies for the evader-defender team were designed using the concepts of impact time guidance strategies. The use of impact time approach widens the applicability of the existing impact time guidance strategies for applications such as aircraft defense. Two different cooperative mechanisms were designed to address the situation when the pursuer's strategy was unavailable to the evader-defender team, which is closer to practice. Design of the cooperative strategies involved using a control scheme that forced the errors to vanish in a fixed-time, independent of the initial error value. In both cases, the evader's strategy was to nullify its LOS rate with respect to the pursuer to lure the latter on the collision path, and thus render it non-maneuvering. This was used as an advantage by the defender, which then intercepted the non-maneuvering pursuer by means of a deviated pursuit law.

In the indirect cooperative scheme, the evader presented itself as a bait, in order to assist the defender in capturing the pursuer. In a resource constrained environment, indirect cooperation is useful as the evader-defender team do not communicate. However, an impact time, sufficiently less than the initial time-to-go for the pursuer-evader engagement, had to be specified for the defender to ensure success of the mission. Contrary to the no-communication indirect scheme, if the evader-defender could communicate between themselves, direct cooperation proved to be an advantageous technique to protect the evader from the pursuer. Direct cooperation did not require explicit impact time to be fed to the defender, provided the evader shared its engagement information with the defender. A time-margin was designated and the defender maintained a constant time difference between the duration of the various pairs of engagements by adjusting its trajectory. This can be thought of as an adaptive scheme, wherein the defender \emph{adapts} to the impact time on-the-go, without requiring the initial time-to-go for the pursuer-evader/defender-evader engagement. The proposed designs may also prove to be computationally cheaper when the number of agents increases.

Introducing two-way cooperation between the evader-defender team may be of interest in future investigations. Further, incorporating multiple pursuers and multiple defenders to safeguard one or more evaders may also be one of the possible future extensions of this work. 

\bibliographystyle{IEEEtran}
\bibliography{references}

\end{document}